\documentclass[11pt]{article}    
\usepackage{times}
\usepackage{epsfig}
\usepackage{amsfonts}
\usepackage{amsmath}
\usepackage{mathbbol}
\makeatletter
\@addtoreset{equation}{section}
\makeatother

\makeatletter
\@addtoreset{figure}{section}
\makeatother

\title{Topological Concepts in Gauge Theories \footnote{Lectures given at the Autumn School ``Topology and Geometry in Physics'', of the Graduiertenkolleg ``Physical systems with  many degrees of freedom'', University of Heidelberg,  Rot an der Rot, September 24-28, 2001}}
\author{Frieder Lenz \\ \\
 Institute for Theoretical Physics III \\
University of Erlangen-N\"urnberg \\
Staudstrasse 7, 91058 Erlangen, Germany\\ \\
FAU-TP3-04/3} 

\begin{document}

\date{}
\maketitle
\begin{abstract} \normalsize
In these lecture notes, an introduction to topological concepts and methods in studies of gauge field theories is presented. The three paradigms of topological objects, the 
Nielsen-Olesen vortex of the abelian Higgs model, the 't Hooft-Polyakov monopole of the non-abelian Higgs model  and the instanton of Yang-Mills theory, are discussed. The common formal elements in their  construction are emphasized  and their different dynamical roles are exposed. The  discussion of  applications of topological methods to Quantum Chromodynamics focuses on confinement. An account is given of various attempts to relate this phenomenon to topological properties of Yang-Mills theory.  The lecture notes also include an introduction to the underlying concept of homotopy with applications from various areas of physics.  \\
\end{abstract}

\maketitle
\newpage
\baselineskip 13pt
\tableofcontents 
\newcommand{\pb}{p\hspace*{-1.0ex}/}
\newcommand{\mat}[3]{\langle\, #1 \,|\, #2 \,|\, #3 \,\rangle}
\newcommand{\skal}[2]{\langle\, #1 \,|\, #2 \,\rangle}
\newcommand{\ket}[1]{|\, #1 \,\rangle}
\newcommand{\bra}[1]{\langle \, #1 \,|}
\newcommand{\olabbr}[1]{{\rm #1}}
\newcommand{\xp}{(x^{\perp})}
\newcommand{\beq}{\begin{equation}}
\newcommand{\eeq}{\end{equation}}
\newcommand{\beqs}{\begin{displaymath}}
\newcommand{\eeqs}{\end{displaymath}}
\newcommand{\bea}{\begin{eqnarray}}
\newcommand{\eea}{\end{eqnarray}}
\newcommand{\beas}{\begin{eqnarray*}}
\newcommand{\eeas}{\end{eqnarray*}}
\newcommand{\vi}{{\bf x}^{\perp}}
\newcommand{\vj}{{\bf x}^{\prime}^{\perp}}
\baselineskip 14pt
\section{Introduction}
In a fragment \cite{GAUS833} written in the year 1833,  C. F. Gau\ss\ describes  a profound topological result which he  derived from the analysis of a physical problem. He considers the work $W_m$ done by transporting a magnetic monopole (ein Element des ``positiven n\"ordlichen magnetischen Fluidums'') with magnetic charge $g$ along a closed path ${\cal C}_1$  in the magnetic field ${\bf B}$ generated by  a current $I$ flowing along a closed loop ${\cal C}_2$. According to the law of Biot-Savart,  $W_m$ is given by
$$W_m = g \oint_{{\cal C}_1} {\bf B}({\bf s}_1)\, d{\bf s}_1 = \frac{4\pi g}{c}\, I\, lk\{{\cal C}_1,{\cal C}_2\} .$$ 
Gau\ss\ recognized that $W_m$  neither  depends on the geometrical details of the current carrying loop ${\cal C}_2$ nor on those of the  closed path ${\cal C}_1$. 
\vskip 1.6cm
\begin{equation}
  \label{gaulink}
\hspace{4cm}\qquad lk\{{\cal C}_1,{\cal C}_2\} = \frac{1}{4\pi}\oint_{{\cal C}_1} \oint_{{\cal C}_2}  \,
\frac{(d{\bf s}_1 \times  d{\bf s}_2)\cdot{\bf s}_{12}}{|{\bf s}_{12}|^3} 
\end{equation}
\vskip .1cm 
$$\hspace{5cm}\;\quad {\bf s}_{12}={\bf s}_{2}-{\bf s}_{1}\, $$
\vskip-4.6cm
\hspace{2cm} 
\vskip-.2cm
\begin{figure}[h]
\epsfig{file=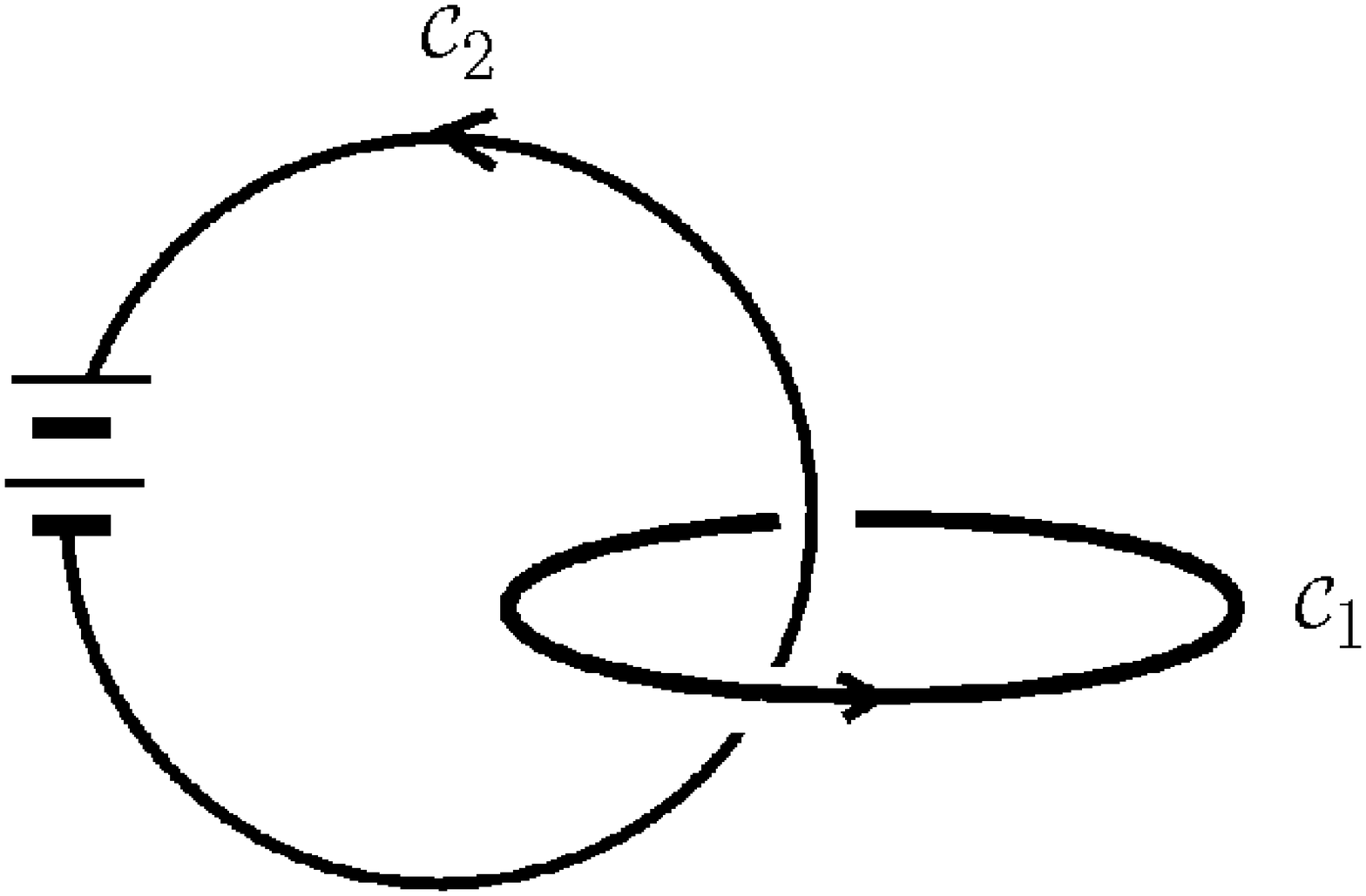,width=0.38\linewidth}
\vskip -.3cm
\caption{Transport of a magnetic charge along ${\cal C}_1$ in the magnetic field generated by a current flowing along ${\cal C}_2$}
\label{link}
\end{figure}
\vskip-3.3cm
\hspace{6.cm} 
\vspace{2.5cm}
\\
Under continuous deformations of these curves, the value of $ lk\{{\cal C}_1,{\cal C}_2\}$, the {\em Linking Number (``Anzahl der Umschlingungen'')},  remains unchanged. This quantity is a topological invariant. It is an integer which counts the (signed) number of intersections of the loop ${\cal C}_1$ with  an arbitrary (oriented) surface in $\mathbb{R}^3$ whose boundary is the loop ${\cal C}_2$ (cf.\ \cite{DFN285,FRAN97}). In the same note, Gau\ss\ deplores the little progress in topology (``Geometria Situs'') since Leibniz's times  who in 1679 postulated ``another analysis, purely geometric or linear which also defines the position (situs), as algebra defines  magnitude''. Leibniz also had in mind applications of this new branch of mathematics to physics.  His attempt to interest a physicist (Christiaan Huygens) in his ideas about topology however was unsuccessful.  Topological arguments made their entrance in   physics with the formulation of the Helmholtz laws of vortex motion (1858) and the circulation theorem by Kelvin (1869) and until today   hydrodynamics  continues to be a fertile field for the development and applications of topological  methods in physics. The success of the topological arguments led Kelvin to seek for  a  description of the  constituents of matter, the atoms at that time in terms of vortices and thereby explain topologically their stability.  Although this  attempt of a topological explanation of the laws of fundamental physics, the first of many to  come,   had to fail,  a classification of knots and links by  P. Tait derived from these efforts \cite{TAIT98}. \\
Today, the use of topological methods in  the analysis of properties of  systems is widespread in physics. Quantum mechanical phenomena such as the Aharonov-Bohm effect or Berry's phase are of topological origin, as is the stability of defects in condensed matter system,  quantum liquids or in cosmology.  
By their very nature, topological methods are insensitive to details of the systems in question. Their application therefore often reveals unexpected links between seemingly very different phenomena. This common basis in the theoretical description  not only refers to obvious topological objects like vortices, which are encountered on almost all scales in physics, it applies also to more abstract concepts. ``Helicity'', for instance, a topological invariant in inviscid fluids, discovered in 1969 \cite{MOFF69}, is closely related to the topological charge in gauge theories. Defects in nematic liquid crystals  are close relatives to defects in certain gauge theories. 
Dirac's work on magnetic monopoles \cite{DIRA31} heralded in 1931 the relevance of topology for field theoretic studies in physics, but it was not until the formulation of non-abelian gauge theories \cite{YAMI54} with their wealth of non-perturbative phenomena that topological methods became a common tool in field theoretic investigations. \\
In these lecture notes, I will give an introduction to topological methods in gauge theories. I will describe excitations with non-trivial topological properties in the abelian and non-abelian Higgs model and  in Yang-Mills theory. The topological objects to be discussed  are   instantons, monopoles, and vortices  which in space-time are respectively singular on a point, a world-line, or a world-sheet. They are solutions to classical non-linear field equations. I will emphasize both their common formal properties and their relevance in physics. The topological investigations of these field theoretic models is based on the mathematical concept of homotopy. These lecture notes include an introductory section on  homotopy with emphasis  on applications. In general, proofs are omitted or replaced by plausibility arguments or illustrative examples from physics or geometry. To emphasize the universal character in the topological analysis of physical systems, I will at various instances display the often amazing connections   between very different physical phenomena which emerge  from such analyses. Beyond the description of the paradigms of topological objects in gauge theories, these lecture notes contain an  introduction to recent applications of topological methods to Quantum Chromodynamics with emphasis on the confinement issue. Confinement of the elementary degrees of freedom is the trademark of Yang-Mills theories. It is a non-perturbative phenomenon, i.e.~the non-linearity of the theory is as crucial here as in the formation of topologically non-trivial excitations. I will describe various ideas and ongoing attempts towards  a topological characterization  of this peculiar property.
\section{Nielsen-Olesen Vortex}
The Nielsen-Olesen vortex \cite{NIOL73} is a topological excitation in the abelian Higgs model. With topological excitation I will  denote in the following a solution to the field equations  with non-trivial topological properties. As in all the subsequent examples, the Nielsen-Olesen vortex owes its existence to vacuum degeneracy, i.e.~to the presence of multiple, energetically degenerate solutions of minimal energy. I will  start with a brief discussion of the abelian Higgs model and its  (classical) ``ground states'', i.e.~ the field configurations with minimal energy.
\subsection{Abelian Higgs Model}
{The abelian Higgs Model is a field theoretic model with important applications  in particle and condensed matter physics. It constitutes  an appropriate   field theoretic framework for the description of phenomena related to superconductivity (cf.\ \cite{DEGE66,TINK75}) (``Ginzburg-Landau Model'') and its topological excitations (``Abrikosov-Vortices''). At the same time, it provides the simplest setting for the mechanism of mass generation operative in the electro-weak interaction. \newline
The abelian Higgs model is a gauge theory. Besides the electromagnetic field it contains a self-interacting scalar  field (Higgs field)  minimally coupled to electromagnetism. From the conceptual point of view, it is advantageous to consider this field theory in $2+1$ dimensional space-time and  to extend it subsequently to $3+1$ dimensions for applications. \\ 
The abelian Higgs model Lagrangian   
 \begin{equation}
   \label{himo}
{\cal L} =   - \frac{1}{4} F_{\mu \nu}
     F^{\mu \nu}+(D_{\mu} \phi)^{*} (D^{\mu} \phi) - V (\phi) 
\end{equation}
contains the   complex (charged), self-interacting scalar field $\phi$.  The Higgs potential  
\begin{equation}
  \label{vp}
V (\phi) = \frac{1}{4} \lambda (| \phi |^{2} - a^{2} )^{2}.  
\end{equation}
  as a function of the real and imaginary part of the Higgs field is shown in  Fig. \ref{hipo}. By construction, this Higgs potential is minimal along a circle $|\phi|=a$ in the complex $\phi$ plane. The constant $\lambda$ controls the strength of the self-interaction of the Higgs field and, for stability reasons, is assumed to be positive
  \begin{equation}
    \label{lab}
    \lambda \ge 0\, .
  \end{equation}
\begin{figure}

\hspace{5cm} 
\epsfig{file=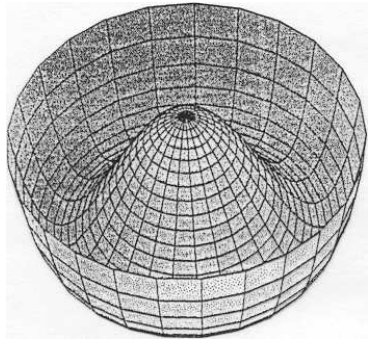, width=0.3\linewidth}
\caption{ Higgs Potential $V(\phi)$  }
\label{hipo}
\end{figure}     
 The Higgs field is  minimally coupled to the radiation field  $A_{\mu}$, i.e.~the partial derivative $\partial_{\mu}$ is replaced by the covariant derivative 

 \begin{equation}
   \label{covde}
D_{\mu}  = \partial _{\mu} + i e A_{\mu}  .
 \end{equation}
Gauge fields and field strengths are related by
$$
F_{\mu \nu} = \partial _{\mu} A_{\nu} - \partial _{\nu} A_{\mu}=
\frac{1}{ie}\left[D_{\mu},D_{\nu}\right] .
$$
 {\em  Equations of motion}
\begin{itemize}
\item
The (inhomogeneous) Maxwell equations are obtained from the principle of  least action,
$$\delta S =\delta \int d^4 x {\cal L} = 0\, ,$$
by variation of $S$ with respect to the gauge fields. 
With 
$$\frac{\delta {\cal L} }{\delta \partial_{\mu}A_{\nu}} = -F^{\mu\nu},\quad\frac{\delta {\cal L} }{\delta A_{\nu}} = -j^{\nu}\, ,$$
we obtain
$$\partial_{\mu} F^{\mu\nu} = j^{\nu},\quad  j_{\nu} = ie (\phi^{\star}\partial _{\nu}\phi-\phi\partial _{\nu}\phi^{\star}) -2e^2\phi^{*}\phi A_{\nu}.$$
\item
  The homogeneous Maxwell equations are not dynamical equations of motion - they are integrability conditions and guarantee that the field strength can be expressed in terms of  the gauge fields. The homogeneous equations follow from the Jacobi identity of the covariant derivative 
$$
\left[D_{\mu},\left[D_{\nu }, D_{\sigma} \right] \right]+ \left[D_{\sigma},\left[D_{\mu }, D_{\nu} \right] \right]+ \left[D_{\nu},\left[D_{\sigma }, D_{\mu} \right] \right]=0 .
$$
   Multiplication with the totally antisymmetric tensor, $\epsilon ^{\mu \nu \rho\sigma},$ yields the homogeneous equations for the   dual field strength $\tilde{F}^{\mu \nu}$
$$
\left[D_{\mu}, \tilde{F}^{\mu \nu}\right] = 0 \quad , \quad \tilde{F}^{\mu \nu} = \frac{1}{2}
\epsilon ^{\mu \nu \rho\sigma} F_{\rho \sigma} .
$$
  The transition 
$$ F \rightarrow  \tilde{F}$$   
corresponds to the following duality relation of electric and magnetic fields 
$${\bf E}\rightarrow {\bf B}\quad , \quad  {\bf B}\rightarrow -{\bf E}.$$ 
\item
 Variation with respect to the charged matter field yields the equation of motion 
$$ D_{\mu} D^{\mu}\phi + \frac{\delta V}{\delta\phi^{*}} =0.$$ 
\end{itemize}  
 Gauge theories contain redundant variables. This redundancy manifests itself in the presence of local symmetry transformations; these ``gauge transformations'' 
 \begin{equation}
   \label{gtqed}
 U(x)  = e ^{ie\alpha (x)}
 \end{equation}
 rotate the phase of the matter field and shift the value of the gauge field in a space-time dependent manner
 \begin{equation}
   \label{gt}   
 \phi \to \phi ^{\,[U]} =  U(x)  \phi (x) \,,\quad A_{\mu} \to A^{\,[U]}_{\mu} = A_{\mu} + U(x)\,\frac{1}{ie}\,\partial_{\mu}\, U^{\dagger}(x)
\,  . \end{equation}
 The covariant derivative   { $D_{\mu}$} has been defined such that  { $D_{\mu} \phi$} transforms covariantly, i.e.~like} the matter field   {  $\phi$} itself.
$$  D_{\mu}\phi(x) \to U(x)\,  D_{\mu}
    \phi (x) .$$
This transformation property together with the invariance of $F_{\mu\nu}$ guarantees invariance of
${\cal L}$ and  of the equations of motion.
A gauge field which is gauge equivalent to $A_{\mu} = 0 $ is called a pure gauge. According to (\ref{gt}) a pure gauge satisfies  
\begin{equation}
  \label{pgqd}
  A_{\mu}^{pg}(x)=  U(x)\frac{1}{ie}\,\partial_{\mu}\, U^{\dagger}(x)= -\partial_{\mu}\,\alpha(x)\, , 
\end{equation}
and the corresponding field strength vanishes.
\vskip .1cm

 { \em Canonical Formalism}\vskip .1cm
In the canonical formalism,   electric and magnetic fields play distinctive dynamical roles.  They are given in terms of the field strength tensor by 
$$ E^{i} = -F^{0i}\,,\quad B^{i}=-\frac{1}{2}\epsilon^{ijk}F_{jk} = (\mbox{rot} A)^{i}.$$ 
Accordingly, 
$$ -\frac{1}{4}F_{\mu\nu} F^{\mu\nu} = \frac{1}{2}\left({\bf E}^{2}-  {\bf B}^{2}\right).$$
The presence of redundant variables complicates the formulation of the canonical formalism and the quantization. Only for independent dynamical degrees of freedom  canonically conjugate variables may be defined and corresponding commutation relations may be associated. In a first step, one has to choose by a ``gauge condition'' a set of variables which are independent. For the development of the canonical formalism there is a particularly suited gauge, the ''Weyl''- or ''temporal'' gauge 
\begin{equation}
  \label{wg}
  A_{0}=0.
\end{equation}
We observe, that the time derivative of $A_{0}$  {  does not appear in $ {\cal L}$, a property which follows from the antisymmetry of the field strength tensor and is shared by  all gauge theories. Therefore in the canonical formalism $A_{0}$ is a constrained variable  and its elimination greatly simplifies the  formulation.  It is easily seen that (\ref{wg}) is a legitimate  gauge condition, i.e.~ that for an arbitrary gauge field a gauge transformation (\ref{gt}) with  gauge function
$$\partial_{0} \alpha(x) = A_{0}(x)\, $$
indeed eliminates $A_{0}.$ With this gauge choice one proceeds straightforwardly with the definition of the canonically conjugate momenta
$$ \frac{\delta {\cal L}}{\delta\partial_{0}A_{i}} = -E^{i}\,,\quad \frac{\delta {\cal L}}{\delta\partial_{0}\phi} = \pi\, ,$$
and constructs via Legendre transformation  the Hamiltonian density
\begin{equation}
  \label{hd}  
 {\cal H} =   \frac{1}{2} (\mbox{\boldmath$E$}^{2}+\mbox{\boldmath$B$}^{2})+ \pi^{*}\pi +(\mbox{\boldmath$D$} \phi) ^{*}(\mbox{\boldmath$D$} \phi )  + V(\phi)\, ,\quad H=\int d^3 x {\cal H}({\bf x})\, .\end{equation}
With the  Hamiltonian density given by a sum of positive definite terms (cf.(\ref{lab})), the energy density of the fields of lowest energy must vanish identically. Therefore, such fields are  static
\begin{equation}
  \label{stat}  
{\bf E}=0\,,\quad \pi=0\, ,
\end{equation}
with vanishing magnetic field
\begin{equation}
  \label{vb}  
{\bf B}=0 \, .
\end{equation}
The following choice of the Higgs field 
\begin{equation}
  \label{mani}  
|\phi|=a,\quad \mbox{i.e.}\quad \phi = ae^{i\beta}
\end{equation}
renders  the potential energy  minimal. 
The ground state is not unique. Rather the  system exhibits a  ``vacuum degeneracy'',  i.e.~it possesses a  continuum of field configurations of minimal energy. It is important to characterize the degree of this degeneracy. We read off from (\ref{mani}) that the  manifold of field configurations of minimal energy is given by the manifold of zeroes of the potential energy. It is  characterized by} $\beta$    and thus  this manifold has the topological properties of a circle  $S^{1}$.
As in other examples to be discussed, this vacuum degeneracy is  the source of the non-trivial topological properties of the abelian Higgs model.  
 \\ To exhibit the physical properties of the system and to study the consequences of the vacuum degeneracy,  we simplify the description by performing a time independent  gauge transformation. Time independent gauge transformations do not alter the  gauge condition (\ref{wg}). In the Hamiltonian formalism, these gauge transformations are implemented as canonical (unitary) transformations which can be regarded as symmetry transformations. We introduce the modulus and phase of the static Higgs field 
$$\phi ({\bf x}) = \rho ({\bf x}) e^{i \theta({\bf x})}\, ,$$
and choose the gauge function
\begin{equation}
  \label{gfug}  
\alpha({\bf x}) = -\theta({\bf x})\, 
\end{equation}
so that in the  transformation (\ref{gt}) to the ``unitary gauge''  the phase of the matter field vanishes
$$
 \phi^{[U]} ({\bf x}) = \rho ({\bf x}) \, ,\quad
\mbox{\boldmath $A$}^{[U]} = \mbox{\boldmath $A$} - \frac{1}{e}
   \mbox{\boldmath $\nabla$} \theta({\bf x})\, , \quad 
(\mbox{\boldmath $D$} \phi)^{[U]}= \mbox{\boldmath $\nabla$} \rho ({\bf x}) - i e \mbox{\boldmath $A$}^{[U]} \rho ({\bf x}) \, .$$
   This results in the following expression for the energy density of the static fields 
   \begin{equation}
     \label{edh}  
\epsilon (\mbox{\boldmath $x$}) = (\mbox{\boldmath $\nabla$} \rho )^{2}
   + \frac{1}{2} \mbox{\boldmath $B$}^{2} + e^{2} \rho ^{2}
   \mbox{\boldmath $A$}^{\prime 2} + \frac{1}{4} \lambda (\rho ^{2} - a^{2})^{2}\,.
 \end{equation}
In this unitary gauge,  the residual gauge freedom in the vector potential  has disappeared  together with the phase of the matter field. In addition to condition (\ref{stat}), fields of vanishing energy must satisfy 
\begin{equation}
  \label{zeun}
{\bf A}=0\,,\quad \rho=a .
\end{equation}
In small oscillations of the gauge field around the ground state configurations (\ref{zeun}) a restoring force appears as a consequence of the non-vanishing value $a$ of the Higgs field $\rho$. Comparison with the energy density of a massive non-interacting scalar field $\varphi$ 
$$\epsilon_{\varphi}({\bf x}) = \frac{1}{2}(\mbox{\boldmath $\nabla$}\varphi)^2 +\frac{1}{2} M^2 \varphi^2$$
 shows that the term quadratic in the gauge field ${\bf A}$ in (\ref{edh}) has to be interpreted as a mass term of the vector field ${\bf A}$. In this Higgs mechanism, the photon has acquired the mass  
\begin{equation}
  \label{phoma}
 M_{\gamma} = \sqrt{2} e a\, ,
\end{equation}
which is determined by the value of the Higgs field. For non-vanishing Higgs field,  the zero energy configuration and the associated small amplitude oscillations describe electrodynamics in the so called Higgs phase, which differs significantly from the familiar Coulomb phase of electrodynamics. In particular, with  photons becoming massive, the system does not exhibit long range forces. This is most directly illustrated by application of the abelian Higgs model to the phenomenon of superconductivity. 
\vskip 0.2cm
{\em Meissner Effect}\\
In this application to condensed matter physics,  one  identifies the energy density (\ref{edh})   with the  free-energy density of a superconductor. This is called the Ginzburg-Landau model.  In this model $ | \phi |^{2} $ is identified with 
 the density of the superconducting  Cooper pairs  (also the electric charge should be replaced  $e \rightarrow e^{\star}=2e$)    and serves as the order parameter to distinguish normal  $a=0$     and superconducting  $a\neq 0$   phases. \vskip .2cm 
 Static solutions (Eq.\ (\ref{stat}))  satisfy the  Hamilton equation (cf.\ Eqs.\ (\ref{hd}), (\ref{edh}))
$$\frac{\delta H }{\delta {\bf A}({\bf x})}= 0\, ,$$
which for a spatially constant scalar field becomes the Maxwell-London  equation
 $$  \mbox{rot}\,{\bf B}=\mbox{ rot}\,\mbox{rot}\,{\bf A} = {\bf j}= 2 e^2 a^2 {\bf A}\, .$$
 The solution to this equation for a magnetic field in the normal conducting phase ($a=0$   for $x<0 $) 
 \begin{equation}
   \label{lon}
{\bf B}(x)= {\bf B}_{0}e^{-x/\lambda_{L}}
 \end{equation}  
 decays  when penetrating into the superconducting region ($a\neq0$ {  for} $x>0 $)   within the penetration or London  depth
 \begin{equation}
   \label{pede}
\lambda_{L} = \frac{1}{M_{\gamma}}\,\,   
 \end{equation}
determined by the photon mass. The expulsion of the magnetic field from the superconducting region is called Meissner effect. \\
Application of the gauge transformation (Eqs.\ (\ref{gt},\ref{gfug})) has been essential for  displaying the physics content of the abelian Higgs model. Its definition requires a  well defined  phase  $\theta(x)$ of the matter field which in turn requires  $\phi(x)\neq 0 $.   At  points where the matter field vanishes, the transformed gauge fields  { ${\bf A}^{\prime}$} are  {  singular}. When approaching the  Coulomb-phase ($a \rightarrow 0$), the Higgs field oscillates around $\phi=0$. In the unitary gauge, the transition from  the  Higgs to the Coulomb phase is therefore expected to be accompanied by the appearance of singular field configurations or equivalently by a  ``condensation'' of  singular points.
\subsection{Topological excitations}
\begin{figure}
\hskip3cm\epsfig{ file=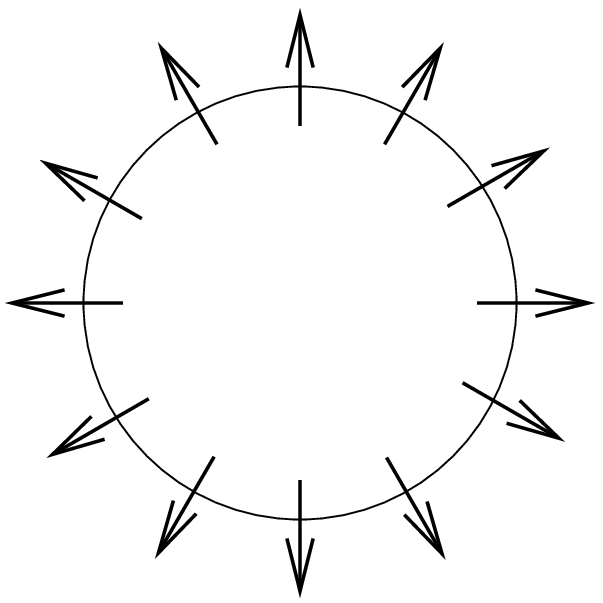, width=0.22\linewidth}\hskip2cm \epsfig{file=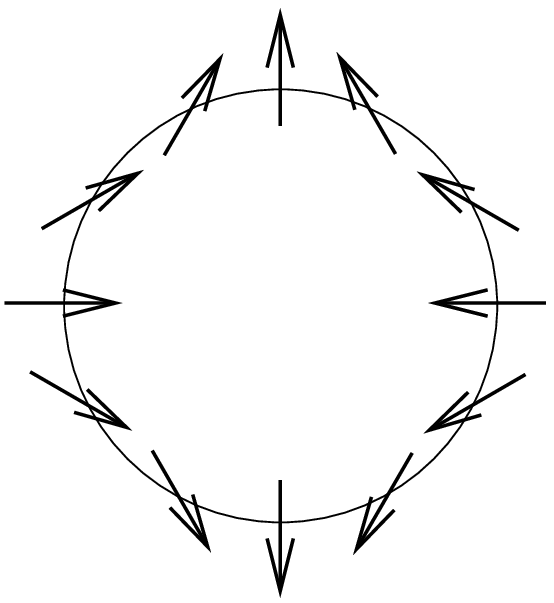, width=0.205\linewidth}
\caption{Phase of a matter field with winding number $n=1$ (left) and $n=-1$ (right)}
\label{phahi}
\end{figure}

In the abelian Higgs model, the manifold of field configurations is a circle $S^1$ parameterized by the angle $\beta$ in Eq.\ (\ref{mani}).  The non-trivial topology  of the manifold of vacuum field configurations is the origin of the topological excitations in the abelian Higgs model as well as in the other field theoretic models to be discussed later. We proceed as in the discussion of the ground state configurations and consider static fields ( Eq.\ (\ref{stat})) but allow for  energy densities  which do not vanish everywhere. As follows immediately from the expression (\ref{hd}) for the energy density, finite energy can result only if  asymptotically ($|{\bf x} | \rightarrow \infty $)
\begin{eqnarray}
 \phi ({\bf x}) &\rightarrow& a e^{i \theta({\bf x})}\nonumber\\
{\bf B}({\bf x}) &\rightarrow& 0 \nonumber\\
{\bf D} \phi({\bf x}) = (\mbox{\boldmath $\nabla$} & -& i e \mbox{\boldmath $A$}({\bf x}))\, \phi ({\bf x}) \rightarrow 0 .
\label{asy}
\end{eqnarray}
For these requirements to be satisfied, scalar and gauge fields have to be correlated asymptotically. According to the last equation, the gauge field is asymptotically given by the phase of the scalar field 
\begin{equation}
  \label{gaph}
\mbox{\boldmath $A$}({\bf x}) = \frac{1}{i e } \mbox{\boldmath$\nabla$} \ln \phi({\bf x}) = \frac{1}{e}\mbox{\boldmath$\nabla$}\theta ({\bf x})\, . 
\end{equation}
 The vector potential is by construction asymptotically a ``pure gauge'' (Eq.\ (\ref{pgqd})) and  no magnetic field strength  is associated with ${\bf A}({\bf x}).$   
\vskip .1cm
 {\em Quantization of magnetic flux}\vskip .1cm
The structure (\ref{gaph})  of the asymptotic gauge field implies that the magnetic flux of field configurations with finite energy is  quantized. Applying Stokes' theorem to a surface $\Sigma$ which is bounded by an asymptotic curve ${\cal C}$ yields
\begin{equation}
  \label{maflu}
 \Phi _{B}^n = \int_{\Sigma}  B\, d^2 x = \oint_{{\cal C}}
\mbox{\boldmath $A$}\cdot d {\bf s}=\frac{1}{e}\oint_{{\cal C}}
\mbox{\boldmath$\nabla$}\theta (x) \cdot d {\bf s}= n \,\frac{2 \pi}{e} \,. 
\end{equation}
Being an integer multiple of the fundamental unit of magnetic flux, $ \Phi^n _{B}$  cannot change as a function of  time, it is a conserved quantity. The appearance of this conserved quantity does not have its origin in an underlying symmetry, rather it is of topological origin.  $ \Phi^n _{B}$ is also considered as a topological invariant since it cannot be changed in a continuous deformation of the asymptotic curve ${\cal C}$. In order to illustrate the topological meaning of this result, we assume the asymptotic curve ${\cal C}$ to be a circle. On this circle, $|\phi|=a$ (cf.\ Eq.\ (\ref{mani})). Thus the scalar field  $\phi({\bf x})$ provides a mapping of the asymptotic circle ${\cal C}$   to the circle of zeroes of the Higgs potential ($V(a)=0$).  
 To study this mapping in  detail, it is convenient to introduce  polar coordinates
$$
\phi({\bf x})=\phi (r, \varphi) _{{\longrightarrow \atop r \to \infty}} 
     a e ^{i \theta (\varphi)}
\quad,\quad e^{i \theta (\varphi + 2 \pi)} = e ^{i \theta (\varphi)}.
$$
 The phase of the scalar field defines a non-trivial mapping of the asymptotic circle 
 \begin{equation}
   \label{map}
\theta: S^{1} \to S^{1} \quad,\quad  \theta (\varphi + 2 \pi) = \theta (\varphi) 
        + 2 n \pi  \end{equation}
to the circle $|\phi| = a$ in the complex plane. These mappings are naturally divided into (equivalence) classes which are characterized by their winding number $n$. This winding number counts how often the phase $\theta$ winds around the circle when the asymptotic circle  ($\varphi$) is traversed once. A formal definition of the winding number is obtained by decomposing a  continuous but otherwise arbitrary $\theta( \varphi)$ into a strictly periodic and a linear function
$$\theta _{n} (\varphi) = \theta^{period} (\varphi) + n \varphi \qquad n = 0, \pm 1, \dots $$
where 
$$\theta^{period} (\varphi+2\pi)=\theta^{period} (\varphi) .$$
The linear functions can serve  as representatives of the equivalence classes. Elements of an equivalence class can be obtained from each other by continuous deformations. The magnetic flux is according to Eq.\ (\ref{maflu}) given by the phase of the Higgs field and is therefore quantized by the winding number $n$ of the mapping (\ref{map}). For instance, for field configurations carrying one unit of magnetic flux, the phase of the Higgs field belongs to the equivalence class  $\theta_{1}$.  Fig. \ref{phahi}  illustrates  the complete turn in the phase when moving around the asymptotic circle. For $n=1$, the phase $\theta({\bf x})$  follows, up to continuous deformations, the polar angle  $\varphi$, i.e.~ 
$\theta(\varphi)= \varphi .$ Note that by continuous deformations the radial vector field can be turned into the velocity field of a vortex $\theta(\varphi)= \varphi +\pi/4 .$ Because of their shape, the $n=-1$ singularities, $\theta(\varphi)= \pi-\varphi$,  are sometimes referred to as ``hyperbolic'' (right-hand side of Fig.\ref{phahi}).
    Field configurations ${\bf A}({\bf x}),\phi({\bf x})$   with  $n\neq 0$    are called    vortices and possess indeed properties familiar from hydrodynamics. The energy density of vortices cannot be zero everywhere with the magnetic flux $\Phi_{B}^n\neq 0$.  Therefore in a  finite region of space  ${\bf B}\neq 0$.  Furthermore, the scalar field must  at least have one zero, otherwise a singularity arises  when contracting the asymptotic circle to a point. Around a zero of $|\phi|$,   the Higgs field displays  a rapidly varying phase $\theta({\bf x})$  similar to the rapid change in  direction of the velocity field close to the center of a vortex in a fluid. However, with the modulus of the Higgs field approaching zero, no infinite energy density is associated with this infinite variation in the phase. In the Ginzburg-Landau theory, the core of the vortex contains no Cooper pairs ($\phi=0$), the system is locally in the ordinary conducting phase containing a magnetic field.  
\vskip .1cm
{\em The Structure of Vortices}\\
The structure of the vortices can be studied in detail by solving the Euler-Lagrange equations of the abelian Higgs model (Eq.\ (\ref{himo})). To this end, it is convenient to  change to dimensionless variables (note that in 2+1 dimensions $\phi, A_{\mu},$ and $e$ are of dimension length$^{-1/2}$) 
\begin{equation}
\label{rscal}
{\bf x} \rightarrow \frac{1}{e a} {\bf x},\quad {\bf A} \rightarrow  \frac{1}{a} {\bf A},\quad \phi \rightarrow  \frac{1}{a}  \phi,\quad \beta = \frac{\lambda}{2e^2} .
\end{equation}
Accordingly, the energy of the static solutions becomes
\begin{equation}
\label{ener}
\frac{E}{a^{2}} = \int d^2 x\left \{ \left | ( \mbox{\boldmath$\nabla$}
    - i {\bf A} ) \phi \right | ^{2} + \frac{1}{2} ( \mbox{\boldmath $\nabla$}\times {\bf A})^{2}
    + \frac{\beta}{2} (\phi \phi^{*} - 1) ^{2} \right \}\, .
\end{equation}    
The static spherically symmetric Ansatz
$$\phi = | \phi (r) | e^{in \varphi},\quad 
{\bf A} = n \frac{\alpha (r)}{r} {\bf e}_{\varphi}\, ,$$ 
converts the equations of motion into a system of (ordinary) differential equations coupling gauge and Higgs fields 
\begin{equation}
\left ( - \frac{d^{2}}{dr^{2}} - \frac{1}{r} \frac{d}{dr} \right ) |\phi| + 
     \frac{n^{2}}{r^{2}} \left ( 1 - \alpha  \right )^{2} |\phi|
    + \beta (|\phi |^{2} - 1 )|\phi| = 0\; ,
\label{eqhi}
\end{equation}
\begin{equation}
\label{eqgf}
\frac{d^{2}\alpha}{dr^{2}} - \frac{1}{r} \frac{d \alpha}{dr} -
    2  (\alpha - 1) | \phi |^{2} = 0\,.
\end{equation}
The requirement of finite energy asymptotically and in the core of the vortex leads to the following boundary conditions
\begin{equation}
 r \to \infty \, :\; \alpha \to 1 \, ,\;
             | \phi | \to 1\, ,\quad \alpha (0) = | \phi (0) | = 0 .
\end{equation}
 From the boundary conditions and the differential equations, the behavior of Higgs and gauge fields is obtained in the core of the vortex 
$$\alpha \sim -2r^2\, ,\quad |\phi|\sim r^n\, ,$$
and asymptotically
$$\alpha-1 \sim \sqrt{r} e^{-\sqrt{2} r},\, \quad |\phi|-1\sim \sqrt{r}e^{-\sqrt{ 2\beta} r} .   $$
The transition from the core of the vortex to the asymptotics occurs on different scales for gauge and Higgs fields. The scale of the variations in the gauge field is the penetration depth $\lambda_L$   determined by the photon mass (cf.\ Eqs.\ (\ref{lon}) and (\ref{pede})). It controls the exponential decay of the magnetic field when reaching into the superconducting phase. The coherence length
\begin{equation}
  \label{cole}
  \xi = \frac{1}{e a \sqrt{2\beta}}=\frac{1}{a\sqrt{\lambda}}
\end{equation}
controls the size of the region of the ``false''  Higgs vacuum ($\phi=0$). In superconductivity, $\xi$ sets the scale for the change in the density of Cooper pairs. The  Ginzburg-Landau parameter  
\begin{equation}
  \label{glp}
\kappa = \frac{\lambda_L}{\xi}=\sqrt{\beta}
\end{equation}
varies with the substance and distinguishes Type I ($\kappa < 1$) from Type II  ($\kappa > 1$) superconductors. When applying the abelian Higgs model to superconductivity, one simply reinterprets the vortices  in 2 dimensional space as 3 dimensional objects by assuming independence of the third coordinate. Often the experimental setting singles out  one of the 3 space dimensions. In such a 3 dimensional interpretation, the requirement of finite vortex energy is replaced by the requirement of finite energy/length, i.e.~finite tension. In Type II superconductors, if the strength of an applied external magnetic field exceeds a certain critical value, magnetic flux is not completely excluded from the superconducting region. It penetrates the superconducting region by exciting one or more vortices each of which carrying a single quantum of magnetic flux $\Phi_B^1$ (Eq.\ (\ref{maflu})). In Type I superconductors, the large coherence length $\xi$ prevents a sufficiently fast rise of the Cooper pair density. In turn the associated shielding currents are not sufficiently strong to contain the flux within the penetration length $\lambda_L$ and therefore no vortex can form.  Depending on the  applied magnetic field  and the temperature, the Type II superconductors exhibit a variety of phenomena related to the intricate dynamics of the vortex lines and display various phases  such as vortex lattices, liquid or amorphous  phases (cf.\ \cite{BFGL94,NELS02}). The  formation of magnetic flux lines inside Type II superconductors by excitation of vortices can be viewed as  mechanism for confining magnetic monopoles. In a Gedankenexperiment we may imagine to introduce  a north  and south magnetic monopole inside a type II superconductor  separated  by a distance $d$. Since the magnetic field will be concentrated in the core of the vortices and will not extend into the superconducting region, the field energy of this system becomes
\begin{equation}
  \label{Vmm}
  V= \frac{1}{2}\int d^3 x\, {\bf B}^2 \propto \frac{4\pi d}{e^2 \lambda_L^2} . 
\end{equation}
Thus, the  interaction energy of the magnetic monopoles grows linearly with their separation. In Quantum Chromodynamics (QCD) one is looking for mechanisms of confinement of (chromo-) electric charges. Thus one attempts to transfer this mechanism by some ``duality transformation'' which interchanges the role of electric and magnetic fields and charges. In view of such applications to QCD, it should be emphasized that formation of vortices does not happen spontaneously. It requires a minimal value of the applied field which depends on the microscopic structure of the material and varies  over three orders of magnitude \cite{POFC95}.  
\vskip .2cm
The point $\kappa=\beta = 1$ in the parameter space of the abelian Higgs model is very special. It separates Type I from Type II superconductors. I will now show that at this point the energy of a  vortex is determined by its charge. To this end, I first derive a bound on the energy  of the topological excitations, 
the  ``Bogomol'nyi bound'' \cite{BOGO76}. Via an integration by parts, the energy (\ref{ener})  can be written in the following form  
\begin{eqnarray*}
\frac{E}{a^2} & = & \int d^{2} x \left | \left[ (\partial _{x} - i A_{x}) 
    \pm i (\partial _{y} - i A_{y}) \right ] \phi \right |^{2}
    + \frac{1}{2} \int d^{2} x \left[B \pm (\phi \phi^{*}-1) \right] ^{2} \\
  & \pm & \int d^{2} xB + \frac{1}{2} (\beta - 1) \int d^{2} x 
     \left[\phi ^{*} \phi -1\right] ^{2}
\end{eqnarray*}
  with the sign chosen according to the sign of the winding number $n$ (cf.\ Eq.\ (\ref{maflu})).  For ``critical coupling'' $ \beta = 1 $ (cf.\ Eq.\ (\ref{rscal})),  the energy is bounded by the third term on the right-hand side, which in turn is given by the winding number 
(\ref{maflu})
$$ E \ge 2 \pi | n | \, .$$
The Bogomol'nyi bound is saturated if the vortex satisfies the following first order differential equations  
$$
\left [ ( \partial _{x} - i A _{x}) \pm i (\partial _{y} - i A _{y} ) 
    \right ] \phi = 0
$$
$$B = \pm (\phi  \phi^{*} - 1) \,.$$

It can be shown that for $\beta=1$ this coupled system of first order differential equations is equivalent to the Euler-Lagrange equations. 
The energy of these particular solutions to the classical field equations is given in terms of the  magnetic charge. Neither the existence of  solutions whose energy is determined by topological properties, nor  the reduction of the equations of motion to a first order system of differential equations is  a peculiar property of the Nielsen-Olesen vortices. We will encounter again the Bogomol'nyi bound and its saturation in our discussion of the 't Hooft monopole and of the instantons. Similar solutions with the energy determined by some charge play also an important role in supersymmetric theories and in string theory.  
\vskip .3cm
A wealth of further results concerning the topological excitations in the abelian Higgs model has been obtained.  Multi-vortex solutions, fluctuations around spherically symmetric solutions, supersymmetric extensions, or extensions to non-commutative spaces have been studied. Finally, one can introduce fermions  by  a Yukawa coupling 
$$ \delta{\cal L} \sim g \phi \bar{\psi}\psi+ e\bar{\psi}A\,\!\!\!\!/ \psi $$
to the scalar and a minimal coupling to the Higgs field. Again one finds what will turn out to be a quite general property. Vortices induce fermionic zero modes  \cite{JARO81,WEIN81}. We will discuss this phenomenon in the context of instantons.   
\section{Homotopy}
\subsection{The Fundamental Group}
In this section I will describe extensions and  generalizations of  the rather intuitive concepts which have been used in the analysis of the abelian Higgs model. From the physics point of view, the vacuum degeneracy is the essential property of the abelian Higgs model which ultimately gives rise to  the quantization of the magnetic flux and the emergence of topological excitations. More formally, one views  fields like the Higgs field as providing a mapping of the asymptotic circle in configuration space to the space of zeroes of the Higgs potential. In this way, the quantization is a consequence of the presence of integer valued topological invariants associated with this mapping. While in the abelian Higgs model these properties are almost self-evident, in the forthcoming applications the structure of the spaces to be mapped are more complicated. In the non-abelian Higgs model, for instance, the space of zeroes of the Higgs potential will be a subset of a non-abelian group. In such situations, more advanced mathematical tools have proven to be helpful for carrying out the analysis. In our discussion and for later applications, the concept of homotopy will be central (cf.\ \cite{NASE83,NAKA90}). It is a concept which is relevant for the characterization of global rather than local properties of spaces and maps (i.e.~fields). In the following we will assume that the spaces are ``topological spaces'', i.e.~sets in which open subsets with certain properties are defined and thereby the concept of continuity  (``smooth maps'') can be introduced (cf.\ \cite{MUNK2000}). In physics, one often requires differentiability of functions. In this case,  the topological spaces must possess  additional properties (differentiable manifolds).  
We start with the formal definition of homotopy.\\
{\bf Definition:} Let $X,Y$ be smooth manifolds and  $f: X\rightarrow Y $ a smooth map between them. A  {\em homotopy} or {\em deformation} of the map $f$ is a smooth map 
\begin{displaymath}
F: X \times I \rightarrow Y \quad \left(I=\left[0,1\right]\right)
\end{displaymath}
with the property
\begin{displaymath}
F\left(x,0\right)=f\left(x\right)
\end{displaymath}
Each of the maps $f_{t}\left(x\right)=F\left(x,t\right)$ is said to be homotopic to the initial map $f_{0}=f$ and the map of the whole cylinder $X\times I$ is called a homotopy. The relation of homotopy between maps is an equivalence relation and therefore allows to divide the set of smooth maps $X \rightarrow  Y$ into equivalence classes,  {\em homotopy classes}. \\
{\bf Definition:} Two maps $f,g$ are called {\em homotopic}, $f \sim g$, if they can be deformed continuously into each other.\\
The mappings 
$$\mathbb{R}^n \to\mathbb{R}^n:\; f(x)=x,\; g(x)=x_0=\mbox{const.}$$
 are homotopic with the homotopy given by
\begin{equation}
  \label{horn}
  F(x,t) = (1-t) x +t x_0 .
\end{equation}
Spaces $X$ in which the identity mapping $1_X$ and the constant mapping  are homotopic, are homotopically equivalent to a point. They are called {\em contractible}.

\begin{figure}
\hskip 6cm\epsfig{ file=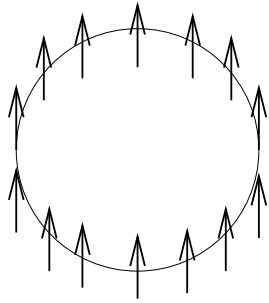, width=0.22\linewidth}
\caption{Phase of matter field with winding number $n=0$}
\label{spgr}
\end{figure}

{\bf Definition:}
Spaces  $X$ and $Y$ are defined to be {\em homotopically equivalent} if continuous mappings exist 
$$f : X  \rightarrow  Y  \quad ,\quad g : Y  \rightarrow  X$$
 such that 
$$g\circ f  \sim  1_X \quad,\quad f\circ g  \sim  1_Y
$$
  An important example is the  equivalence of the $n-$sphere and the punctured  $\mathbb{R}^{n+1}$ (one point removed) 
 \begin{equation}
   \label{stp}
S^ n = \{{\bf x}\in\mathbb{R}^{n+1} | x_1^2 + x_2^2 + \ldots + x_{n+1}^2 = 1\}
\sim \mathbb{R}^{n+1}\backslash\{0\} .   
 \end{equation}
 which can be proved by stereographic projection. It shows that with regard to homotopy, the essential property of a circle is the hole inside. Topologically identical ({\em homeomorphic}) spaces, i.e.~spaces which can be mapped continuously and bijectively onto each other,  possess the same connectedness properties and are therefore homotopically equivalent. The converse is not true.
\vskip .1cm
In physics, we often can identify the parameter $t$ as time. Classical fields, evolving continuously in time are examples of homotopies. Here the restriction to  continuous functions follows from energy considerations. Discontinuous changes of fields are in general connected with infinite energies or energy densities. For instance, a homotopy of the ``spin system'' shown in Fig. \ref{spgr} 
is provided by a spin wave connecting some initial  $F(x,0)$ with some final configuration  $F(x,1)$. Homotopy theory classifies the different sectors (equivalence classes) of field configurations. Fields of a given sector  can evolve into each other as a function of time.  One might be interested, whether the configuration of spins in Fig.\,\ref{phahi} can evolve with time from the ground state configuration shown in Fig. \ref{spgr}.\vskip.2cm 
\newpage
{\em The Fundamental Group}
\vskip.1cm 
The fundamental group characterizes connectedness properties of spaces related to properties of loops in these spaces. The basic idea is to detect defects- like a hole in the plane - by letting loops shrink to a point. Certain defects will provide a topological  obstruction to such attempts. Here one considers arcwise (or path) connected spaces, i.e.~spaces where any pair of points can be connected by some path.\\
A {\bf loop} (closed path) through $x_0$ in $M$ is formally defined as a map
$$
\alpha : [0,1]\rightarrow M \qquad\mbox{with}\qquad
\alpha(0) = \alpha(1) = x_0\, .
$$
A product of two loops is defined  by 
$$
\gamma = \alpha * \beta , \qquad\qquad
\gamma(t) = \left \{
\begin{array}{l@{\quad , \quad}r@{\ \le t \le\ }l}
\alpha(2t)  & 0 & \displaystyle{\frac{1}{2}}\\
\beta(2t-1) & \displaystyle{\frac{1}{2}} & 1\\
\end{array}
\right\}\, ,
$$
and corresponds to traversing the loops consecutively.
Inverse and constant loops are given by 
$$
\alpha^{-1}(t) = \alpha(1-t),\qquad 0 \le t \le 1
$$
and 
$$\qquad c(t) = x_0$$
respectively.
The inverse corresponds to traversing a given loop in the opposite direction.\\
\textbf{Definition:}
Two loops through $x_0 \,\in M$ are said to be homotopic, $\alpha\sim\beta$,
if they can be continuously deformed into each other, i.e.~if a mapping $H$ exists, 
$$
H : [0,1]\times [0,1] \rightarrow M\, ,
$$
with the properties
\begin{eqnarray}
  \label{hst}
 H(s,0) & =&  \alpha(s) ,\quad 0 \le s \le 1\, ;\quad H(s,1)  =
\beta(s)\nonumber \\
 H(0,t) & = & H(1,t) = x_0 ,\quad 0 \le t \le 1 . 
\end{eqnarray}
Once more, we may interpret  $t$ as time and the homotopy $H$ as a time-dependent evolution  of loops into each other. \\
{\bf Definition:}
$\pi_1(M,x_0)$ denotes the set of equivalence classes (homotopy classes) of loops through $x_0\in M$. 
\vskip 0.15cm
The product of equivalence classes is defined by the product of their representatives. It can be easily seen that this definition does not depend on the loop chosen to represent a certain class. In this way, $\pi_1(M,x_0)$ acquires a group structure with the constant loop representing the neutral element. Finally, in an arcwise connected space $M$, the equivalence classes  $\pi_1(M,x_0)$ are independent of the base point $x_{0}$ and one therefore denotes  with $\pi_1(M)$ the {\em fundamental group} of $M$.\\ 
For applications, it is important that the fundamental group (or more generally the homotopy groups) of homotopically equivalent spaces $X$,$Y$
are identical
$$\pi_1(X) = \pi_1(Y). $$
\vskip 0.3cm
\newpage 
{\em Examples and Applications}
 { 
\vskip 0.2cm
Trivial topological spaces as far as their connectedness is concerned are {\em simply connected} spaces.\\
{\bf Definition:} A topological space $X$ is said to be {\em simply connected} if any loop in
$X$ can be continuously shrunk to a  point.\\
The set of equivalence classes consists of one element, represented by the constant loop and one writes 
$$\pi_1 = 0 .$$
Obvious examples are the spaces $\mathbb{R}^n$. \\
Non-trivial connectedness properties are the source of the peculiar properties of the abelian Higgs model. The phase of the Higgs field $\theta$ defined on  a loop at infinity, which can continuously be deformed into a circle at infinity,  defines  a mapping 
$$\theta : S^{1}\rightarrow S^{1}.$$}
An arbitrary phase $\chi$ defined on $S^1$  has the properties
\begin{equation}
  \label{phi0}
 \chi(0) = 0 \,,\quad  \chi(2\pi) = 2\pi m \, .
\end{equation}
It can be continuously deformed into the linear function $m\varphi$.   
The mapping 
  $$
H(\varphi,t)  =  (1-t)\,\chi(\varphi) + t\,\varphi\,\frac{\chi(2\pi)}{2\pi}$$\\
with the properties
   $$ H(0,t) = \chi(0)    = 0\quad ,\quad H(2\pi,t) =  \chi(2\pi)\, , $$
is a homotopy and thus  
$$\chi(\varphi)  \sim  m\varphi .$$
The equivalence classes are therefore characterized by integers $m$ and since these winding numbers are additive when traversing two loops  
 \begin{equation}
   \label{pi1}
\pi_1(S^1)\sim \mathbb{Z} .   
 \end{equation}
Vortices are defined on $\mathbb{R}^2\backslash\{0\}$ since the center of the vortex, where $\theta({\bf x})$ is ill-defined, has to be removed. The homotopic equivalence of this space to $S^1$ (Eq.\ (\ref{stp})) implies that a vortex with winding number $N\ne 0$ is stable; it cannot evolve with time into the homotopy class of the ground-state configuration  where up to continuous deformations, the phase points everywhere into the same direction. \\
This argument also shows that the (abelian) vortex is not topologically stable in higher dimensions. In $\mathbb{R}^n\backslash\{0\}$ with $n\ge3$, by continuous deformation, a loop can always avoid the origin and can therefore be shrunk to a point. Thus
\begin{equation}
  \label{pi1n}
  \pi_1(S^n)= 0 \, ,\; n\ge 2\, ,
\end{equation}
i.e.~$ n-$spheres with $n>1$ are simply connected. In particular, in 3 dimensions a ``point defect'' cannot be detected by the fundamental group. On the other hand, if we remove a line from the $\mathbb{R}^3$, the fundamental group is again characterized by the winding number and we have 
\begin{equation}
  \label{pi1nm}
  \pi_1(\mathbb{R}^3\backslash\mathbb{R})\sim \mathbb{Z} \, . 
\end{equation}
This result can also be seen as a consequence of the general homotopic equivalence 
\begin{equation}
  \label{hote}
 \mathbb{R}^{n+1}\backslash\mathbb{R} \sim S^{n-1} \, .
\end{equation}
The result (\ref{pi1n}) implies that stringlike objects in 3-dimensional spaces can be detected by loops and that their topological stability is determined by the non-triviality of the fundamental group.
For constructing pointlike objects in higher dimensions, the fields must assume values in spaces with different connectedness properties. \\
The fundamental group of a product of spaces $X,Y$ is isomorphic to the product of their fundamental groups
\begin{equation}
  \label{prod}
  \pi_1(X\otimes Y) \sim \pi_1(X)\otimes\pi_1(Y)\, .
\end{equation}
For a torus $T$ and a cylinder $C$ we thus have
\begin{equation}
  \label{tozy}
\pi_1(T) \sim \mathbb{Z}\otimes \mathbb{Z},\quad \pi_1(C)=\mathbb{Z}\otimes\{0\}\, .  
\end{equation}
\vskip.2cm
\subsection{Higher Homotopy Groups}
The fundamental group displays the properties of loops under continuous deformations and thereby characterizes topological properties of the space in which the loops are defined. With this tool only a certain class of non-trivial topological properties can be detected. We have already seen above that a point defect cannot be detected by loops in dimensions higher than two and therefore the concept of homotopy groups must be generalized to higher dimensions. Although in $\mathbb{R}^3$ a circle cannot enclose a pointlike defect, a 2-sphere can. 
The higher homotopy groups are obtained by suitably defining higher dimensional analogs of the (one dimensional) loops. For technical reasons, one does not choose directly spheres and starts with $n-$cubes which are defined as 
$$
I^n = \{ (s_1,\ldots,s_n)\,|\, 0 \le s_i \le 1\quad \mbox{all } i\}
$$
whose boundary is given by
$$
\partial I^n 
= \{ (s_1,\ldots,s_n)\in I^n\,|\, s_i=0\quad \mbox{or} \;  s_i=1 \quad
\mbox{for at least one}\;  i\}.
$$
Loops are curves with the initial and final points identified. Correspondingly, one considers continuous maps from the $n-$cube to the topological space $X$
$$\alpha:\; I^n \rightarrow X $$
with the properties that the image of the boundary is one point in $X$
$$
\alpha : I^n\rightarrow X \qquad,\qquad \alpha(s) = x_0
\quad\mbox{for}\quad s\in \partial I^n.
$$
$\alpha(I^n)$ is called an $n-$loop in $X$. Due to the identification of the points on the boundary  these $n-$loops are topologically equivalent to $n-$spheres. 
One now proceeds as above and introduces a homotopy, i.e.~continuous deformations of $n-$loops 
$$ F:I^n\times I \rightarrow X$$
and requires
\begin{eqnarray*}
F(s_1,s_2,\ldots,0) & = & \alpha(s_1,\ldots,s_n)\\
F(s_1,s_2,\ldots,1) & = &  \beta(s_1,\ldots,s_n)\\[1ex]
F(s_1,s_2,\ldots,t) & = & x_0\quad,\quad\mbox{for}\quad 
                          (s_1,\ldots,s_n)\in\partial I^n
                          \qquad\Rightarrow\qquad\alpha\sim\beta
\end{eqnarray*}
The  homotopy establishes an equivalence relation between the $n-$loops. The space of $n-$loops is thereby partitioned into disjoint classes. The set of equivalence classes is, for arcwise connected spaces (independence of $x_0$), denoted by  
$$\pi_{n}(X)= \left\{\alpha|\alpha: I^n\rightarrow X,\;
  \alpha(s \in \partial I^n ) = x_0 \right\}.$$
As  $\pi_1$, also $\pi_n$ can be equipped with an algebraic structure. To this end one  defines a product of maps $\alpha,\beta$ by connecting them along a common part of the boundary, e.g. along the part given by  $s_1$=1  
\begin{eqnarray*}
\alpha\circ\beta(s_1,s_2,\ldots,s_n) & = & \left \{
\begin{array}{l@{\quad , \quad}r@{\ \le s_1 \le\ }l}
\alpha(2s_1,s_2,\ldots,s_n) & 0 & \displaystyle{\frac{1}{2}}\\
\beta(2s_1-1,s_2,\ldots,s_n)& \displaystyle{\frac{1}{2}} & 1\\
\end{array}
\right.\\
\alpha^{-1}(s_1,s_2,\ldots,s_n) & = & \alpha(1-s_1,s_2,\ldots,s_n) \, .
\end{eqnarray*}
After definition of the unit element and the inverse respectively
$$ e(s_{1},s_{2}...s_{n})= x_{0}\,, \qquad \alpha^{-1}(s_{1},s_{2}...s_{n})=\alpha(1-s_{1},s_{2}...s_{n})$$
$\pi_n$ is seen to be a group. 
The algebraic structure of the higher homotopy groups is  simple
\begin{equation}
  \label{abho}
\pi_{n}(X) \;\mbox{is abelian for}\; n \, > \, 1\, .  
\end{equation}
The fundamental group, on the other hand, may be non-abelian, although most of the applications in physics deal with abelian fundamental groups. An example of  a non-abelian fundamental group will be discussed below (cf.\ Eq.\ (\ref{quat})).\\  
The mapping between spheres is of relevance for many applications of homotopy theory. The following result holds  
\begin{equation}
  \label{pin}
\pi_{n}(S^{n}) \sim \mathbb{Z} .  
\end{equation}
In this case the integer $n$ characterizing the mapping generalizes the winding number of mappings between circles. By introducing polar coordinates $\theta,\varphi$ and   $\theta^{\prime},\varphi^{\prime}$ on two spheres,  under the mapping
$$\theta^{\prime} = \theta,\; \varphi^{\prime} = \varphi,$$
the sphere  $S^{2\prime}$ is covered once if $\theta$ and $\varphi$ wrap the sphere $S^2$ once. This 2-loop belongs to the class $k=1 \in \pi_2(S^2)$. Under the mapping
$$\theta^{\prime} = \theta,\; \varphi^{\prime} = 2\varphi$$
$S^{2\,\prime}$ is covered twice and the 2-loop belongs to the  class   $k= 2 \in \pi_2(S^2)$ .
Another important result is
\begin{equation}
  \label{hskn}
\pi_{m}(S^n)=0 \quad  m<n ,  
\end{equation}
a special case of which ($\pi_1(S^2)$) has been discussed above. There are no simple intuitive arguments concerning the homotopy groups $\pi_{n}(S^m)$ for $n>m$, which in general are non-trivial. A famous example (cf.\ \cite{DFN285}) is
\begin{equation}
  \label{hoin} 
\pi_{3}(S^2)\sim \mathbb{Z}\, ,
\end{equation}
a result which is useful in the study of Yang-Mills theories in a certain class of gauges (cf.\ \cite{JAHN00}). The integer $k$ labeling the equivalence classes has a geometric interpretation. Consider two points $y_1,y_2 \in S^2$, which are regular points in the (differentiable) mapping $$ f:\, S^3 \rightarrow S^2\, $$
i.e.~the differential $df$ is 2-dimensional in $y_1$ and $y_2$.
The pre images of these points $M_{1,2}= f^{-1}(y_{1,2})$ are curves ${\cal C}_1, {\cal C}_2$ on $S^3$; the integer  $k$ is the linking number $lk\{{\cal C}_1,{\cal C}_2\}$ of these curves, cf.\ Eq.\ (\ref{gaulink}). It is called the {\em Hopf invariant}.         \vskip 0.2cm
\subsection{Quotient Spaces}
 Topological spaces  arise in very different fields of physics and are frequently of complex structure. Most commonly, such non-trivial topological spaces are obtained by identification of certain points which are elements of simple topological spaces. The mathematical concept behind such identifications is that of a {\em quotient space}. The identification of points is formulated as an equivalence relation between them. \\
{\bf Definition} 
Let $X$ be a topological space and $\sim$ an equivalence relation on $X$. Denote by
\begin{displaymath}
\left[x\right] = \left\{y\in X | y\sim x \right\} 
\end{displaymath}
the equivalence class of $x$ and with  { $X/\hspace{-.14cm}\sim$}\,  the set of equivalence classes; the projection taking each  { $x \in X$} to its equivalence class be denoted by 
\begin{displaymath}
\pi \left(x\right) = \left[x\right] .
\end{displaymath}
$X/\hspace{-.14cm}\sim$\,  is then called quotient space of $X$ relative to the relation   $\sim$. The quotient space is a topological space with subsets $V\subset X/\hspace{-.14cm}\sim$ \, defined to be open if $\pi^{-1}(V)$ is an open subset of $X$.
\begin{itemize}
\item  
An elementary example of a quotient space is a circle. It is obtained by an equivalence relation of points in $\mathbb{R}$ and therefore owes its non-trivial topological properties to this identification. 
Let the equivalence relation be defined by :
\begin{displaymath}
X=\mathbb{R},\; x,y \in \mathbb{R},\quad x \sim y \quad \mbox{if} \quad  x-y \, \in  \, \mathbb{Z}.
\end{displaymath}
$\mathbb{R}/\hspace{-.14cm}\sim$\, can be identified with 
\begin{displaymath}
S^{1}= \left\{z \in \mathbb{C}| |z|=1\right\}\, ,
\end{displaymath}
the unit circle in the complex plane and the projection is  given by
\begin{displaymath}
\pi \left(x\right) =  e^{ 2 i\pi x} . 
\end{displaymath}
The circle is the topological space in which the phase of  the Higgs field or of the wave function of a superconductor lives.  Also  the orientation of the spins  of magnetic substances with  restricted to a plane can be specified by points on a circle. In field theory such models are called $O(2)$ models. If the spins can have an arbitrary direction in 3-dimensions ($O(3)$ models), the relevant manifold representing such spins is  the surface of a sphere, i.e.~$S^2$. 
\item 
Let us consider
\begin{displaymath}
X= \mathbb{R}^{n+1} \backslash \left\{0\right\}\, ,
\end{displaymath}
i.e.~the set of all (n+1) tuples  { $x=(x^{1}, x^{2},...,x^{n+1})$}  except  { $(0,0,...,0)$},  and define 
\begin{displaymath}
x \sim y  \quad \mbox{if for real } t\neq 0 \quad  (y^{1}, y^{2},...,y^{n+1}) =  (t x^{1}, t x^{2},...,t x^{n+1})\, .
\end{displaymath}
The equivalence classes  { $ \left[x\right]$} may be visualized as lines through the origin. The resulting quotient space   is called  the {\em real projective space} and denoted by  {$\mathbb{R}P^{n} $}; it is a differentiable manifold of dimension $n$. Alternatively, the projective spaces can be viewed as spheres with antipodal points identified
  \begin{equation}
    \label{prsp}
\mathbb{R}P^{n}=\{x|x\in S^n, x\sim -x\}.
 \end{equation}
These topological spaces are important in condensed matter physics. These are the topological spaces of  the  degrees of freedom of (nematic) liquid crystals. Nematic liquid crystals consist of  long rod-shaped molecules which spontaneously orient themselves like spins of a magnetic substances. Unlike spins, there is no distinction between head and tail. Thus, after identification of head and tail, the $n-$spheres relevant for the degrees of freedom of  magnetic substances, the spins,  turn into the projective spaces relevant for the degrees of freedom of liquid crystals, the {\em directors}.  

\item  The $n-$spheres are the central objects of homotopy; physical systems in general are  defined in the $\mathbb{R}^n$. In order to apply homotopy arguments, often the space $\mathbb{R}^n$ has to be replaced by  $S^n$. Formally this is possible by adjoining the point $\{\infty \}$ to $\mathbb{R}^n$
  \begin{equation}
    \label{Alex}
   \mathbb{R}^n \cup \{ \infty \} = S^n \, .
  \end{equation}
 This procedure is called the {\em one-point (or Alexandroff) compactification } of $\mathbb{R}^n$ (\cite{GAGR99}). Geometrically this is achieved by the stereographic projection with the infinitely remote points being mapped to the north-pole of the sphere. For this to make sense, the fields which are defined in $\mathbb{R}^n$ have to approach a constant with $|{\bf x}| \to \infty $. Similarly the process of compactification of a disc $D^2$ or equivalently a square to $S^2$ as shown in Fig. \ref{dis2} requires the field (phase and modulus of a complex field) to be constant along the boundary.
\begin{figure}
\hspace{4.5cm}\epsfig{file=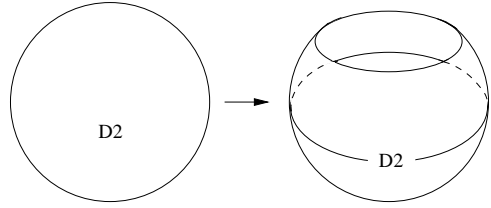, width=.4\linewidth}
\caption{ Compactification of a disc $D^2$ to $S^2$ can be achieved by deforming  the disc and finally adding a point, the north-pole }
\label{dis2}
\end{figure}   
\end{itemize}
\subsection{Degree of  Maps}
For mappings between closed oriented  manifolds $X$ and $Y$ of  equal dimension ($n$),  a homotopy invariant, the {\em degree} can be introduced \cite{DFN285,FRAN97}. Unlike many other topological invariants, the degree possesses an integral representation, which is extremely useful for actually calculating the value of topological invariants. If $y_{0} \in Y $ is a regular value of $f$, the set $f^{-1}\left(y_{0}\right)$ consists of only a finite number of points $x_{1},...x_{m}$ . Denoting  with $x_{i}^{\beta},y_{0}^{\alpha}$ the local coordinates,   the Jacobian defined by
\begin{displaymath}
J_{i}=\det\left(\frac{\partial y_{0}^{\alpha}}{\partial x_{i}^{\beta}}\right)\neq 0
\end{displaymath}
is non-zero.\\
{\bf Definition:} The {\em degree } of $f$ with respect  to $y_{0} \in  Y$ is defined as
\begin{equation}
\label{deg}
\mbox{deg} f = \sum_{x_{i}\in f^{-1}\left(y_{0}\right)} \mbox{sgn} \left(J_{i}\right)\, .
\end{equation}
The degree has the important property of being independent of the choice of the regular value $y_{0}$ and to be invariant under homotopies, i.e.~the degree can be used to classify homotopic classes. In particular, it can be proven that a pair of smooth maps from a closed oriented $n$-dimensional  manifold $X^{n}$  to the $n$-sphere $S^{n}$,  $f,g: X^{n}\rightarrow S^{n}$,  are homotopic {\em iff} their degrees coincide.\\
For illustration, return to our introductory example and consider maps from the unit circle to the unit circle $S^{1}\rightarrow S^{1}$. As we have seen above, we can picture the unit circle as arising from $\mathbb{R}^1$  by identification of the points  $x+2n\pi$  and $y+2n\pi$ respectively.  We consider a map with the property 
\begin{displaymath}
f\left(x+2\pi\right)=f\left(x\right)+2k\pi\, ,
\end{displaymath}
i.e.~if $x$ moves around {\em once} the unit circle, its  image $y=f\left(x\right)$ has turned around $k$ times. In this case, every $y_{0}$ has at least $k$ pre images with slopes (i.e.~values of the Jacobian)  of the same sign. 
For the representative  of the $k$-th homotopy class, for instance,  
\begin{displaymath}
f_{k}\left(x\right)= k\cdot x
\end{displaymath}
and with the choice $y_0=\pi$ we have $f^{-1}(y_0) = \{\frac{1}{k}\pi, \frac{2}{k}\pi,...\pi\}$. Since $\partial y_0/\partial x \big|_{x=l/(k\pi)} =1, $ the degree is  $k$. 
 Any continuous deformation can only add  pairs of pre-images with slopes of opposite signs which do not change the degree.   The degree can  be rewritten in the  following integral form:
\begin{displaymath}
\mbox{deg} f = k = \frac{1}{2\pi}\int_{0}^{2\pi} dx \left(\frac{df}{dx}\right)\, .
\end{displaymath}
Many of the homotopy invariants appearing in our discussion can actually be calculated after identification with the degree of an appropriate map and its evaluation by  the integral representation of the degree. In the Introduction we have seen that the work of transporting a magnetic monopole around a closed curve in the magnetic field generated by circular current is given by the linking number $lk$ (\ref{gaulink}) of these two curves. The topological invariant $lk$ can be identified with the degree of the following map  \cite{ARKH98}
$$T^2\to S^2:\quad (t_1,t_2)\to \hat{\bf s}_{12} = \frac{{\bf s}_1(t_1)-{\bf s}_2(t_2)}{|{\bf s}_1(t_1)-{\bf s}_2(t_2)|}\, .$$
The generalization of the above integral representation of the degree is usually formulated in terms of differential forms as 
\begin{equation}
  \label{indeg}
  \int_{X} f^*\omega =  \mbox{deg} f \, \int_Y \omega
\end{equation}
where $f^*$ is the induced map (pull back) of differential forms of degree $n$ defined on $Y$. In the $\mathbb{R}^n$ this reduces to the formula for changing the variables of integrations over some function $\chi$
$$ \int_{f^{-1}(U_{i})} \chi(y(x)) \det\left(\frac{\partial y_{0}^{\alpha}}{\partial x_{i}^{\beta}}\right) dx_1 ...dx_n = \mbox{sgn} \det\left(\frac{\partial y_{0}^{\alpha}}{\partial x_{i}^{\beta}}\right)\int_{U_i} \chi(y) dy_1...dy_n $$
where the space is represented as a union of  disjoint neighborhoods $U_i$ with $y_0\in U_i$ and with non-vanishing Jacobian determinant.    
\subsection{Topological Groups}
In many application of topological methods to physical systems, the relevant degrees of freedom are described by fields which take values in  topological groups like the Higgs field in the abelian or non-abelian Higgs model or link variables and Wilson loops in gauge theories. In condensed matter physics an important  example is the order parameter in superfluid $^3$He in the ``A-phase'' in which the pairing of the Helium atoms occurs in p-states with the spins coupled to 1. This pairing mechanism is the source of a variety of different phenomena and gives rise to the  rather complicated manifold of the order parameter  $SO(3)\otimes S^2/\mathbb{Z}_2$ (cf.\ \cite{THOU98}). 
\vskip .1cm    
 {\em  SU(2) as topological space}\vskip .1cm
The group $SU(2)$ of unitary transformations is of fundamental importance for many applications in physics. It can be generated by the Pauli-matrices 
\begin{equation}
  \label{pauli}
\tau^{{1}}=\left (\begin {array}{cc} 0&1\\\noalign{\medskip}1&0
\end {array}\right )\, ,\,\, \tau^{{2}}=\left (\begin {array}{cc} 0&-i
\\\noalign{\medskip}i&0\end {array}\right )\, ,\,\,  
\tau^{{3}}=\left (\begin {array}{cc} 1&0\\\noalign{\medskip}0&-1 \, .
\end {array}\right )\end{equation}
Every element of  $SU(2)$ can be parameterized in the following way 
\begin{equation}
  \label{su2pa}
 U = e^{i \mbox{\boldmath$\phi$} \cdot \mbox{\boldmath$\tau$}} = \cos \phi 
     + i \mbox{\boldmath$\tau$} \cdot \hat{\phi} \sin \phi  
     = a + i \mbox{\boldmath$\tau$} \cdot \mbox{\boldmath $b$}.\end{equation}
Here {\boldmath$\phi$} denotes an arbitrary vector in  internal (e.g. isospin or color) space and we do not explicitly write the neutral element $e$. This vector is parameterized by the 4 (real) parameters $a, {\bf b}$ subject to the unitarity constraint 
$$UU^{\dagger}  =  (a + i {\bf b}\cdot\mbox{\boldmath$\tau$})(a - i
 {\bf b}\cdot\mbox{\boldmath$\tau$})= a^2 + b^2 = 1 \, .$$ 
This parameterization establishes  the topological equivalence (homeomorphism) of $SU(2)$ and $S^3$ 
\begin{equation}
  \label{su2s3}
SU(2)\sim S^{3} .  
\end{equation}
This homeomorphism together with the results (\ref{pin}) and (\ref{hskn}) shows 
 \begin{equation}
   \label{pisu2}
\pi_{1,2}\Big(SU(2)\Big) =0,\quad\pi_3\Big(SU(2)\Big) = \mathbb{Z}.   
 \end{equation}
One can show more generally the following properties of homotopy groups
$$\pi_k\Big(SU(n)\Big) =0\, \; k<n\, .$$ 
The triviality of the fundamental group of $SU(2)$ (Eq.\ (\ref{pisu2})) can be verified by constructing an explicit homotopy between the loop
\begin{equation}
  \label{lpsu2}
  u_{2n}(s) =\exp \{i 2 n \pi s \tau^3\}
\end{equation}
and the constant map
\begin{equation}
  u_c(s) =1\, .
\end{equation}
The mapping
$$H(s,t)=\exp \Big\{ -i \frac{\pi}{2}t \tau^1\Big\}\,\exp \Big\{ i \frac{\pi}{2}t( \tau^1\cos 2\pi ns + \tau^2\sin 2\pi ns )\Big\}$$
has the desired properties (cf.\ Eq.\ (\ref{hst}))
$$H(s,0)= 1,\quad H(s,1)= u_{2n}(s),\quad H(0,t)=H(1,t)= 1\, ,$$
as can be verified with the help of the identity (\ref{su2pa}). After continuous deformations and proper choice of the coordinates on the group manifold, any loop can be parameterized in the form \ref{lpsu2}.

Not only  Lie groups but also quotient spaces formed from them appear in important physical applications. The presence of the group structure suggests the following construction of quotient spaces.  Given any subgroup $H$ of a group $G$, one defines an equivalence between  two arbitrary elements $g_1,g_2 \in G$ if they are identical up to multiplication by elements of H
\begin{equation}
  \label{equi}
  g_1\sim g_2\;\mbox{iff}\; g^{-1}_1 g_2\, \in H \, .
\end{equation}
The set of elements in $G$ which are equivalent to $g\in G$ is called the left coset (modulo $H$) associated with $g$ and is denoted by 
\begin{equation}
  \label{equi2}
  g\,H = \{ gh\,| h\in H\}\, .
\end{equation} 
The space of cosets is called the coset space and denoted by 
\begin{equation}
  \label{equi3}
  G/H= \{ gH\,| g\in G\}\, .
\end{equation}
If $N$ is an {\em invariant} or {\em normal} subgroup, i.e.~if $gNg^{-1} = N $ for all $g\,\in G$, the coset space is actually a group with the product defined by $(g_1 N)\cdot (g_2 N)= g_1g_2 N$. It is called the {\em quotient} or {\em factor} group of $G$ by $N$. \\
As an example we consider the group of translations in $\mathbb{R}^{3}$. Since this is an abelian group, each subgroup is normal and can therefore be used to define factor groups. Consider $N=T_{x}$ the subgroup of translations in the $x-$direction. The cosets are translations in the y-z plane followed by an arbitrary translation in the $x-$direction. The factor group consists therefore of translations with unspecified  parameter for the translation in the $x-$direction.   As a further example consider rotations $R\left(\varphi\right)$ around a point in the $x-y$ plane. The two elements 
\begin{displaymath}
e=R\left(0\right) \quad, \quad r=R\left(\pi\right)
\end{displaymath}
form a normal subgroup $N$ with the factor group given by  
\begin{displaymath}
G/N=\left\{R\left(\varphi\right) N | 0\leq \varphi < \pi\right\}\, .
\end{displaymath}
Homotopy groups of coset spaces can  be calculated with the help of the following two identities for  connected and simply connected Lie-groups such as $SU(n)$. With $H_0$ we denote the component of $H$ which is connected to the neutral element $e$. This component of $H$ is  an invariant subgroup of $H$. To verify this, denote with $\gamma(t)$ the continuous curve which connects the unity $e$ at $t=0$ with an arbitrary element $h_0$ of $H_0$. With $\gamma(t)$ also $h\gamma(t)h^{-1}$ is part of $H_0$ for arbitrary $h\in H$.  Thus $H_0$ is a normal subgroup of $H$ and the coset space $H/H_0$ is a group. One extends the definition of the homotopy groups and defines 
\begin{equation}
  \label{pi0}
 \pi_0(H)=H/H_0\, .  
\end{equation}
The following identities hold (cf.\ \cite{STEE51}, \cite{MORA92})
\begin{equation}
  \label{th0}
 \pi_1(G/H)= \pi_0(H)\, , 
\end{equation}
and
\begin{equation}
  \label{th2}
 \pi_2(G/H)= \pi_1(H_0)\, .   
\end{equation}
Applications of these identities to coset spaces of $SU(2)$ will be important in the following. We first observe that, according to the parameterization (\ref{su2pa}), together with the neutral element $e$ also $-e$ is an element of $SU(2)$ $\big(\phi=0, \pi$ in Eq.\ (\ref{su2pa})$\big)$. These 2 elements  commute with all elements of SU(2) and form a subgroup , the {\em center} of $SU(2)$
\begin{equation}
  \label{center}
Z\big(SU(2)\big)= \{\,e,-e\,\}\sim \mathbb{Z}_2\, . 
\end{equation} According to the identity (\ref{th0}) the fundamental group of the factor group is non-trivial
\begin{equation}
  \label{pi1suz}
  \pi_1\Big(SU(2)/Z(SU(2))\Big)= \mathbb{Z}_2\, .
\end{equation}
As one can see from the following argument, this result implies that the group of rotations in 3 dimensions $SO(3)$ is not simply connected. Every  rotation matrix  $R_{ij} \in SO(3)$ can be represented in terms of  $SU(2)$ matrices (\ref{su2pa})
$$ R_{ij}[U] = \frac{1}{4}\,\mbox{tr}\,\Big\{U\tau^i U^{\dagger}\tau^{j}\Big\}\, .$$ 
The $SU(2)$ matrices  $U$ and $-U$   represent the same $SO(3)$ matrix.  Therefore,   
 \begin{equation} 
  \label{sou}
SO(3)\sim SU(2)/\mathbb{Z}_2\,
\end{equation}
and thus 
\begin{equation} 
  \label{so3}
\pi_1\Big(SO(3)\Big)= \mathbb{Z}_2\, ,  
\end{equation}
i.e.~$SO(3)$ is not simply connected.\\
We have verified above that the loops $u_{2n}(s)$ (\ref{lpsu2}) can be shrunk in SU(2) to a point. They also can be shrunk to a point on $SU(2)/\mathbb{Z}_2$. The loop 
\begin{equation}
  \label{lpsu2z2}
  u_1(s)= \exp\{i\pi s\tau^3\}
\end{equation}
connecting antipodal points however is topologically stable on  $SU(2)/\mathbb{Z}_2$, i.e.~it cannot be deformed continuously to a point,   while its square, $u_1^2(s)=u_2(s)$ can be.\\
The identity (\ref{th2}) is important for the spontaneous symmetry breakdown with a remaining $U(1)$ gauge symmetry. Since the groups $SU(n)$  are simply connected, one obtains
\begin{equation}
  \label{th2su}
 \pi_2\Big(SU(n)/U(1)\Big)= \mathbb{Z} .   
\end{equation}
\vskip .2cm
\subsection{Transformation Groups}
Historically, groups arose as collections of permutations or one-to-one transformations of a set $X$ onto itself with composition of mappings as the group product. If $X$ contains just $n$ elements, the collection $S\left(X\right)$ of all its permutations is the symmetric group with $n!$ elements.
In F. Klein's approach, to each geometry is associated a group of transformations of the underlying space of the geometry. For example, the group $E(2)$  of Euclidean plane geometry is the subgroup of $S\left(E^{2}\right)$ which leaves the distance $d\left(x,y\right)$ between two arbitrary points in the plane ($E^2$)  invariant, i.e.~a transformation 
\begin{displaymath}
T:E^{2}\rightarrow E^{2}
\end{displaymath}
is in the group iff
\begin{displaymath}
d\left(Tx,Ty\right)=d\left(x,y\right).
\end{displaymath}
The group $E(2)$ is also called the group of {\em rigid motions}. It  is generated by translations, rotations, and reflections. Similarly, the general Lorentz group is the group of Poincar$\acute{\mathrm{e}}$ transformations which leave the (relativistic) distance between two space-time points invariant. The interpretation of groups as transformation groups is very important in physics. Mathematically, transformation groups are defined in the following way (cf.\cite{MILL72}):\\
{\bf Definition :} A Lie group $G$ is represented as a group of transformations of a manifold $X$ (left action on X) if there is associated with each $g \in  G$ a diffeomorphism of $X$ to itself
\begin{displaymath}
x\rightarrow T_{g}\left(x\right), \quad x\in X \quad \mbox{with} \quad T_{g_1g_2}=T_{g_1}T_{g_2}
\end{displaymath}
("right action" $T_{g_1g_2}=T_{g_2}T_{g_1}$) and if $T_{g}\left(x\right)$ depends smoothly on the arguments $g,x$. 

If $G$ is any of the Lie groups $GL\left(n,R\right), O\left(n,R\right), GL\left(n,C\right), U\left(n\right)$ then $G$ acts in the obvious way on the manifold $\mathbb{R}^{n}$ or $C^{n}$.

The {\it orbit} of $x\in X$ is the set 
\begin{equation}
\label{orbit}
G x=\left\{T_{g}\left(x\right)|g\in G\right\}\subset X 
\end{equation}
The action of a group $G$ on a manifold $X$ is said to be {\it transitive} if for every two points $x,y \in X$ there exists $g\in G$ such that $T_{g}\left(x\right)=y$, i.e.~if the orbits satisfy $Gx=X$ for every $x\in X$ . Such a manifold is called a {\it homogeneous} space of the Lie group. 
The prime example of a homogeneous space is $\mathbb{R}^{3}$ under translations; every two points can be connected by translations. Similarly, the group of  translations acts transitively on the $n-$dimensional torus $T^{n}=\left(S^{1}\right)^{n}$ in the following way: 
\begin{displaymath}
T_{y}\left(z\right)=\left(e^{2i\pi\left(\varphi_{1}+t_{1}\right)},...,e^{2i\pi\left(\varphi_{n}+t_{n}\right)}\right)
\end{displaymath}
with
\begin{displaymath}
y=\left(t_{1},...,t_{n}\right) \in\mathbb{R}^{n},\quad  z=\left(e^{2i\pi\left(\varphi_{1}\right)},...,e^{2i\pi\left(\varphi_{n}\right)}\right) \in T^{n}\, .
\end{displaymath}
If the translations are given in terms of integers, $t_{i}=n_{i}$, we have $ T_{{\bf n}}\left(z\right)=z$. This is a subgroup of the translations and is defined more generally:\\
{\bf Definition :} The {\it isotropy group} $H_{x}$ of the point $x\in X$ is the subgroup of all elements of $G$ leaving $x$ fixed and is defined by 
\begin{equation}
\label{isot}
H_{x}=\left\{g \in G |T_{g}\left(x\right)=x\right\}\, .
\end{equation}

The group $O\left(n+1\right)$ acts transitively on the sphere $S^{n}$ and thus $S^{n}$ is a homogeneous space for the Lie group $O\left(n+1\right)$ of orthogonal transformations of $\mathbb{R}^{n+1}$. The isotropy group of the point $x=\left(1,0,...0\right) \in S^{n}$ is comprised by all matrices of the form
\begin{displaymath}
\left( \begin{array}{cc}
1&0 \\ 0 & A \end {array} \right) \quad , \quad A\,\in O\left(n\right)
\end{displaymath}
describing rotations around the $x_{1}$ axis.\\
Given a  transformation group  $G$ acting on a manifold $X$,  we define orbits as the equivalence classes, i.e.~
\begin{displaymath}
x\sim y \quad \mbox{if for some g}\,\in\, G \quad y=g\, x .
\end{displaymath}
  For  $X= \mathbb{R}^{n}$ and   $G= O(n)$ the  orbits are concentric spheres and thus in one to one correspondence with real numbers   $r \geq 0$. This is a homeomorphism of    $\mathbb{R}^{n}/O \left(n\right)$ on the ray   $0 \leq r \leq\infty $ (which is  almost a manifold).\\
If one defines points on $S^2$  to be equivalent if they are  connected by a rotation around a fixed axis, the $z$ axis, the resulting quotient space $S^2/O(2)$ consists of all the points on $S^2$ with fixed azimuthal angle, i.e.~the quotient space is a segment
\begin{equation}
  \label{s2o2}
  S^2/O(2)=\{\theta\,|\, 0\le\theta \le \pi\}\,. 
\end{equation}
Note that in the integration over the coset spaces $\mathbb{R}^n/O(n)$ and $ S^2/O(2)$ the radial volume element $r^{n-1}$ and the volume element of the polar angle $\sin \theta$ appear respectively. \\
The quotient space  $X/G$    needs not be a manifold, it is then called an {\em orbifold}. If $G$ is a discrete group,  the fixed points in $X$ under the action of $G$ become singular  points on $X/G$ . For instance, by identifying  the points  ${\bf x}$ and $-{\bf x}$ of a plane, the fixed point ${\bf 0} \in\mathbb{R}^2$ becomes the tip of the cone $\mathbb{R}^2/\mathbb{Z}_2$. \vskip .1cm
Similar concepts are used for a  proper description of  the topological space of the degrees of freedom in gauge theories. Gauge theories contain redundant variables, i.e.~variables which are related to each other by gauge transformations. This suggests to define an  equivalence relation in the space of gauge fields $\big($cf.Eqs.\ (\ref{gt}) and (\ref{FGT})$\big)$  
\begin{equation}
  \label{gaeq}
  A_{\mu}\sim \tilde{A}_{\mu} \quad \mbox{if}\quad \tilde{A}_{\mu}= A_{\mu}^{[U]}\quad \mbox{ for some}\; U\, ,
\end{equation}
i.e.~elements of an equivalence class can be transformed into each other by  gauge transformations $U$, they are {\em gauge copies} of a chosen representative.   The equivalence classes 
\begin{equation}
  \label{GO}
O= \left\{A^{\left[U\right]} | U \in G \right\}  
\end{equation}
 with $A$ fixed and $U$ running over the set of gauge transformations  are called the  {\em gauge  orbits}.
Their elements  describe the  same physics. 
Denoting with  { ${\cal A}$}  the space of gauge configurations and with $G$ the space of gauge transformations, the coset space  of gauge orbits is denoted with  ${\cal A}/{\cal G}$. It is this space  rather than ${\cal A}$ which defines the physical configuration space of the gauge theory. As we will see later, under suitable assumptions concerning the asymptotic behavior of gauge fields, in Yang-Mills theories, each  gauge orbit is labeled by a topological invariant,  the {\em topological charge}. 
\subsection{Defects in Ordered Media}
\begin{figure}
\hspace{3cm} \epsfig{file=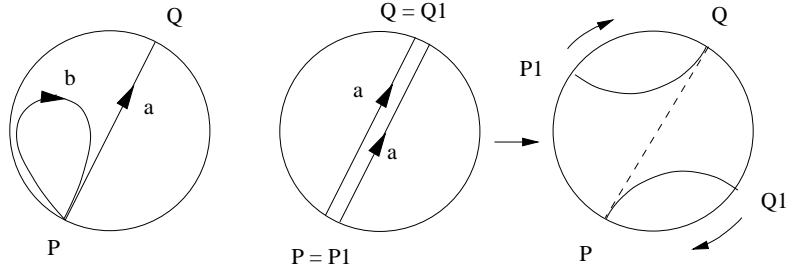, width=.65\linewidth}
\caption{ The left figure shows loops a, b which on $\mathbb{R}P^2$ can (b) and cannot (a) be shrunk to a point. The two figures on the right demonstrate  how two loops of the type a can be shrunk to one point.  By moving the point $P1$ together with its antipodal point $Q1$ two shrinkable loops of the type b are generated}
\label{rp2}
\end{figure}

\begin{figure}
\hspace{5cm} \epsfig{file=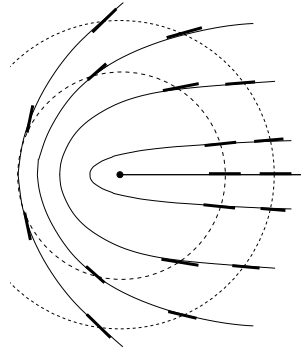, width=.3\linewidth}
\caption{Line defect in $\mathbb{R}P^2$. In addition to the directors also the  integral curves are shown }
\label{nemat2}
\end{figure}

\begin{figure}
\hspace{4cm}  \epsfig{file=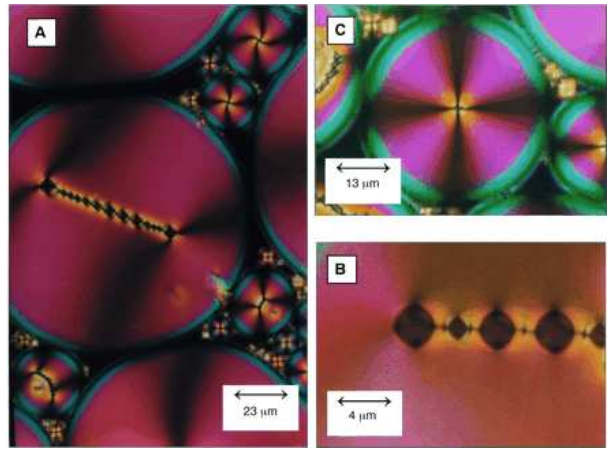, width=.5\linewidth}
\caption{Nematic drops (A) containing 1 (C) or more water droplets (B) (the Figure is taken from  \cite{PSLW97}). The distance between the defects is about 5 $\mu$m}
\label{nemat3}
\end{figure}

In condensed matter physics, topological methods find important applications in the investigations of properties of defects occurring in ordered media \cite{MERM79}. For applying topological arguments, one has to specify the topological space $X$ in which the fields describing the degrees of freedom are defined and the topological space $M$ (target space) of the values of the fields. In condensed matter physics the (classical) fields $\psi({\bf x})$ are called the order parameter and $M$ correspondingly the order parameter space. A system of spins or directors may be defined on lines, planes or in the whole space, i.e.~$X=\mathbb{R}^n$ with $n=1,2$ or $3$. The fields or order parameters describing spins are spatially varying unit vectors with arbitrary orientations: $M=S^2$ or if restricted to a plane $M=S^1$. The target spaces of directors  are the corresponding projective spaces $\mathbb{R}P^{n}$. A defect is a point, a line or a surface on which the order parameter is ill-defined. The defects are  defined accordingly as {\em point defects (monopoles), line defects (vortices, disclinations), or surface defects (domain walls)}. Such defects are topologically stable if they cannot be removed by a continuous change in the order parameter. Discontinuous changes require in physical systems of e.g. spin degrees of freedom substantial changes in a large number of the degrees of freedom and therefore  large energies. The existence of singularities alter the topology of the space $X$. Point and line defects induce respectively the following changes in the topology:  $X=\mathbb{R}^3 \to  \mathbb{R}^{3}\backslash\{0\}\sim S^2$ and $X=\mathbb{R}^3 \to  \mathbb{R}^{3}\backslash\mathbb{R}^1\sim S^1$.  Homotopy provides the appropriate tools to study the stability of defects. To this end, we proceed as in the abelian Higgs model and investigate the order parameter  on a circle or a 2-sphere sufficiently far away from the defect. In this way, the order parameter defines a mapping $\psi: S^n \rightarrow M$ and  the stability of the defects is guaranteed if the homotopy group $\pi_n(M)$ is non-trivial. Alternatively one may study the defects by removing from the space X the manifold on which the order parameter becomes singular.  The structure of the homotopy group has important implications for the dynamics of the defects. If the asymptotic circle encloses two defects, and if the homotopy group is abelian, than in a merger of the two defects the resulting defect is specified by the sum of the two integers characterizing the individual defects. In particular, winding numbers $\big(\pi_1(S^1)\big)$ and monopole charges  $\big(\pi_2(S^2)\big)$\, \big(cf.\ Eq.\ (\ref{pin})\big) are additive. \\
I conclude this discussion by  illustrating some of the results using the examples of magnetic systems represented by spins and nematic liquid crystals represented by directors, i.e.~ spins with indistinguishable heads and tails (cf.\ Eq.\ (\ref{prsp}) and the following discussion). 
If 2-dimensional spins ($M=S^1$)  or directors ($M=\mathbb{R}P^{1}$) live on a plane ($X=\mathbb{R}^2$), a defect is topologically stable. The punctured plane obtained by the removal of the defect is  homotopically equivalent to a circle (\ref{stp}) and  
  the topological stability follows from the non-trivial homotopy group  $\pi_{1}(S^{1})$ for magnetic substances. The argument applies to nematic substances as well since   identification of antipodal points of a circle  yields again a circle  
$$\mathbb{R}P^{1}\sim S^{1}\, . $$
On the other hand, a point defect  in a system of 3-dimensional spins $M=S^2$ defined on a plane $X=\mathbb{R}^2$  - or equivalently a line defect in $X=\mathbb{R}^3$ -   is not stable. Removal of the defect manifold generates once more a circle. The triviality of $\pi_1(S^2)$ (cf.\ Eq.\ (\ref{pi1n})) shows that the defect can be continuously deformed into a configuration where all the spins point into the same direction.   On   $S^{2}$ a loop can always be shrunk to a point (cf.\ Eq.\ (\ref{pi1n})). 
In nematic substances, there are stable point and line defects for $X=\mathbb{R}^2$ and $X=\mathbb{R}^3$, respectively, since
 $$\pi_{1}(\mathbb{R} P^2 )=\mathbb{Z}_{2} .$$ 
Non-shrinkable loops on $RP^2$ are  obtained by  connecting a given  point on $S^2$ with its antipodal one. Because of the identification of antipodal points, the line connecting the two points cannot be contracted to a point. In the identification, this line on $S^2$ becomes a non-contractible loop on $RP^2$. Contractible and non-contractible loops   on $RP^2$ are  shown in Fig. \ref{rp2} . 
This Figure also demonstrates that connecting  two antipodal points with two different lines  produces a contractible loop.  Therefore the space of loops contains only two inequivalent classes. 
More generally, one can show (cf.\ \cite{MUNK2000})
\begin{equation}
  \label{hopr1}
 \hspace{0.1cm} \pi_1(\mathbb{R}P^n)=\mathbb{Z}_2\, ,\quad \hspace{.8cm} n\ge 2\, ,
\end{equation}
and (cf.\ \cite{NAKA90})
\begin{equation}
  \label{hoprn}
  \pi_n(\mathbb{R}P^m)=  \pi_n(S^m)\,,\quad n\ge 2\, .
\end{equation}
Thus, in 3-dimensional nematic substances  point defects (monopoles), also present in magnetic substances, and line defects (disclinations), absent in magnetic substances, exist. In Fig. \ref{nemat2} the topologically stable line defect is shown. The circles around the defect are mapped by $\theta(\varphi)=\frac{\varphi}{2}$ into $\mathbb{R}P^2$. Only due to the identification of the directions $\theta\sim\theta+\pi$  this mapping is continuous. For magnetic substances, it would be discontinuous along the $\varphi=0$ axis.\\
Liquid crystals  can be considered with regard to their underlying dynamics as close relatives to some of the fields of particle physics. They exhibit spontaneous orientations, i.e.~form ordered media with respect to  'internal' degrees of freedom not joined by formation of  a crystalline structure. Their topologically stable defects are also encountered in gauge theories as we will see later.  Unlike the fields in particle physics, nematic substances can be manipulated and, by their birefringence property,  allow for a beautiful visualization of the structure and dynamics of defects (for a thorough discussion of the physics of liquid crystals and their defects (cf.\ \cite{CHAN92,DGPR93}). 
These substances offer the opportunity to study on a macroscopic level, emergence of monopoles and their dynamics. For instance, by enclosing a water droplet in a nematic liquid drop, the boundary conditions on the surface of the water droplet and 
 on the surface of the nematic drop cooperate to  generate  a monopole (hedgehog) structure which, as Fig. \ref{nemat3}C  demonstrates, can be observed via its peculiar birefringence properties, as a four armed star of alternating bright and dark regions.  If more water droplets are dispersed in a nematic drop, they form chains (Fig. \ref{nemat3}A) which consist of the water droplets alternating with hyperbolic defects of the nematic liquid (Fig. \ref{nemat3}B). The non-trivial topological properties stabilize these objects for as long as  a couple of weeks \cite{PSLW97}. 
\vskip .1cm  
In all the examples considered so far, the relevant fundamental groups were abelian. In nematic substances the ``biaxial nematic phase'' has been identified (cf.\ \cite{CHAN92}) which is characterized by a non-abelian fundamental group. The elementary constituents of this phase can be thought of as rectangular boxes rather than rods which, in this phase are aligned. Up to $180^{\,\circ}$  rotations around the 3 mutually perpendicular axes ($R_i^{\pi}$), the orientation of such a box is specified by an element of the rotation group $SO(3)$. The order parameter space of such a system is therefore given by 
$$M=SO(3)/{\mathbb D_2},\quad {\mathbb D}_2=\{{\mathbb 1}, R_1^{\pi} , R_2^{\pi} , R_3^{\pi} \} .$$ 
By representing the rotations  by elements of $SU(2)$  (cf.\ Eq.\ (\ref{sou})), the group ${\mathbb D_2}$ is extended to the group of 8 elements, containing the Pauli matrices (\ref{pauli}), 
$${\mathbb Q}=\{\pm {\mathbb 1}, \pm \tau^1,\pm \tau^2,\pm \tau^3 \}\, ,$$
the group of {\em quaternions}. 
With the help of the identities (\ref{pi0}) and (\ref{th0}),  we derive
\begin{eqnarray}
  \label{quat}
\pi_1(SO(3)/{\mathbb D_2})\sim \pi_1(SU(2)/{\mathbb Q})\sim {\mathbb Q}\, .  
\end{eqnarray}
In the last step it has been used that in a discrete group the connected component of the identity contains the identity only. 

The non-abelian nature of the fundamental group has been predicted to have important physical consequences for the behavior of defects in the nematic biaxial phase. This concerns in  particular the coalescence of defects and the possibility of entanglement of disclination lines \cite{CHAN92}.

\section{Yang-Mills Theory}
 In this introductory section I review concepts, definitions, and  basic properties of gauge theories.  \\ 
{\em Gauge fields} 
\\
In non-abelian gauge theories, {gauge fields} {  are matrix-valued functions of space-time. In SU(N) gauge theories they can be represented by the   generators of the corresponding Lie algebra, i.e.~gauge fields and their color components are related by 
\begin{equation}
\label{gfla}
A_{\mu}(x)=A_{\mu}^{a}(x)\frac{\lambda^{a}}{2}\, ,
\end{equation}
where the color sum runs over the {$N^{2}-1$} generators. The  generators are hermitian, traceless $N\times N$ matrices whose commutation relations  are specified by the structure constants  {$f^{abc}$ 
$$
\left[ \frac{\lambda^{a}} {2}, \frac{\lambda^{b}}{2} \right]=i f^{abc}\frac{\lambda^{c }}{2} \, .
$$ 
The normalization is chosen as 
$$
\mbox{tr}\left( \frac{\lambda^{a}} {2} \cdot  \frac{\lambda^{b}}{2} \right)= \frac{1}{2}\delta_{ab} .
$$
Most of our applications will be concerned with {$SU(2)$} gauge theories; in this case the generators are the Pauli matrices  (\ref{pauli}) 
$$\lambda^{a} =\tau^a\, ,$$
with structure constants
 $$f^{abc}=\epsilon^{abc}.$$
Covariant derivative, field strength tensor, and its color components are respectively defined by 
\begin{equation}
D_{\mu} = \partial_{\mu}+igA_{\mu}, 
\end{equation}
\begin{equation}
\label{fs}
F^{\mu\nu} = \frac{1}{ig}[D_{\mu},D_{\nu}],\quad F_{\mu\nu}^a = \partial_\mu A_\nu^a - \partial_\nu A_\mu^a 
             - gf^{abc} A_\mu^b A_\nu^c .
\end{equation}
The definition of  { electric and magnetic fields} in terms of the field strength tensor is the same as in electrodynamics
 \begin{equation}
\label{eb}
 E^{i a}\left(x\right)=-F^{0i a}\left(x\right) \quad , \quad  
B^{i a}\left(x\right)=-\frac{1}{2}\epsilon^{ijk}F^{jk a}\left(x\right) \ .\end{equation}
 The dimensions  of gauge field and field strength in  {4} dimensional space-time are  
$$ [A] = \ell^{-1} ,\quad [F] = \ell^{-2}  $$
and therefore in  absence of a scale
$$ A_{\mu}^{a} \sim M^{a}_{\mu\nu}\frac{x^{\nu}}{x^2}\, ,$$
with arbitrary constants $M^{a}_{\mu\nu}$. 
In general, the action associated with these fields exhibits infrared and ultraviolet  logarithmic divergencies.     {
In the following we will discuss 
\begin{itemize}
 \item  { Yang-Mills Theories}\\
Only gauge fields are present. The Yang-Mills Lagrangian is 
{ \begin{equation} 
\label{LYM}
{\cal L}_{YM} =  - \frac{1}{4} F^{\mu\nu a} F_{\mu\nu}^a = - \frac{1}{2}\mbox{tr}\, \left\{F^{\mu\nu} F_{\mu\nu}\right\}= \frac{1}{2}({\bf E}^2-{\bf B}^2) .\end{equation}} 
\item  {Quantum Chromodynamics}\\
QCD contains besides the gauge fields (gluons), fermion fields (quarks). Quarks are in the fundamental representation, i.e.~in {SU(2)} they are represented by {2-component} color spinors. The QCD Lagrangian is (flavor dependences suppressed) 
 \begin{equation} {\cal L}_{QCD} = {\cal L}_{YM} +{\cal L}_{m},\quad   {\cal L}_{m}={\bar \psi}\left(i \gamma^{\mu}D_{\mu}-m\right)\psi, 
\label{matt}
\end{equation}
with the action of the covariant derivative on the quarks given by
$$(D_{\mu}\psi)^{i} = (\partial_{\mu}\delta^{ij}+ig A_{\mu}^{ij})\,\psi^{j}\quad i,j =1\ldots N\, . $$
\item  {Georgi-Glashow Model}\\
In the Georgi-Glashow model \cite{GEGL72} (non-abelian Higgs model), the gluons are coupled  to a scalar, self-interacting ($V(\phi)$)  (Higgs) field in the adjoint representation. The Higgs field has the same representation in terms of the generators as the gauge field (\ref{gfla}) and can be thought of as a 3-component color vector in $SU(2)$. Lagrangian and action of the covariant derivative are respectively
 \begin{equation} {\cal L}_{GG} = {\cal L}_{YM} +{\cal L}_{m},\quad   {\cal L}_{m}=\frac{1}{2} D_{\mu} \phi D^{\mu} \phi - V (\phi)\, ,
\label{GEGL}
\end{equation}
\begin{equation} 
\label{dphi}
(D_{\mu}\phi)^{a}= \left[ D_{\mu},\phi \,\right]\, ^{a}= (\partial_{\mu}\delta^{ac}-gf^{abc} A_{\mu}^{b})\phi^{c}\, .
\end{equation}
\end{itemize}
{\em Equations of Motion}\\
The principle of least action 
$$ \delta S =0,\quad S=\int d^4 x {\cal L} $$
yields when varying the gauge fields 
\begin{eqnarray*}
\delta S_{YM}&=&-\int d^{4}x \,\mbox{tr}\left\{F_{\mu \nu}\delta F^{\mu \nu}\right\} =-\int d^{4}x \,\mbox{tr}\left\{F_{\mu \nu} \frac{2}{ig} \left[D^{\mu},\delta A^{\nu}\right] \right\}\\
&=& 2\int d^{4}x\, \mbox{tr}\left\{\delta A^{\nu} \left[D^{\mu}, F_{\mu \nu}\right]\right\} 
\end{eqnarray*}
the  {inhomogeneous} field equations 
\begin{equation}
  \label{emga}
\left[D_{\mu}, F^{\mu \nu}\right] = j^{\nu}\, ,    
\end{equation}
with  $ j^{\nu}$ the {color current} associated with the matter fields 
 \begin{equation}j^{a\nu} = \frac{\delta {\cal L}_{m}}{\delta A_{\nu}^{a}}.
\label{jmat}
\end{equation} 
For QCD and the Georgi-Glashow model, these currents are given respectively by  
 \begin{equation}
 j^{a \nu} = g\bar{\psi}\gamma ^{\nu}\frac{\tau^{a}}{2} \psi \,,\quad j^{a \nu} = g f^{abc}\phi^{b}(D^{\nu} \phi)^{c}\, .
\end{equation} 
As in electrodynamics, the  {homogeneous} field equations for the Yang-Mills field strength 
$$
\left[D_{\mu}, \tilde{F}^{\mu \nu}\right] = 0 \, ,$$
with the dual field strength tensor 
\beq
\label{DFD}
\tilde{F}^{\mu \nu} = \frac{1}{2}\,
\epsilon ^{\mu \nu \sigma \rho} F_{\sigma \rho}\, ,
\eeq
are obtained as the Jacobi identities of the covariant derivative
$$ [D_{\mu},[D_{\nu},D_{\rho}]]+[D_{\nu},[D_{\rho},D_{\mu}]]+[D_{\rho},[D_{\nu},D_{\mu}]]=0 \, ,$$
i.e.\,they follow from the mere fact that the field strength is represented in terms of gauge potentials. \\
 { \em Gauge Transformations}
\\
 Gauge transformations change the color orientation of the matter fields locally, i.e.~in a space-time dependent manner, and are defined as 
\beq
\label{lgt}
U \left(x\right)= \exp\left\{ig \alpha \left(x\right)\right\}=\exp\left\{ig \alpha^{a} \left(x\right) \frac{\tau ^{a}}{2}\right\} \ ,
\eeq
with the arbitrary gauge function $\alpha^{a} \left(x\right)$.
 {Matter fields transform  {covariantly} with} $U$ 
\begin{equation}
\label{matr}
\psi \rightarrow U\psi\,,\quad \phi \rightarrow  U\phi U^{\dagger}.\end{equation}
The transformation property of $A$ is chosen such that the covariant derivatives of the matter fields $D_{\mu}\psi$  and $D_{\mu}\phi$ { transform as} the matter fields $\psi$ and $\phi$ respectively.   As in electrodynamics, this requirement makes the gauge fields transform  {inhomogeneously} 
 \begin{equation}
\label{FGT}
A_{\mu} \left(x\right) \rightarrow U \left(x\right) \left(A_{\mu}\left(x\right)+\frac{1}{ig} \partial _{\mu} \right) U^{\dagger} \left(x\right)= A^{\, \left[\,U\right] }_{\mu} \left(x\right)
\end{equation}
resulting in a covariant transformation law for the field strength
\begin{equation}
\label{ufu} 
F_{\mu\nu} \rightarrow  UF_{\mu\nu} U^{\dagger} .
\end{equation}
  Under  {infinitesimal} gauge transformations  ($|g\alpha^{a} \left(x\right)| \ll 1$)
\begin{equation}
  \label{ifgt}
A_{\mu} ^{a}\left(x\right) \rightarrow  A_{\mu}^{a} \left(x\right)  - \partial _{\mu}\alpha^{a} \left(x\right) - g f^{abc}\alpha^{b} \left(x\right) A_{\mu} ^{c}\left(x\right) .
\end{equation}
As in electrodynamics, gauge fields which are gauge transforms of $A_{\mu}=0$ are called pure gauges (cf.\ Eq.\ (\ref{pgqd})) and are, according to (\ref{FGT}), given by
\begin{equation}
  \label{puga}
A_{\mu}^{pg} \left(x\right)= U \left(x\right)\frac{1}{ig}\, \partial _{\mu} \,U^{\dagger} \left(x\right)\, . 
\end{equation}
Physical observables must be independent of the choice of gauge (coordinate system in color  space).  
Local quantities such as the Yang-Mills action density  \mbox{tr}\,$F^{\mu\nu}(x)F_{\mu\nu}(x)$   {or matter field bilinears like} $\bar{\psi}(x)\psi(x), \phi^{a}(x)\phi^{a}(x)$   {are {gauge invariant}, i.e.~their value does not change under local gauge transformations. 
One also introduces {non-local} quantities which, in generalization of the transformation law (\ref{ufu}) for the field strength, change homogeneously under gauge transformations. In this construction a basic building block is the} { path ordered integral}
\begin{eqnarray}
\Omega\left(x,y,{\cal C}\right)&=& P \exp\left\{-ig \int_{s_0}^{s} d\sigma \frac{dx^{\mu}}{d\sigma}A_{\mu} \Big(x(\sigma)\Big)\right\}= P \exp\left\{-ig \int_{{\cal C}} dx^{\mu}A_{\mu} \right\} .\nonumber\\
\label{PI}
\end{eqnarray} 
It describes a  {gauge string} between the space-time points $x=x(s_0)$ and $y=x(s)$.
 $\Omega$ satisfies the differential equation  
\begin{equation}
  \label{dipo}
\frac{d\Omega}{ds}= -ig \frac{dx^{\mu}}{ds}A_{\mu} \Omega .
\end{equation}
 Gauge transforming this differential equation yields the transformation property of $\Omega$
\begin{equation}
  \label{gaom}
  \Omega\left(x,y,{\cal C}\right) \rightarrow  U\left(x\right)\Omega\left(x,y,{\cal C}\right) U^{\dagger}\left(y\right) .\end{equation}
With the help of $\Omega$,  non-local, gauge invariant  quantities like 
$$ \mbox{tr} F^{\mu\nu}(x)\Omega\left(x,y,{\cal C}\right)F_{\mu\nu}(y)\,, \quad  \bar{\psi}(x)\Omega\left(x,y,{\cal C}\right)\psi(y),$$ 
 or  closed gauge strings -   $\big($SU(N)\big) Wilson loops 
 \begin{equation}
   \label{wlop}
W_{\cal C}= \frac{1}{N} \mbox{tr}\, \Omega\left(x,x,{\cal C}\right)    
 \end{equation}
 can be constructed. For pure gauges (\ref{puga}), the differential equation (\ref{dipo}) is solved by 
 \begin{equation}
   \label{wlpg}
\Omega^{pg}\left(x,y,{\cal C}\right) = U(x)\, U^{\dagger}(y) .   
 \end{equation}
{While  $\bar{\psi}(x)\Omega\left(x,y,{\cal C}\right)\psi(y)$ is an operator which connects the vacuum with  {meson states} for   {$SU(2)$} and {$SU(3)$},  {fermionic baryons} appear only in {$SU(3)$} in which gauge invariant states containing an odd number of fermions can be constructed.
In  {SU(3)} a point-like gauge invariant baryonic state is obtained  by creating three quarks in a color antisymmetric state at the same space-time point
$$\psi(x) \sim \epsilon^{abc}\psi^{a}(x)\psi^{b}(x)\psi^{c}(x) .$$
Under gauge transformations}, 
\begin{eqnarray*}
\psi(x) &\rightarrow& \epsilon^{abc}U_{a\alpha}(x)\psi^{\alpha}(x)U_{b\beta}(x)\psi^{\beta}(x)U_{c\gamma}(x)\psi^{\gamma}(x)\\
&=& \det\big(U(x)\big)\epsilon^{abc}\psi^{a}(x)\psi^{b}(x)\psi^{c}(x)\, .
\end{eqnarray*}
Operators that create finite size baryonic states must contain appropriate  gauge strings as given by the following expression
$$\psi(x,y,z) \sim\epsilon^{abc}[\Omega(u,x,{\cal C}_{1}) \psi(x)]^a\, [\Omega(u,y,{\cal C}_{2}) \psi(y)]^b\, [\Omega(u,z,{\cal C}_{3}) \psi(z)]^c \, .$$ }  The presence of these gauge strings makes $\psi$ gauge invariant as is easily verified with the help of the transformation property (\ref{gaom}). Thus, gauge invariance is enforced by  color exchange processes taking place between the quarks. 
\vskip .1cm
{\em  Canonical Formalism}
\vskip .1cm
The canonical formalism is developed in the  same way as in electrodynamics. Due to  the antisymmetry of $F_{\mu\nu}$, the Lagrangian (\ref{LYM}) does not contain the time derivative of $A_0$ which, in the canonical formalism, has to be treated as a constrained variable. In the Weyl gauge \cite{WEYL,JACK}
 \begin{equation}
   \label{WEYL}
  A_{0}^{a} = 0,\quad a=1....N^2-1 ,  
 \end{equation}
these constrained variables are eliminated and the standard procedure of canonical quantization can be employed.
 In a first step,  the canonical momenta of gauge and matter fields (quarks and Higgs fields) are identified 
$$ \frac{\delta {\cal L}_{YM}}{\partial_{0}A_{i}^{a}} = -E^{a\,i}\,,\quad \frac{\delta {\cal L}_{mq}}{\partial_{0}\psi^{\alpha}} = i\psi^{\alpha\,\dagger}\,,\quad \frac{\delta {\cal L}_{mH}}{\partial_{0}\phi^{a}} = \pi^{a}\, .$$
By Legendre transformation,  one obtains the  {Hamiltonian density of the gauge fields} 
\begin{equation} 
\label{YMH}
{\cal H}_{YM} =   \frac{1}{2} (\mbox{\boldmath$E$}^{2}+\mbox{\boldmath$B$}^{2}),\end{equation}
 {and of the matter fields }
 \begin{equation}
   \label{HQCD}
\mbox{ QCD}:\quad {\cal H}_{m} =\psi^{\dagger}\left(\frac{1}{i}  \gamma^{0}\gamma^{i}D_{i}+\gamma^{0}m\right)\psi,   
 \end{equation}
 \begin{equation}
   \label{HGEGL} 
\hspace{-3.2cm} \mbox{ Georgi-Glashow model}:\quad {\cal H}_{m}=\frac{1}{2}\pi^{2} +\frac{1}{2}(\mbox{\boldmath$D$} \phi)^{2} + V(\phi) \, . \end{equation}
The gauge condition (\ref{WEYL}) does not fix the gauge  uniquely, it still allows for time-independent gauge transformations $U({\bf x})$, i.e.~gauge transformations which are generated by time-independent gauge functions $\alpha({\bf x})$ (\ref{lgt}). As a consequence the Hamiltonian  exhibits a local symmetry
\begin{equation}
  \label{hlsy}
  H= U({\bf x})\, H\,U({\bf x})^{\dagger}
\end{equation}
This residual gauge symmetry is taken into account by requiring  physical states  $|\Phi\rangle $ to satisfy the Gau\ss\ law, i.e.~the 0-component of the equation of motion (cf.\ Eq.\ (\ref{emga}))
$$\big([D_{i},E^i]+j^0\big)|\Phi\rangle = 0 .$$
In general,  the non-abelian Gau\ss\  law cannot be implemented in closed form which severely limits the applicability of the canonical formalism. A complete canonical formulation  has been given in axial gauge \cite{LeNT94} as will be discussed below. The connection of  canonical to path-integral quantization is discussed in  detail in \cite{LEWO01}.  
\section{`t Hooft-Polyakov Monopole}
The t'Hooft-Polyakov monopole \cite{THOO74,POLY74} is a topological excitation in the non-abelian Higgs or Georgi-Glashow model ($SU(2)$ color). We start with a brief discussion of the properties of this model with emphasis on  ground state configurations and their topological properties.
\subsection{Non-Abelian Higgs Model}
The Lagrangian (\ref{GEGL}) and  the equations of motion (\ref{emga}) and (\ref{jmat}) of the non-abelian Higgs model have been discussed in the previous section. For the following discussion we specify the self-interaction, which as in the abelian Higgs model is assumed to be a fourth order polynomial in the fields with the normalization chosen such that its minimal value is zero
\begin{equation}
  \label{SEIN}
 V(\phi) = \frac{1}{4}\lambda(\phi^{2}-a^{2})^{2}, \quad \lambda > 0\, .
\end{equation}
Since $\phi$ is a vector in color space and gauge transformations rotate the color direction of the  Higgs field (\ref{matr}),  $V$ is gauge invariant
\begin{equation}
  \label{vgin}
V(g \phi) = V (\phi)\, . 
\end{equation}
We have used the notation 
$$ g\phi =  U \phi U^{\dagger}\, , \quad g\in G=SU(2) .$$
The analysis of this model parallels that of the abelian Higgs model. Starting point is the energy density of static solutions, which in the Weyl gauge is given by 
\big(Eqs.\ (\ref{YMH}, \ref{HGEGL})\big) 
\begin{equation}
  \label{naende}
\epsilon({\bf x}) =  \frac{1}{2} {\bf B}^{2} + \frac{1}{2} ({\bf D}\phi)^{2}+ V (\phi).  
\end{equation}
The choice
\begin{equation}
  \label{mifi}
{\bf A}=0,  \quad \phi=\phi_0=\mbox{const.}\, , \quad V (\phi_0)=0    
\end{equation}minimizes the energy density. Due to the presence of the local symmetry of the Hamiltonian \big(cf.\ Eq.\ (\ref{hlsy})\big), this choice is not unique. Any field configuration connected to (\ref{mifi}) by a time-independent gauge transformation will also have vanishing energy density. Gauge fixing conditions by which the Gau\ss\  law constraint is implemented remove these gauge ambiguities; in general a global gauge symmetry remains (cf.\cite{LeNT940,LeNT94}). Under a  space-time independent gauge transformation 
\begin{equation}
  \label{coro}
g= \exp\left\{ig \alpha^{a} \frac{\tau ^{a}}{2}\right\} 
\,,\quad \alpha=\mbox{const} \, , 
\end{equation}
applied to a configuration (\ref{mifi}), the gauge field is unchanged as is the modulus of the Higgs field. The transformation rotates the spatially constant $\phi_0$. 
In such a ground-state configuration, the Higgs field exhibits  a spontaneous orientation analogous to the spontaneous magnetization of a ferromagnet,  
$$\phi=\phi_{0}
\quad,\quad |\phi_{0}| = a\, .   $$
This appearance of a phase with spontaneous orientation of the Higgs field is a consequence of a  vacuum degeneracy completely analogous to the vacuum degeneracy of the  abelian Higgs model with its  spontaneous orientation of the phase of the Higgs field.\\
Related to the difference in the topological spaces of the abelian and non-abelian Higgs fields, significantly different phenomena occur in  the spontaneous symmetry breakdown.  In the Georgi-Glashow model, the loss of rotational symmetry in color space is not complete. While the configuration (\ref{mifi}) changes under the (global) color rotations (\ref{coro}) and does therefore not reflect the invariance of the Lagrangian or Hamiltonian of the system, it remains invariant under  rotations around the axis in the direction of the Higgs field $\alpha \sim \phi_0$. These transformations form a subgroup of the group of rotations (\ref{coro}), it is the  isotropy group (little group, stability group)  \big(for the definition cf.\ (\ref{isot})\big) of transformations which leave $\phi_0$ invariant   
\begin{equation}
  \label{lgro}
H_{\phi_{0}} = \left \{ h \in SU(2) | h \phi _{0} = \phi _{0} \right \}.  
\end{equation}
The space of the zeroes of $V$, i.e.~the space of vectors $\phi$ of fixed length $a$, is $S^2$ which is a homogeneous space \big(cf.\ the discussion after Eq.\ (\ref{orbit})\big) with all elements being generated by application of arbitrary transformations $g\in G$ to a (fixed) $\phi_0$. 
The space of zeroes of  { $V$} and the coset space    $G/H_{\phi_0}$ are  mapped onto each other by 
$$ F_{\phi_0}: G/H_{\phi_0} \rightarrow \{\phi|V(\phi)=0\} \, ,\quad  F_{\phi_0}(\tilde{g})= g\phi_{0}=\phi $$
with $g$ denoting a representative of the coset $\tilde{g}$.
This mapping is bijective. The space of zeroes is homogeneous and therefore all zeroes of $V$ appear as an image of some $\tilde{g}\in G/H_{\phi_0}$.  This mapping is injective since  $\tilde {g}_1\phi_{0}=\tilde {g}_2\phi_{0}$  implies  ${g}_1^{-1}\phi_{0}g_2 \in H_{\phi_0}$ with $g_{1,2}$ denoting representatives of the corresponding cosets  $\tilde{g}_{1,2}$ and therefore  the two group elements belong to the same equivalence class \big(cf.\ Eq.\ (\ref{equi})\big) i.e.~$\tilde {g}_1= \tilde {g}_2$. Thus, these two spaces are homeomorphic 
\begin{equation}
  \label{ghs2}
G/H_{\phi_0} \sim S^2\, .  
\end{equation}
It is instructive to compare  the topological properties of the abelian and non-abelian Higgs model.
\begin{itemize}
\item
  In the abelian Higgs model, the gauge group is 
 $$G = U(1)$$
and by the requirement of gauge invariance, the self-interaction is of the form
$$  V(\phi) = V (\phi^{*} \phi) .$$ 
The vanishing of $V$  determines the modulus of $\phi$ and leaves the phase undetermined   
 $$V=0 \quad \Rightarrow  | \phi_{0} | = a e^{i\beta} .$$
After choosing the phase $\beta$, no residual symmetry is left, only  multiplication with  $1$   leaves  $\phi_{0} $ invariant, i.e.~
\begin{equation}
  \label{hu1}
H =\{e\}\, ,  
\end{equation}
and thus
\begin{equation}
  \label{fcu1}
G/H = G\sim S^{1}.  
\end{equation}
\item
 In the non-abelian Higgs model, the gauge group is 
  $$ G =SU(2),$$
and by the requirement of gauge invariance, the self-interaction is of the form
$$ V(\phi) = V (\phi^{2})\, ,\quad  \phi^2= \sum_{a=1,3}  \phi^{a\, 2}.$$ 
The vanishing of $V$ determines the modulus of $\phi$ and leaves the orientation  undetermined
  $$V=0 \quad \Rightarrow \quad  \phi_{0}=a\hat{\phi}_0 .$$
After choosing the orientation $\hat{\phi}_0$,  a residual symmetry persists, the invariance of $\phi_{0}$  under (true) rotations  around the  { $\phi_{0}$} axis and under  multiplication with an element of the center of $SU(2)$ \big(cf.\ Eq.\ (\ref{center})\big)
\begin{equation}
  \label{hsu2}
 H=U(1)\otimes \mathbb{Z}_2\, ,  
\end{equation}
 and thus
 \begin{equation}
   \label{fcsu2}
G/H=SU(2)/\big(U(1)\otimes \mathbb{Z}_2\big)\sim S^{2}.   
 \end{equation}
\end{itemize}}
\subsection{The Higgs Phase}
To display the physical content of the Georgi-Glashow model we consider small oscillations around the ground-state configurations (\ref{mifi}) - the normal modes of the classical system and the particles of the quantized system.  The analysis of the normal modes simplifies greatly if the gauge theory is represented in  the unitary gauge, the gauge  which makes the particle content manifest. In this gauge, components of the Higgs  field rather than those of the gauge field (like the longitudinal gauge field in Coulomb gauge) are eliminated as redundant variables. The Higgs field is used to define the coordinate system in  internal space
  \begin{equation}
    \label{ugc}
  \phi(x) = \phi^{a}(x) \frac{\tau^{a}}{2}=
  \rho(x) \frac{\tau^{3}}{2} .  
\end{equation} 
Since this gauge condition does not affect the gauge fields, the Yang-Mills part of the Lagrangian (\ref{LYM}) remains unchanged and
the contribution of the Higgs field (\ref{GEGL}) simplifies
\begin{equation}
  \label{gugg}
   {\cal L} = -\frac{1}{4}F^{a\mu\nu}F_{\mu\nu}^a+ 
\frac{1}{2}\partial_{\mu}\rho\partial^{\mu}\rho +g^{2}\rho^{ \,
2}A^{-}_{\mu} A^{+\,\mu}-V(|\rho|)\, ,
\end{equation}
with the ''charged'' components of the gauge fields defined by
\begin{equation}
  \label{spba}
 A^{\pm}_{\mu}\,  =
  \frac{1}{\sqrt{2}}(A_{\mu}^{1}\mp i A_{\mu}^{2}).  
\end{equation}
For small oscillations we expand  the Higgs field $\rho(x)$  around the value in the zero-energy configuration (\ref{mifi})
\begin{equation}
  \label{smam}
\rho(x) = a +\sigma(x),\quad |\sigma|\ll a  \, .  
\end{equation}
To leading order, the interaction with the  Higgs field makes the charged components (\ref{spba}) of the gauge fields massive with the value of the mass given by the  value of $\rho(x)$ in the zero-energy configuration
\begin{equation}
  \label{hima}
M^2=g^2 a^2\, .  
\end{equation}
The fluctuating Higgs field $\sigma(x)$ acquires its mass through the self-interaction
\begin{equation}
  \label{hima2}
m_{\sigma}^2= V^{\prime\prime}_{\rho=a}= 2\,a^2\, .  
\end{equation}
The neutral vector particles $A_{\mu}^3$,  i.e.~the color component of the gauge field along the Higgs field,  remains massless. This is a consequence of the survival  of the non-trivial isotropy group $H_{\phi_0}\sim U(1)$  \big(cf.\ Eq.\ (\ref{lgro})\big) in the  symmetry breakdown of the gauge group $SU(2)$. By coupling to a second Higgs field, with expectation value  pointing in a color direction different from $\phi_0$, a further  symmetry breakdown can be achieved which is complete up to the discrete $\mathbb{Z}_2$ symmetry \big(cf.Eq.\ (\ref{fcsu2})\big). In such a system  no massless vector particles can be present \cite{NIOL73,VISH94}.\\
Superficially it may appear that the emergence of massive vector particles in the Georgi-Glashow model happens almost with necessity. The subtleties of the procedure are connected to the gauge choice (\ref{ugc}). Definition of a  coordinate system in the internal color space via the Higgs field requires 
$$\phi\ne 0.$$
 This requirement can be enforced by the  choice of form (controlled by $a$) and strength $\lambda$ of the Higgs potential $V$ (\ref{SEIN}). Under appropriate circumstances, quantum or thermal fluctuations will only rarely give rise to configurations where   $\phi(x)$ vanishes at certain points and singular gauge fields (monopoles) are present. On the other hand, one expects at fixed $a$ and $ \lambda$   with increasing temperature the occurrence of a phase transition to a gluon-Higgs field plasma. Similarly, at $T=0$ a ``quantum phase transition'' ($T=0$ phase transition induced by variation of external parameters, cf.\ \cite{SGCS97})  to a confinement phase is expected to happen  when decreasing $a,\lambda$ . In the unitary gauge, these phase transitions should be accompanied  by a condensation of singular fields. When approaching either the  plasma or the  confined phase,  the dominance of the equilibrium positions  $\phi=0$ prohibits  a proper definition of a coordinate system in color space based on the the color direction of the Higgs field.\\
The fate of the discrete $\mathbb{Z}_2$ symmetry is  not  understood in detail. As will be seen, realization of the center symmetry indicates confinement. Thus, the $\mathbb{Z}_2$ factor should not be part of the isotropy group (\ref{hsu2}) in the Higgs phase. The gauge choice (\ref{ugc}) does not break this symmetry. Its breaking is a dynamical property of the symmetry. It must occur spontaneously. This $\mathbb{Z}_2$ symmetry must be restored in the quantum phase transition to the confinement phase and will remain broken in the transition to the high temperature plasma phase.        
\subsection{Topological Excitations} 
As in the abelian Higgs model, the non-trivial topology ($S^2$) of the manifold of vacuum field configurations of the Georgi-Glashow model  is the origin of the topological excitations.  We proceed as above and discuss field configurations  of finite energy  which differ in their topological properties from the ground-state configurations. As follows  from the expression (\ref{naende}) for the energy density, finite energy can result only if  asymptotically, $|{\bf x} | \rightarrow \infty$
$$\phi ({\bf x}) \rightarrow a \phi_0({\bf x})$$
$$ \hspace{-0.5cm}{\bf B} \rightarrow 0 $$
\begin{equation}
 \left[D_i \phi({\bf(x)})\right]^a = [\partial_i \delta^{ac} -  g \epsilon^{abc}A_i^b({\bf x})] \phi^c ({\bf x}) \rightarrow 0 \, ,
\label{asyna}
\end{equation} 
where $ \phi_0({\bf x})$ is a unit vector specifying the color direction of the Higgs field.
The last equation correlates asymptotically  the gauge and the Higgs field. In terms of the scalar field, the asymptotic gauge field is given by 
 \begin{equation}
   \label{asya}
    A_{i}^{a}\rightarrow {1 \over g a^2} \epsilon^{abc} \phi^b \partial_{i}
\phi^c +
{1 \over a} \phi^a {\cal A}_{i},
 \end{equation}
where ${\cal A}$ denotes the component of the gauge field along the Higgs field. It is arbitrary since Eq.\ (\ref{asyna}) determines only the components perpendicular to $\phi$. The asymptotic field strength associated with this gauge field \big(cf.\ Eq.\ (\ref{fs})\big) has only a color component parallel to the Higgs field - the ``neutral direction'' \big(cf.\ the definition of the charged gauge fields in Eq.\ (\ref{spba})\big) and we can write
\begin{equation}
  \label{asyf}
   F^{aij} = {1 \over a} \phi^a  F^{ij},\quad \mbox{with}\quad F^{ij} = {1 \over g a^3} \epsilon^{abc} \phi^a \partial^{i}
\phi^b
\partial^{j}\phi^c + \partial^{i} {\cal A}^{j} - \partial^{j} {\cal A}^{i}\, .
\end{equation}
One easily verifies that the Maxwell equations 
\begin{equation}
  \label{max}
\partial_{i} F^{ij} = 0  
\end{equation}
are satisfied.  These results confirm the interpretation of  $F_{\mu\nu}$ as a legitimate 
field strength related to the unbroken $U(1)$ part of the gauge symmetry. As the magnetic flux in the abelian Higgs model, the magnetic charge in the non-abelian Higgs model is quantized.  Integrating over  the asymptotic surface  $S^2$ which encloses the system and using the integral form of the degree (\ref{indeg}) of  the map defined by the scalar field 
(cf.\ \cite{RAJA82}) yields 
\begin{equation}
  \label{mgch}
m= \int_{S^2} {\bf B}  \cdot  d \mbox{\boldmath $\sigma$} = -{1 \over 2 g a^3}
\int_{S^2}
\epsilon^{ijk} \epsilon^{abc}
\phi^a \partial^j \phi^b \partial^k \phi^c d\sigma^{i}=- {4 \pi N \over g} \, . 
\end{equation}
No contribution  to the magnetic charge arises from $\mbox{\boldmath$\nabla$}\times {\bf \cal A}$ when integrated over a surface without boundary. The existence of a winding number associated with the Higgs field is a direct consequence of the topological properties discussed above. The Higgs field $\phi$ maps the asymptotic $S^2$ onto the space of zeroes of $V$ which topologically is $S^2$ and  has been shown \big(Eq.\ (\ref{ghs2})\big) to be homeomorph to the coset space $G/H_{\phi_0}$ . Thus, asymptotically, the map  
\begin{equation}
  \label{wnu2}
  \phi:\quad S^2 \rightarrow S^2\sim G/H_{\phi_0}
\end{equation}
is characterized by the homotopy group $\pi_2(G/H_{\phi_0})\sim \mathbb{Z}$. Our discussion provides an illustration of the general relation   (\ref{th2})
$$ \pi_2\left(SU(2)/U(1)\otimes \mathbb{Z}_2\right)= \pi_1\left(U(1)\right)\sim \mathbb{Z}\, .$$
The non-triviality of the homotopy group guarantees the stability of topological excitations of finite energy. \\
An important example is the spherically symmetric  { hedgehog} configuration
  $$
\phi ^{a} (\mbox{\boldmath$r$}) _{{\longrightarrow \atop r \to \infty
    }}\phi_{0}^{a}(\mbox{\boldmath$r$})= a \cdot \frac{x^{a}}{r}
$$
which on the asymptotic sphere covers the space of zeroes of $V$ exactly once. Therefore, it describes a monopole with the  asymptotic field strength (apart from the ${\cal A}$ contribution) given, according to Eq.\ (\ref{asyf}),  by
\begin{equation}
  \label{mob}
  F^{ij} = \epsilon^{ijk} \frac{x^k}{g r^3}\, ,\quad {\bf B} = -\frac{{\bf r}}{g\, r^3}\, .
\end{equation}

 {\em Monopole Solutions}\vskip .1cm
The asymptotics of Higgs and gauge fields suggest  the following spherically symmetric Ansatz for monopole solutions
 { \begin{equation}
 \phi^a = a \frac{x^a}{r} H(agr)\, ,\quad A_i^a  = \epsilon^{aij} \frac{x^j}{gr^{2}} [1 - K(agr)] 
\label{moposu} 
\end{equation}}
with the boundary conditions at infinity 
$$ H(r) _{{\longrightarrow \atop r \to \infty}}1,\quad  K(r) _{{\longrightarrow \atop r \to \infty}} 0\, .$$
As in the abelian Higgs model, topology forces the Higgs field to have a zero. Since the winding  of the Higgs field $\phi$ cannot be removed by continuous deformations, $\phi$ has to have a zero.  This  defines the center of the monopole.  The boundary condition
$$H(0)=0\, ,\quad K(0)=1 $$
in the core of the monopole guarantees continuity of the solution.
As in the abelian Higgs model, the changes in the Higgs and gauge field are occurring on  two different length scales. Unlike at asymptotic distances,  in the core  of the monopole also charged vector fields are present. The core of the monopole represents the perturbative phase of the Georgi-Glashow model, as the core of the vortex is made of normal conducting material and ordinary gauge fields. \\ 
With the Ansatz (\ref{moposu}) the equations of motion are  converted into a coupled system of ordinary differential equations for the unknown functions $H$ and $K$ which  allows for analytical solutions only in certain limits. Such a limiting case is obtained by saturation of the Bogomol'nyi bound. As for the abelian Higgs model, this bound is obtained by rewriting the total energy of the static solutions 
   \begin{eqnarray*} 
E = \int d^3 x  \left[\frac{1}{2} {\bf B}^{2} + \frac{1}{2} ({\bf D}\phi)^{2}+ V (\phi)\right] =  \int d^3 x \left[\frac{1}{2} ( {\bf B} \pm {\bf D}\phi)^2 + V (\phi)\mp {\bf B}{\bf D}\phi\right]\, ,
\end{eqnarray*}
and by expressing the last term via an integration by parts \big(applicable for covariant derivatives due to  antisymmetry of the structure constants in the definition of D in (\ref{dphi})\big) and with the help of the equation of motion  ${\bf D}{\bf B}=0$ by the magnetic charge \big(Eq.\ (\ref{mgch})\big) 
 { $$ \int d^3 x  {\bf B}\,{\bf D}\phi = a  \int_{S^2} {\bf B}\, d \mbox{\boldmath $\sigma$} = a\,m  .  $$} 
The energy satisfies the Bogomol'nyi bound
$$E\ge | m |\, a   .$$  
For this bound  to be   saturated, the strength of the Higgs potential has to approach zero
$$ V=0,  \quad \mbox{{ i.e.}}\quad \lambda = 0,  $$
and the fields have to satisfy the first order equation
$$ {\bf B}^a \pm ({\bf D}\phi)^a = 0 .$$
In the  approach to vanishing $\lambda$, the asymptotics of the Higgs field $|\phi|_{\longrightarrow \atop r \to \infty} a$ must remain unchanged.
The solution to this system of  first order differential equations is known as the  Prasad-Sommerfield monopole} 
$$ H(agr)= \coth agr-\frac{1}{agr}\, ,\quad K(agr)= \frac{\sinh agr} {agr}. $$
In this limiting case of saturation of the  Bogomol'nyi bound, only one length scale exists $\big((ag)^{-1}\big)$. The energy of the excitation, i.e.~the mass of the monopole is given in terms of the mass  of the charged vector particles \big(Eq.\ (\ref{hima})\big) by
$$E = M \frac{4\pi}{g^2}\, .$$   
As for  the Nielsen-Olesen vortices, a wealth of further results have been obtained concerning properties and generalizations of the 't Hooft-Polyakov monopole solution. Among them I mention  the ``Julia-Zee'' dyons \cite{JUZE75}. These  solutions of the field equations are obtained using the  Ansatz (\ref{moposu}) for the Higgs field and the spatial components of the gauge field but admitting a  non-vanishing time component of the form
$$A_0^a= \frac{x^a}{r^2}J(agr) .$$
This time component  reflects the presence of a source of electric charge q.  Classically the  electric  charge of the dyon can assume any value, semiclassical arguments suggest quantization of the charge  in units of g \cite{TWGS76}. \\
As the vortices of the Abelian Higgs model,  't Hooft-Polyakov monopoles induce zero modes if massless fermions are coupled to the gauge and Higgs fields of the monopole
 \begin{equation}  {\cal L}_{\psi}=i{\bar \psi} \gamma^{\mu}D_{\mu}\psi -g\phi^a\bar{\psi}\frac{\tau^a}{2}\psi\, . 
\label{femo}
\end{equation}
The number of zero modes is given by the magnetic charge $|m|$ (\ref{mgch}) \cite{CALL78}. Furthermore, the  coupled system of a t'Hooft-Polyakov monopole and  a fermionic zero mode behaves   as a boson if the fermions belong to the fundamental representation of $SU(2)$ \big(as assumed in (\ref{femo})\big)  while for isovector fermions the coupled system behaves as a fermion. Even more puzzling, fermions can be generated by coupling  bosons in the fundamental representation to the 't Hooft-Polyakov monopole. The origin of this conversion of isospin into spin \cite{JARE76a,JARE76b,HATH76} is the correlation  between angular and isospin dependence of Higgs and gauge fields in solutions of the form (\ref{moposu}). Such  solutions do not transform covariantly under spatial rotations generated by ${\bf J}$. Under combined spatial and isospin rotations (generated by  ${\bf I}$) 
\begin{equation}
  \label{moam}
  {\bf K} ={\bf J}+{\bf I}\, ,
\end{equation}
monopoles of  the type $(\ref{moposu})$ are invariant.  ${\bf K}$ has to be identified with the angular momentum operator.  If  added to this  invariant monopole,   matter fields determine  by their spin and isospin the  angular momentum ${\bf K}$ of the coupled system.\\ 
Formation of monopoles is not restricted to the particular model. The Georgi-Glashow model  is the simplest  model in which this phenomenon occurs. With the topological arguments at hand, one can easily see the general condition for the existence of monopoles. If we assume  electrodynamics to appear  in the process of symmetry breakdown from a simply connected topological group $G$, the  isotropy group $H$ (\ref{isot}) must contain a $U(1)$ factor. According to  the identities (\ref{th2}) and  (\ref{prod}),   the resulting non trivial second homotopy group of the coset space  
\begin{equation}
  \label{exmo}
  \pi_2(G/[\tilde{H}\otimes U(1)]) = \pi_{1}(\tilde{H})\otimes \mathbb{Z}
\end{equation}
guarantees the existence of monopoles. This prediction is independent of the group $G$, the details of the particular model, or  of the process of the symmetry breakdown. It applies to Grand Unified Theories in which the structure of the standard model ($SU(3)\otimes SU(2)\otimes U(1)$) is assumed to originate from  symmetry breakdown. The fact that  monopoles cannot be avoided has posed a serious problem to the standard model of cosmology.  The predicted abundance of monopoles created in the symmetry breakdown occurring in the early universe is in striking conflict with observations. Resolution of this problem is offered by the inflationary model of cosmology \cite{KOTU90,PEAC99}.
\newpage 
\section{Quantization of Yang-Mills Theory}
{\em Gauge copies}
\vskip .1cm
{Gauge theories are formulated in terms of redundant variables. Only in this way, a covariant, local representation of the dynamics of gauge degrees of freedom is possible. For quantization of the theory both canonically or in the path integral, redundant variables have to be eliminated. This procedure is called  gauge fixing. It is not unique and the implications of a particular  {choice} are  generally not well understood. In the path integral one performs a sum over all field configurations. In gauge theories this procedure has to be modified by making use of the decomposition of the space of gauge fields into equivalence classes, the gauge orbits \big(Eq.\ (\ref{GO})\big).  
Instead of summing in the path integral  over formally different but 
physically equivalent fields, the integration is performed over the  {equivalence classes} of such fields, i.e.~over the
corresponding gauge orbits. The value of the  {action} is gauge invariant, i.e.~the same for all members of a given gauge 
orbit, 
 $$
S \left[A^{\left[U\right]} \right] = S \left[A\right]\, .
$$
Therefore, the action is seen to be a functional defined on classes (gauge orbits). 
Also  the integration measure
$$
d\left[A^{\left[U\right]}\right] =  d\left[A\right] \quad , \quad d\left[A\right] = \prod_{x,\mu,a}dA^{a}_{\mu} \left(x\right)  .
$$ 
is gauge invariant since shifts and rotations of an integration variable do not change the value of an integral. Therefore,  in the naive path integral 
$$
\tilde{Z}=\int d\left[A\right] e^{i S \left[A\right]} \propto \int \prod_{x}dU\left(x\right)  .
$$
a ``volume'' associated with the gauge transformations  { $ \prod_{x}dU\left(x\right)$} can be factorized and thereby the integration be performed over the gauge orbits. To turn this property into a working algorithm,  redundant variables are eliminated by imposing a  { gauge condition}
\begin{equation}
  \label{GC}
f[A]=0 ,  
\end{equation}
which is supposed to eliminate all gauge copies of a certain field configuration  {$A$}. 
In other words, the functional {$f$} has to be chosen such that  for arbitrary field configurations the equation
$$ f[ A^{\, \left[\,U\right]}\,]=0$$ 
determines uniquely the gauge transformation  {$U$}. If successful, the set of all gauge equivalent fields, the gauge orbit, is represented by exactly  {one representative}.
In order to write down an integral over gauge orbits, we insert into the integral the gauge-fixing   {$\delta$-functional}
$$
\delta\left[f\left(A\right)\right] = \prod_{x}\prod_{a=1}^{N^{2}-1}
\delta\left[f^{a}\left(A\left(x\right)\right)\right] .
$$
This modification of the integral however changes the value depending on  the representative chosen, as the following elementary identity shows
\beq \label{GPD}
\delta\left(g\left(x\right)\right) = \frac{\delta\left(x-a\right)}{|g^{\prime} \left(a\right)|} \quad ,\quad g\left(a\right)=0.
\eeq
This difficulty is circumvented with the help of the  { Faddeev-Popov determinant} { $\Delta_{f} \left[A\right]$} defined implicitly by
 $$
\Delta_{f} \left[A\right] \int d\left[U\right]\delta\left[f\left(A^{\left[U\right]}\right)\right]=1.   
$$ 
Multiplication of the path integral  $\tilde{Z}$  with the above  {``1''} and taking into account the gauge invariance of the various factors } yields 
\begin{eqnarray*}
\tilde{Z}& = & \int d\left[U\right] \int d\left[A\right] e^{i S \left[A\right]}\Delta_{f} \left[A\right]
 \delta\left[f\left(A^{\left[U\right]}\right)\right]\\ 
&=& \int d\left[U\right] \int d\left[A\right] e^{i S \left[A^{\left[U\right]}\right]}\Delta_{f} \left[A^{\left[U\right]}\right]
 \delta\left[f\left(A^{\left[U\right]}\right)\right] =   \int d\left[U\right] \, Z .
\end{eqnarray*}
{The gauge volume has been factorized  and, being independent of the dynamics, can be dropped. In summary, the  final definition of the  {generating functional} for gauge theories
   \begin{equation}
     \label{GFP}
Z \left[J\right]= \int d\left[A\right]\Delta_{f}\left[A\right]
\delta\left(f\left[A\right]\,\right) e^{i S \left[A\right]+i \int d^{4}x J^{\mu}  A_{\mu}}
 \end{equation}
is given in terms of  a sum over gauge orbits. 
\vskip .1cm
{\em Faddeev-Popov determinant}\vskip .1cm 
For the calculation of  {$\Delta_{f}\left[A\right]$}, we first consider the change of the gauge condition  { $f^{a} \left[A\right]$}  under infinitesimal gauge transformations . Taylor expansion
\begin{eqnarray*}
f^{a}_{x} \left[A^{\left[U\right]}\right] &\approx&  f^{a}_{x} \left[A\right]
+\int d^{4}y 
\sum _{b,\mu} \frac{\delta  f^{a}_{x} \left[A\right]}{\delta A^{b}_{\mu} \left(y\right)}\delta A^{b}_{\mu} \left(y\right) 
 \nonumber \\
&=&  f^{a}_{x} \left[A\right] + \int d^{4}y \sum _{b }M\left(x,y;a,b\right)\alpha^{b} \left(y\right) \,   
\end{eqnarray*}
with  $\delta A^{a}_{\mu}$ given by  infinitesimal gauge transformations (\ref{ifgt}),  yields
\begin{equation}
  \label{M}
M\left(x,y;a,b\right) = \left(\partial _{\mu} \delta ^{b,c}+g f^{bcd} A^{d}_{\mu} \left(y\right) \right)
\frac{\delta  f^{a}_{x} \left[A\right]}{\delta A^{c}_{\mu} \left(y\right)}\, .  
\end{equation}
In the second step, we compute the integral  
 $$
\Delta_{f}^{-1} \left[A\right] =\int d\left[U\right]\delta\left[f\left(A^{\left[U\right]}\right)\right]    
$$ 
by expressing the integration  {$d\left[U\right]$} as an  integration over the  gauge functions  {$\alpha$}. We  finally  change to the variables  $\beta=M \alpha$ 
$$
\Delta_{f}^{-1} \left[A\right] = |\det M|^{-1} \int  d \left[\beta\right] \delta\left[f\left(A\right)-\beta\right] 
$$ 
and arrive at the final expression for the Faddeev-Popov determinant
\begin{equation}
  \label{FPD}
\Delta_{f}\left[A\right] = |\det M| \, .    
\end{equation}
Examples 
\begin{itemize}
\item    Lorentz gauge 
\begin{eqnarray}
f^{a}_{x}\left(A\right)& = & \partial ^{\mu}A^{a}_{\mu} \left(x\right) - \chi ^{a} \left(x\right) \nonumber \\
M\left(x,y;a,b\right) & = & - \left( \delta^{ab}\Box - g f^{abc}A^{c}_{\mu}
\left(y\right)\partial^{\mu}_{y}\right) \delta ^{ \left(4\right)} \left( x-y\right) 
  \label{loga}
\end{eqnarray}
\item   Coulomb gauge 
\begin{eqnarray}
f^{a}_{x}\left(A\right) & = & \mbox{div}  {\bf A}^{a} \left(x\right) - \chi ^{a} \left(x\right)\nonumber \\
M\left(x,y;a,b\right) & = & \left( \delta^{ab}\Delta  + g f^{abc}{\bf A}^{c}
\left(y\right)\mbox{\boldmath$\nabla$}_{y}\right) \delta ^{ \left(4\right)} \left( x-y\right) 
  \label{coga}
\end{eqnarray}
\item   Axial gauge 
\begin{eqnarray}
f^{a}_{x}\left(A\right)& = & n^{\mu}A^{a}_{\mu} \left(x\right)- \chi ^{a} \left(x\right)\nonumber \\
M\left(x,y;a,b\right) &=& -\delta^{ab}n_{\mu}\partial^{\mu}_{y} \delta ^{ \left(4\right)}
 \left( x-y\right)  
  \label{axga}
\end{eqnarray}
\end{itemize}}
{We note that in axial gauge, the Faddeev-Popov determinant does not depend on the gauge fields and therefore changes the generating functional only by an irrelevant factor.
\vskip .1cm
{\em Gribov-Horizons}\vskip .1cm 
 As the elementary example (\ref{GPD}) shows, a  vanishing  Faddeev-Popov determinant \big($g^{\prime}(a)=0$\big)  indicates  the gauge condition to exhibit a quadratic or higher order zero. This implies that at this point in function space, the gauge condition is satisfied by at least two gauge equivalent configurations, i.e.~vanishing of {$ \Delta_{f}\left[A\right]$} implies the existence of zero modes associated with  {$M$ (Eq.\ (\ref{M}))
$$
M \chi_{0} = 0
$$
and therefore the  {gauge choice is ambiguous}. 
The (connected) spaces of gauge fields which make the gauge choice ambiguous
 $${\cal M}_{H} = \left\{A \big|\,\det M = 0 \right\} $$   
are called   {Gribov horizons} \cite{GRIB78}. Around Gribov horizons, pairs of  infinitesimally close gauge equivalent fields exist which satisfy the gauge condition. 
If on the other hand two gauge fields satisfy the gauge condition and are separated by an infinitesimal gauge transformation, these two fields are separated by a Gribov horizon.
  The region beyond the horizon thus contains  {gauge copies} of fields inside the horizon. In general,  one therefore needs additional conditions to select exactly one representative of the gauge orbits. The structure of Gribov horizons and of the space of fields which contain no Gribov copies depends on the choice of the gauge. Without  specifying further the procedure, we  associate an  {infinite potential energy}  {${\cal V}[A]$} with every gauge copy of a configuration which already has been taken into account, i.e.~after gauge fixing, the action is supposed to contain implicitly this potential energy
  \begin{equation}
    \label{V}
S[A] \rightarrow S[A] -\int d^4 x\,{\cal V}[A] .     
  \end{equation}
With the above expression, and given a reasonable gauge choice, the generating functional is written as an integral over gauge orbits and can serve as starting point for further formal developments such as the  {canonical formalism} \cite{LEWO01} or applications e.g.} perturbation theory.
\vskip .1cm 
{The occurrence of  {Gribov horizons} points to a more general problem in the gauge fixing procedure. Unlike in electrodynamics,  {global gauge conditions} may not exist in non-abelian gauge theories \cite{SING78}. In other words, it may not be possible to formulate a condition which in the whole space of gauge fields selects exactly one representative. This difficulty of imposing a global gauge condition is similar to the problem of a global coordinate choice on e.g.  {$S^2$}.
In this case, one either has to resort to some  {patching procedure} and use more than one set of coordinates (like for the Wu-Yang treatment of the Dirac Monopole \cite{WUYA75}) or deal with singular fields arising from these gauge ambiguities (Dirac Monopole).  Gauge singularities} are analogous to the coordinate singularities on non-trivial manifolds (azimuthal angle on north pole). \\
The appearance of Gribov-horizons poses severe technical problems in analytical studies of  non-abelian gauge theories. Elimination of redundant variables is necessary for proper definition of the path-integral of infinitely many variables. In the gauge fixing procedure it must be ascertained that every gauge orbit is represented by exactly one  field-configuration. Gribov horizons may make this task impossible. On the other hand, one may regard  the existence of global gauge conditions in QED and its non-existence in QCD as an expression of a fundamental  difference in the structure of  these two theories which ultimately could be responsible for their  vastly different physical 
properties. 
\section{Instantons}
\subsection{Vacuum Degeneracy}
Instantons are solutions of the classical Yang-Mills field equations with distinguished  topological properties \cite{BELA75}. 
Our discussion of  instantons follows the pattern of that of the Nielsen-Olesen vortex or the 't Hooft-Polyakov monopole and starts with a discussion of configurations of vanishing energy (cf.\ \cite {BJOR,JACK,SCHW93}). As follows from the Yang-Mills Hamiltonian (\ref{YMH}) in the Weyl gauge (\ref{WEYL}),
 static zero-energy solutions  of the equations of motion (\ref{emga}) satisfy  
$$ {\bf E} = 0 \, ,\quad {\bf B} = 0 ,$$ 
and  therefore   are pure gauges (\ref{puga})
\begin{equation}
  \label{PG}
{\bf A} = \frac{1}{ig} \; U ({\bf x}) \mbox{\boldmath$\nabla$} U^{\dagger} ({\bf x}).  
\end{equation}
 In the Weyl gauge, pure gauges in electrodynamics are gradients of  time-independent  scalar functions.  In $SU(2)$ Yang-Mills theory, the manifold of zero-energy solutions is according to (\ref{PG}) given by the set of all  $U({\bf x}) \in SU(2) $. Since topologically $SU(2) \sim  S^{3}$ \big(cf.\ Eq.\ (\ref{su2s3})\big), each $U({\bf x})$ defines a mapping from the base space $\mathbb{R}^3$ to $S^3$.  
We impose the  requirement that at infinity, $U({\bf x})$ approaches a unique value independent of the direction  of ${\bf x}$  
\begin{equation}
  \label{pginf}
U({\bf x})\rightarrow \mbox{const.} \quad \mbox{for}\,\,| {\bf x}| \to \infty  .
\end{equation}
Thereby, the configuration space becomes compact $\mathbb{R}^{3}\rightarrow S^{3}$  \big(cf.\ Eq.\ (\ref{Alex})\big) and pure gauges define a map 
\begin{equation}
  \label{s3s3}
U({\bf x}): \; S^{3} \longrightarrow S^{3}  
\end{equation}
to which, according to Eq.\ (\ref{pin}), a winding number can be assigned. This winding number counts how many times the 3-sphere of gauge transformations $U({\bf x})$ is covered if ${\bf x}$ covers once the  3-sphere of the compactified configuration space. Via the degree of the map (\ref{indeg}) defined by $U({\bf x})$, this winding number can be calculated  \cite{RAJA82,JACK85}  and expressed in terms of the gauge fields 
\begin{eqnarray}
\label{wnu}
n_{w}
& = & \frac{g^{2}}{16 \pi ^{2}} \; \int \; d^{3} \, x \; \epsilon_{ijk} \, 
\left ( A^{a}_{i} \, \partial _{j} \, A^{a}_{k} - \frac{g}{3} \; \epsilon^{abc} \; 
A^{a}_{i} \; A^{b}_{j} \; A^{c}_{k} \right ).
\end{eqnarray}
The expression on the right hand side yields an integer only if $A$ is a pure gauge. 
Examples of gauge transformations giving rise to  non-trivial winding (hedgehog solution for  $n=1$) are 
 \begin{equation}
   \label{UN}
 U_{n}({\bf x}) = \exp\{i \pi n \; \frac{{\bf x}\mbox{\boldmath$\tau$}}{\sqrt{x^{2} + p^{2}}}\}  
 \end{equation}
with winding number  $ n_{w} = n $ \big(cf.\ Eq.\ (\ref{su2pa}) for verifying the asymptotic behavior (\ref{pginf})\big). Gauge transformation which change the winding number $n_{w}$ are called {\em large} gauge transformations. Unlike {\em small} gauge transformations, they cannot be deformed continuously to $U=1$. \\
These topological considerations show that Yang-Mills theory considered as a classical system possesses an infinity of different lowest energy ($E=0$) solutions which can be labeled by an integer $n$. They are connected to each other by gauge fields which cannot be pure gauges and which therefore produce a finite  value of the magnetic field, i.e.~of the potential energy. The schematic plot of the potential energy in Fig. \ref{perpot} 
 shows that the ground state of QCD can be expected to exhibit similar properties  as that of a particle moving in a periodic potential. In the quantum mechanical  case too, an  infinite degeneracy is present with the winding number in gauge theories replaced by the integer characterizing the equilibrium positions of the particle.
\begin{figure}
\hspace{3cm} 
\epsfig{file=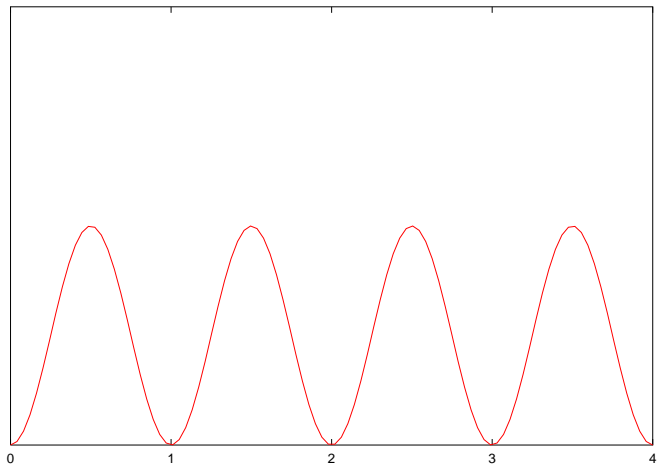, width=0.4\linewidth,angle=-90}
\caption{ Schematic plot of the potential energy $  V[A] = \int d^3\, x {\bf B}[A]^2 $ as a function of the winding number (\ref{wnu}) }
\label{perpot}
\end{figure}   
\subsection{Tunneling}
``Classical vacua'' are states with  values of the coordinate of a mechanical system {$ x=n $} given by the equilibrium positions. Correspondingly, in gauge theories the classical vacua, the ``$n$ - vacua'' are given by the  pure gauges \big(Eqs.\ (\ref{PG}) and (\ref{UN})\big).  To proceed from here to a description of the quantum mechanical ground state,  tunneling processes have to be included which, in such a semi-classical approximation, connect classical  vacua with each other.}  Thereby  the quantum mechanical ground state becomes a linear superposition of classical vacua. Such tunneling solutions are most easily obtained by changing to imaginary time with a concomitant change in the time component of the gauge potential 
\begin{equation}
  \label{eucl}
 t \to - i t \, ; \quad A_{0} \to - i A_{0}\, . 
\end{equation}
The metric becomes Euclidean and there is no distinction between covariant and contravariant indices. 
Tunneling solutions are solutions of the classical field equations derived from the Euclidean action $S_E$, i.e.\,the Yang-Mills action \big(cf.\,Eq.\ (\ref{LYM})\big) modified by the substitution (\ref{eucl}). 
 We proceed in a by now familiar way and derive the Bogomol'nyi bound for topological excitations in Yang-Mills theories. To this end we rewrite the action   \big(cf.Eq.\ (\ref{DFD})\big)
\begin{eqnarray}
\label{BB}
S_{E}& =& \frac{1}{4} \, \int \, d^{4} x \; F^{a}_{\mu \nu} \, F^{a}_{\mu \nu} =  \frac{1}{4} \, \int \, d^{4} x \; \left ( \pm F^{a}_{\mu \nu} \; \tilde{F}^{a}_{\mu \nu} + \frac{1}{2} \, (F^{a}_{\mu \nu} \mp \tilde{F}^{a}_{\mu \nu} )^{2} \right )   \\
&   \geq &\; \pm\frac{1}{4} \; \int \, d^{4} x \; F^{a}_{\mu \nu} \; \tilde{F}^{a}_{\mu \nu} 
\end{eqnarray}
This bound for  {$ S_{E} $} (Bogomol'nyi bound)   is determined by the
topological charge $\nu$ , i.e.~it can be rewritten as a surface term  
\begin{equation}
  \label{TPC}
\nu = \frac{g^{2}}{32 \pi ^{2}} \; \int \, d^{4} x \, F_{\mu\nu}^a \, \tilde{F}_{\mu\nu}^a = 
\int \, d \, \sigma_{\mu} \, K^{\mu} 
\end{equation}
of the topological current 
\begin{equation}
  \label{TPK}
K^{\mu} = \frac{g^{2}}{16 \pi ^{2}} \, \epsilon ^{\mu \alpha \beta \gamma} \; 
\left ( A^{a}_{\alpha} \, \partial _{\beta} \, A^{a}_{\gamma} - \frac{g}{3} \, 
\epsilon ^{a b c} \, A^{a}_{\alpha} \, A^{b}_{\beta} \, A^{c}_{\gamma} \right ).
  \end{equation} 
Furthermore, if we assume  $ K $ to vanish at spatial infinity so that
  \begin{equation}
    \label{WN} 
\nu = \int^{+ \infty}_{- \infty} \; d t \, \frac{d}{dt} \, \int \; K^{0} \, d^{3} \, x = n_{w} \, (t = \infty ) - n_{w} (t=  - \infty)\, ,    
  \end{equation}
the charge $\nu$ is seen to be quantized as a  difference of two winding numbers.  \\
I first discuss the formal implications of this result. The topological charge has been obtained as a difference of winding numbers of pure (time-independent) gauges (\ref{PG}) satisfying the condition (\ref{pginf}). With the winding numbers, also $\nu$  is a topological invariant. It characterizes the  space-time dependent gauge fields $A_{\mu}(x)$. Another and more  direct approach to the topological charge (\ref{TPC}) is provided by the study of cohomology groups. Cohomology groups characterize connectedness properties of topological spaces by properties  of differential forms and their integration via  Stokes' theorem  (cf.\ Chapt. 12 of \cite{GRSW87} for an introduction).   \\
 Continuous deformations of gauge fields cannot change the topological charge. This implies that  $\nu$ remains unchanged under continuous gauge transformations. In particular, the $\nu =0$ equivalence class of gauge fields containing $A_{\mu}=0$ as an element cannot be connected to gauge fields with non-vanishing topological charge. Therefore, the gauge orbits can be labeled by $\nu$.
Field configurations with $ \nu \neq 0 $  connect vacua (zero-energy field configurations)  with different winding number \big(Eqs.\ (\ref{WN}) and (\ref{wnu})\big). Therefore, the solutions to the classical Euclidean field equations with non-vanishing topological charge are the tunneling solutions needed for the construction of the semi-classical Yang-Mills ground state.\\ 
Like in the examples discussed in the previous sections, the field equations simplify if the Bogomol'nyi bound is saturated. In the case of  Yang-Mills theory, the equations of motion can then be solved in closed form. Solutions with  topological charge $\nu = 1\, (\nu=-1)$ are called instantons (anti-instantons). Their action is given by 
$$
S_{E} = \frac{8 \pi ^{2}}{g^{2}} \, .
$$
By construction, the action of any other field configuration  with $|\nu|=1$ is larger. 
Solutions with action $S_{E} = 8 \pi ^{2} |\nu|/ g^{2}$ for $|\nu| > 1$ are called multi-instantons. 
According to Eq.\ (\ref{BB}), instantons satisfy 
\begin{equation}
  \label{SD}
F_{\mu \nu} = \pm \tilde{F} _{\mu \nu}.  
\end{equation}
The  interchange 
$  F_{\mu\nu}\leftrightarrow   \tilde{F}_{\mu\nu}$ corresponding in Minkowski space to the interchange $ {\bf E}\rightarrow {\bf B},\;{\bf B}\rightarrow - {\bf E}$
is a duality transformation and fields  satisfying (\ref{SD}) are said to be selfdual $(+)$ or anti-selfdual ($-$) respectively.
A spherical Ansatz yields the solutions 
\begin{equation}
  \label{RINS}
A^{a}_{\mu} = -\frac{2}{g} \; \frac{\eta _{a\mu \nu } x_{\nu}}{x^{2} 
+ \rho ^{2}} \qquad F^{2}_{\mu \nu} = \frac{1}{g^{2}} \; \; 
\frac{192 \rho ^{4}}{(x ^{2} + \rho ^{2}) ^{4}}\, ,
  \end{equation}

with the 't Hooft symbol \cite{THOO76}
$$
\eta_{a \mu \nu} = \left \{ 
\begin{array}{ll} 
\epsilon _{a\mu \nu}   &   \mu, \nu = 1, 2, 3   \\
\delta_{a \mu}          &    \nu = 0    \\
- \delta _{a \nu}       &   \mu = 0\qquad .
\end{array}
\right.
$$
The size of the instanton $\rho$ can be  chosen freely. Asymptotically, gauge potential and field strength behave as  
$$
A \; _{{\longrightarrow  \atop | x | \to \infty }} \;  \frac{1}{x} \qquad  
F \; _{{\longrightarrow \atop | x | \to \infty }} \; \frac{1}{x^{4}}.
$$
The unexpectedly strong decrease in the field strength is the result of a partial cancellation of abelian and non-abelian contributions to $F_{\mu\nu}$ \big(Eq.\ (\ref{fs})\big). For instantons, the asymptotics of the gauge potential is actually gauge dependent. By a gauge transformation, the asymptotics  can be changed to $x^{-3}$. Thereby the gauge fields develop a singularity at $x=0$, i.e.~in the center of the instanton.   In this ``singular'' gauge, the gauge potential is given by
\begin{equation}
  \label{singins}
A^{a}_{\mu} = -\frac{2\rho^2}{gx^2} \; \frac{\bar{\eta} _{a\mu \nu } x_{\nu}}{x^{2} + \rho ^{2}},\quad \bar{\eta}_{a\mu\nu}= \eta_{a\mu\nu}(1-2\delta_{\mu,0})(1-2\delta_{\nu,0})\, .
\end{equation}

\vskip .1cm
\subsection{Fermions in Topologically Non-Trivial Gauge Fields}
Fermions are severely affected by the presence of gauge fields with non-trivial topological properties. A dynamically very important phenomenon is the appearance of fermionic zero modes in certain gauge field configurations. For a variety of low energy hadronic properties,  the existence of such zero modes appears to be fundamental.  
Here I will not enter a detailed discussion of non-trivial fermionic properties induced by topologically non-trivial gauge fields (cf.\,\cite{ZINN02}). Rather I will try to indicate the origin of the induced topological fermionic properties  in the context of a simple system. I will consider massless  fermions in 1+1 dimensions moving in an external (abelian) gauge field. The Lagrangian of this system is \big(cf.\ Eq.\ (\ref{matt})\big)
\begin{equation}
  \label{qed2}
{\cal L}_{YM} =  - \frac{1}{4} F^{\mu\nu } F_{\mu\nu}+ {\bar \psi}i \gamma^{\mu}D_{\mu}\psi, \end{equation}
with the covariant derivative $D_{\mu}$  given in (\ref{covde}) and $\psi$ denoting a 2-component spinor. The Dirac algebra of the $\gamma$ matrices 
$$\{\gamma^{\mu},\gamma^{\nu}\}=g^{\mu\nu}$$
can be satisfied by the following choice in terms of Pauli-matrices \big(cf.\ (\ref{pauli})\big) 
$$\gamma^0 =\tau^1,\; \gamma^1 =i \tau^2,\;\gamma^5 =-\gamma^0\gamma^1 = \tau^3.$$
In Weyl gauge, $A_0=0$,  the Hamiltonian density \big(cf.\ Eq.\ (\ref{HQCD})\big) is given by 
\begin{equation}
  \label{ham1+1}
{\cal H}= \frac{1}{2}E^2+\psi^{\dagger}{\cal H}_f \psi\, ,
\end{equation}
with 
\begin{equation}
  \label{hf}
{\cal H}_f= (i\partial_1 -eA_1) \, \gamma^5 \, .  
\end{equation}
The application of topological arguments is greatly simplified if the spectrum of the fermionic states is discrete. We assume the fields to be defined on a circle and impose antiperiodic boundary conditions for the fermions 
$$\psi(x+L) = -\psi(x)\, .$$ The (residual) time-independent gauge transformations  are given by Eqs.\ (\ref{gtqed}) and (\ref{gt}) with the Higgs field $\phi$ replaced by the fermion field $\psi$. On a circle, the gauge functions $\alpha(x)$  have to satisfy \big(cf.\ Eq.\ (\ref{gtqed})\big)
$$ \alpha(x+L)= \alpha(x) +  \frac{2n\pi }{e} .$$
The winding number $n_w$ of the mapping 
$$U: S^1 \rightarrow S^1$$
partitions gauge transformations into equivalence classes with representatives given by the gauge functions
\begin{equation}
  \label{gfn}
\alpha_n(x)= d_n x,\quad d_n= \frac{2\pi n }{eL}.  
\end{equation}
Large gauge transformations  define pure gauges 
\begin{equation}
  \label{pg1+1}
A_1= U(x)\, \frac{1}{ie}\, \partial_1 \, U^{\dagger}(x)\, ,  
\end{equation}
which inherit the winding number \big(cf.\ Eq.\ (\ref{wnu})\big).  For 1+1 dimensional electrodynamics the winding number of a pure gauge is given by 
\begin{equation}
  \label{wn1+1}
  n_w = -\frac{e}{2\pi}\int_0^{L} dx A_1(x)\, .
\end{equation}
As is easily verified, eigenfunctions and eigenvalues of ${\cal H}_f$ are given by 
\begin{equation}
  \label{eig}
\psi_n(x)=e^{-ie\int_0^x A_1 dx -iE_n(a) x} u_{\pm} \,,\quad E_n(a)= \pm \frac{2\pi}{L}(n+\frac{1}{2}-a)\, , 
\end{equation}
with the positive and negative chirality eigenspinors $u_{\pm}$ of $\tau^3$ and the zero mode of the gauge field
$$a = \frac{e}{2\pi}\int _0^{L} dx A_1(x)\, .$$ 
We now consider a  change of the external gauge field $A_1(x)$ from $A_1(x)=0$ to a pure gauge of winding number $n_w$. The change is supposed to be adiabatic, such that the fermions  can adjust at each instance to the changed value of the external field. In the course of this change, $a$ changes continuously from $0$ to $n_w$. Note that  adiabatic  changes of   $A_1$ generate finite field strengths and therefore do not correspond to gauge transformations.  As a consequence we have 
\begin{equation}
  \label{sfl}
  E_n(n_w)= E_{n- n_w}(0) .
\end{equation}
As expected, no net change of the spectrum results from this adiabatic changes between two gauge equivalent fields $A_1$. However, in the course of these changes the labeling of the eigenstates has changed.  $n_w$ negative eigenenergies of a certain chirality have become positive and  $n_w$ positive eigenenergies of the opposite  chirality have become negative. This is called the {\em spectral flow} associated with this family of Dirac operators. The spectral flow is determined by the winding number of pure gauges and therefore a topological invariant. The presence of pure gauges with non-trivial winding number implies the occurrence of zero modes in the process of adiabatically changing the gauge field. In mathematics, the existence of zero modes of Dirac operators has become  an important tool in topological investigations of manifolds (\cite{ESPO98}). In physics, the spectral flow of the Dirac operator and the appearance of zero modes induced by topologically non-trivial gauge fields is at the origin  of important phenomena like the formation of condensates or the  existence of chiral anomalies.         
\vskip 0.1cm
\subsection{Instanton Gas}
\vskip 0.1cm
In the semi-classical approximation, as sketched above, the non-perturbative QCD ground state is assumed to be given by  topologically distinguished pure gauges and the instantons connecting the different classical vacuum configurations. 
In the instanton model for the description of low-energy strong interaction physics, one replaces  the QCD partition function \big(Eq.\ (\ref{GFP})\big), i.e.~the weighted sum over all gauge fields by a sum over (singular gauge) instanton fields (\ref{singins})
\begin{equation}
  \label{IE}
A_{\mu} = \sum _{i=1}^{N} \; U (i) \; A_{\mu} (i) \; U^{+} \, (i)\, ,
\end{equation}
with
 $$
 A_{\mu} (i) = \, - \bar{\eta} _{a\mu \nu}\, \frac{2 \rho^2}{g[x-z(i)]^2} \frac{x_{\nu} - z_{\nu} (i)}{[x - z(i)]^{2}+ \rho ^{2}}\,\tau^{a} \;  .
$$
The gauge field is composed of  $N$  instantons with their centers located at the positions  {$z(i)$} and color orientations specified by the  {$SU(2)$} matrices $U(i)$. The instanton ensemble for calculation of $n-$point functions is obtained by summing over these positions and color orientations  
$$
Z [J] = \int \; \prod _{i=1}^{N} \; [ d U (i ) d z(i) ] \; 
e^{- S_{E} [A] + i \int  d ^{4} x \, J  \cdot A \,  }.
$$
Starting point of hadronic phenomenology in terms of instantons are the fermionic zero modes induced by the non-trivial topology of instantons. The zero modes are concentrated around each individual instanton and  can be constructed in closed form 
$$  D \hspace{-.25cm}/ \,\psi_{0}=0,$$
$$\psi_{0}=\frac{\rho}{\pi\sqrt{x^2}}\frac{\gamma x}{(x^2+\rho^2)^{\frac{3}{2}} }\frac{1+\gamma_{5}}{2}\,\varphi_{0},$$
where $\varphi_0$ is an appropriately chosen  constant spinor.       
 In the instanton model, the functional integration over quarks  is truncated as well and replaced by a sum over the zero modes in a given configuration of non-overlapping instantons. A successful description of low-energy hadronic properties has been achieved \cite{SCSH96} although a dilute gas of instantons does  {not confine}  quarks and  gluons. It appears that the low energy-spectrum of QCD is dominated by the chiral properties of QCD which in turn seem to be properly accounted for by the instanton induced fermionic zero modes.
The failure of the instanton model in generating confinement will be analyzed later and related  to a deficit of the model in properly accounting for  the 'center symmetry' in the confining phase. \\
To describe confinement, merons have been proposed \cite{CADG78}  as the relevant field configurations. Merons are singular solutions of the classical equations of motion \cite{DAFF76}. They are  literally half-instantons, i.e.\,up to a factor of $1/2$ the meron gauge fields  are identical to the instanton fields in the ``regular gauge''  \big(Eq.\ (\ref{RINS})\big)
   \begin{eqnarray*}
A_{\mu}^{a \,M}(x) &=&\frac{1}{2} A_{\mu}^{a \,I}(x)= -\frac{1}{g} \; \frac{\eta _{a\mu \nu } x_{\nu}}{x^{2} }\, ,
\end{eqnarray*}
and carry half a unit of topological charge.  
By this  change of normalization, the cancellation between abelian and non-abelian contributions to the field strength is upset and therefore,  asymptotically {
$$ A\sim \frac{1}{x},\quad F\sim \frac{1}{x^2} \, .$$
The action 
$$ S\sim \int d^4 x \frac{1}{x^4}$$
exhibits a logarithmic infrared divergence in addition to the ultraviolet divergence. 
Unlike instantons in singular gauge ($A\sim x^{-3}$),  merons always overlap. A dilute gas limit  of an ensemble of merons does not exist, i.e.\,merons  are strongly interacting. The absence of a dilute gas limit has prevented development of a quantitative meron model of QCD. Recent investigations \cite{LENT04} in which this strongly interacting system of merons is treated numerically indeed  suggest that merons are appropriate effective degrees of freedom for describing the confining phase.
\subsection{Topological Charge and Link Invariants}
Because of its wide use in the topological analysis of physical systems, I will discuss the topological charge and related topological invariants in the concluding paragraph on instantons.\\
 The quantization of the topological charge $\nu$ is a characteristic property of the Yang-Mills theory in 4 dimensions and has its origin in the non-triviality of the mapping (\ref{s3s3}). Quantities closely related to $\nu$ are of topological relevance  in other fields of physics. In electrodynamics topologically non-trivial gauge transformations in 3 space dimensions do not exist \big($\pi_3(S^1) = 0$\big) and therefore the topological charge is not quantized. Nevertheless,  with
$$\tilde{K}^0 =\epsilon^{0ijk} A_i\partial_j A_k $$
the charge 
 \begin{equation}
   \label{toqed}
   h_{{\bf B}}= \int d^3 x\tilde{K}^0 =\int  d^3 x \,{\bf A}\cdot{\bf B}
 \end{equation}
describes topological properties of fields. 
For illustration  we consider two linked magnetic flux tubes (Fig. \ref{matu}) with the axes of the flux tubes forming closed curves ${\cal C}_{1,2}$. Since  $h_{{\bf B}}$ is gauge invariant (the integrand is not, but the integral over the scalar product of the   transverse magnetic field and the (longitudinal) change in the gauge field vanishes), we may assume the vector potential to satisfy the Coulomb gauge condition
$$\mbox{div} {\bf A} = 0 \, ,$$
which allows us to invert the curl operator
\begin{equation}
  \label{cuin}
( \mbox{\boldmath$\nabla$}\times\;)^{-1} = - \mbox{\boldmath$\nabla$}\times \, \frac{1}{\Delta}\,   
\end{equation}
and to express $\tilde{K}^0$ uniquely in terms of the magnetic field
$$ \tilde{K}^0 = -\big( \mbox{\boldmath$\nabla$}\times \, \frac{1}{\Delta}{\bf B}\big)\cdot {\bf B}= \frac{1}{4\pi} \int d^3 x \int d^3 x^{\prime} \big ({\bf B}({\bf x})\times {\bf B}({\bf x}^{\prime})\big)\cdot \frac{{\bf x}^{\prime}-{\bf x}}{|{\bf x}^{\prime}-{\bf x}|^3}  \, . $$ 
For single  field lines, 
$${\bf B}({\bf x})= b_1\frac{d{\bf s}_1}{dt}\delta( {\bf x}-{\bf s}_1(t))+ b_2\frac{d{\bf s}_2}{dt}\delta( {\bf x}-{\bf s}_2(t))$$
the above integral is given by  the linking number of the curves  ${\cal C}_{1,2}$ 
 \big(cf.Eq.\ (\ref{gaulink})\big). Integrating finally over the field lines, the result becomes proportional to the magnetic fluxes $\phi_{1,2}$
 \begin{equation}
   \label{hb}
 h_{{\bf B}}= 2\, \phi_1\phi_2 \, lk\{{\cal C}_1,{\cal C}_2\}\, .   
 \end{equation}

\begin{figure}
\hspace{5.5cm} 
\epsfig{file=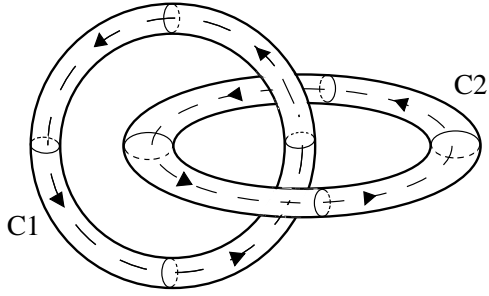, width=0.4\linewidth}
\caption{ Linked magnetic flux tubes }
\label{matu}
\end{figure} 
This result indicates that the charge $ h_{{\bf B}}$, the ``magnetic helicity'', is a topological invariant. For an arbitrary magnetic field, the helicity $h_{{\bf B}}$ can be interpreted as an average linking number of the magnetic field lines \cite{ARKH98}. The helicity $h_{\mbox{\boldmath $\scriptstyle{\omega}$}}$ of vector fields has actually been introduced in hydrodynamics \cite{MOFF69} with the vector potential  replaced by the velocity field ${\bf u}$ of a fluid and the magnetic field by the vorticity $\mbox{\boldmath $\omega$} =  \mbox{\boldmath$\nabla$}\times {\bf u}$. The helicity measures the alignment of velocity and vorticity. The prototype of a ``helical'' flow \cite{MOTS92}  is 
$$  {\bf u}= {\bf u}_0 +\frac{1}{2}\mbox{\boldmath $\omega$}_0\times {\bf x} .$$
The helicity density is constant for constant velocity  ${\bf u}_0$ and vorticity $\mbox{\boldmath $\omega$}_0 $. For parallel  velocity and vorticity, the streamlines of the fluid are-right handed helices. In  magnetohydrodynamics, besides  $h_{{\bf B}}$ and $h_{\mbox{\boldmath $\scriptstyle{\omega}$}}$, a further topological invariant the ``crossed'' helicity can be defined. It characterizes the linkage of $\mbox{\boldmath $\omega$}$  and ${\bf B}$ \cite{DAVI01}. \\
Finally, I would like to mention the role of the topological charge in the connection between  gauge theories and topological invariants \cite{WITT88,KAUF91}. The starting point is the expression (\ref{toqed}) for the helicity, which we use as action of the 3-dimensional abelian gauge theory \cite{POLY88}, the abelian ``Chern-Simons'' action
$$S_{CS} = \frac{k}{8\pi}\int_M d^3x \,{\bf A}\cdot {\bf B}\, ,$$ 
where $M$ is a 3-dimensional manifold and $k$ an integer. One calculates the expectation value of a product of circular Wilson loops
$$W_N =\prod_{i=1}^N \exp\Big\{i\int_{{\cal C}_i}{\bf A}\,d{\bf s}\Big\} .$$
The Gaussian path integral 
$$ \langle W_N\rangle= \int D[A] e^{iS_{CS}}\, W_N$$
can be performed after inversion of the curl operator (\ref{cuin}) in the space of transverse gauge fields. The calculation proceeds along the line of the calculation of $h_{\bf B}$ (\ref{toqed}) and one finds 
$$ \langle W_N\rangle \propto \exp\Big\{\frac{2i\pi}{k} \sum_{i\ne j=1}^{N} lk\{{\cal C}_i,{\cal C}_j\}\Big\}. $$
The path integral for the Chern-Simons theory leads to a representation of a topological invariant. The key property of the Chern-Simons action is its invariance under general coordinate transformations. $S_{CS}$ is itself a topological invariant. As in other evaluations of expectation values of Wilson loops, determination of the proportionality  constant in the expression for $\langle W_N\rangle$ requires regularization of the path integral due to the linking  of each curve with itself (self linking number). In the extension to non-abelian (3-dimensional) Chern-Simons theory, the very involved analysis starts with $K^0$ \big(Eq.\ (\ref{TPK})\big) as  the non-abelian Chern-Simons Lagrangian. The final result is the Jones-Witten invariant associated with the  product of circular Wilson loops \cite{WITT88}.
\section{Center Symmetry and Confinement}
 Gauge theories exhibit, as we have seen, a variety  of non-perturbative phenomena which are naturally analyzed by topological methods. The common origin of all the topological excitations which I have discussed is  vacuum degeneracy, i.e.~the existence of a continuum or a discrete set of classical fields of minimal energy. The phenomenon of confinement, the trademark of non-abelian gauge theories, on the other hand,  still remains mysterious  in spite of  large efforts undertaken to confirm or disprove the many proposals for its explanation.  In particular, it remains unclear whether confinement is related to the vacuum degeneracy associated with the existence of large gauge transformations or more generally whether  classical or semiclassical arguments are at all appropriate for its explanation. In the absence of quarks, i.e.\,of matter in the fundamental representation, $SU(N)$ gauge theories exhibit a residual gauge symmetry, the center symmetry, which is supposed to distinguish between confined and deconfined phases \cite{SVET86}.  
Irrespective of the details of the dynamics which give rise to
confinement, this symmetry must be realized in the confining phase and spontaneously broken in the ``plasma'' phase. Existence of a residual gauge symmetry  implies certain non-trivial topological properties akin to the non-trivial topological properties emerging in the incomplete spontaneous breakdown of gauge symmetries discussed above. In this and the  following chapter I will describe  formal considerations and discuss physical consequences related to the  center symmetry properties of $SU(2)$ gauge theory.  To properly formulate the center symmetry and to construct explicitly the corresponding symmetry transformations and the order parameter associated with the symmetry, the gauge theory has to be formulated on space-time with (at least) one of the space-time directions being compact, i.e.~one has to study gauge theories at finite temperature or finite extension. 
\subsection{Gauge Fields at Finite Temperature and Finite Extension}
When heating a system described  by a field theory or enclosing it by making a spatial direction compact new phenomena occur which to some extent can be analyzed by topological methods. In relativistic field theories systems at finite temperature and systems at finite extensions with an appropriate choice of boundary conditions are copies of each other. In order to display the physical consequences of this equivalence we consider the Stefan-Boltzmann law for the energy density and pressure for a non-interacting scalar field with the corresponding quantities appearing in the Casimir effect, i.e.\,the energy density of the system if it is enclosed in one spatial direction by walls. I assume the scalar field to satisfy periodic boundary conditions on the enclosing walls. The comparison \vskip .1cm
  
\hskip 3.5 cm Stefan-Boltzmann \hskip3.8cm  {Casimir}
$$\hspace{.5cm} \epsilon=\frac{\pi^2}{15}T^4 \hskip4.0cm {p=-\frac{\pi^2}{15}L^{-4}}$$ 
\begin{equation}
  \label{SBC}
\hspace{.7cm} p=\frac{\pi^2}{45}T^4 \hskip4.0cm {\epsilon=-\frac{\pi^2}{45}L^{-4}}\, .
\end{equation}
expresses a quite general relation  between thermal and quantum fluctuations in relativistic field theories  \cite{Toms,LETH98}. This connection is easily established by considering the partition function given in terms of the Euclidean form \big(cf.\ Eq.\ (\ref{eucl})\big) of the Lagrangian 
$$Z=\int_{\mbox{period.}} D[...]e^{- \int_{0}^{\beta}dx_{0}\int dx_1 dx_2 dx_3 {\cal L}_E[...]}$$
 {which describes a system of infinite extension  at } {temperature $T= \beta^{-1}.$}
\vskip.3cm
The partition function 
$$Z=\int_{\mbox{period.}} D[...]e^{- \int_{0}^{L}dx_{3}\int dx_0 dx_1 dx_2 {\cal L}_E[...]}$$ 
describes  the same dynamical system  in its ground state ($T=0$) at finite extension $L$ in 3-direction. As a consequence, by interchanging the coordinate labels in the Euclidean, one easily derives allowing for both finite temperature and finite extension
$$
Z(\beta,L)=\;\;Z(L,\beta)$$
\begin{equation}
  \label{EP}
\hspace{.3cm}\epsilon(\beta,L)=- p(L,\beta) .  
\end{equation}
These relations hold irrespective of the dynamics of the system. They apply to non-interacting systems \big(Eq.(\ref{SBC})\big) and, more interestingly, they  imply that any phase transition taking place when heating up an interacting system has as  counterpart a phase transition occurring when compressing the system ({\em Quantum phase transition}  \cite{SGCS97} by variation of the size parameter $L$).  Critical temperature and critical length are related by  
$$T_c= \frac{1}{L_c} .$$
 For QCD with its supposed phase transition at  about 150 MeV, this relation predicts the existence of a phase transition when compressing the system beyond 1.3 fm. \\
Thermodynamic quantities can be calculated as ground state properties of the same system at the corresponding finite extension. This enables us to  apply the canonical formalism and  with it the standard tools of analyzing the system by symmetry considerations and topological methods.  Therefore, in the following a spatial direction, the 3-direction, is chosen to be compact  and of extension $L$ 
$$ 0\leq x_{3}\leq L \quad x=(x_{\perp}, x_{3}),$$
with 
$$ x_{\perp} = (x_0,x_1,x_2).$$
Periodic boundary conditions for gauge and bosonic matter fields
\begin{equation}
  \label{abc}
A_{\mu}(x_{\perp},x_{3}+L) =  A_{\mu}(x_{\perp},x_{3})\, ,\quad 
 \phi(x_{\perp},x_{3}+L) =  \phi(x_{\perp},x_{3})  
\end{equation}  
are imposed, while 
fermion fields are subject to antiperiodic boundary conditions
\begin{equation}
  \label{pbc}
\psi(x_{\perp},x_{3}+L)= - \psi(x_{\perp},x_{3}) .  
\end{equation}
In finite temperature field theory, i.e.\,for $T=1/L$, only this choice of boundary conditions defines the correct partition functions \cite{KAPU89}. The difference in  sign of fermionic and bosonic  boundary conditions reflect the difference in the quantization of the two fields by commutators and anticommutators respectively. The negative sign, appearing when going around the compact direction is akin to the change of sign in a $2\pi$ rotation of a spin $1/2$ particle. 
   
At finite extension or finite temperature, the fields are defined on $S^1\otimes\mathbb{R}^3$ rather than on $\mathbb{R}^4$ if no other compactification is assumed. Non-trivial topological properties therefore emerge in connection with the $S^1$ component. $\mathbb{R}^3$ can be contracted to a point \big(cf.\ Eq.\ (\ref{horn})\big) and therefore the cylinder is homotopically equivalent to a circle
\begin{equation}
  \label{srho}
S^1\otimes\mathbb{R}^n \sim S^1\, .   
\end{equation}
Homotopy properties of fields defined on a cylinder (mappings from $S^1$ to some target space) are therefore given by the fundamental group of the target space.  
This is illustrated in Fig. \ref{cyl2} 
\begin{figure}
\hspace{5cm} 
\epsfig{file=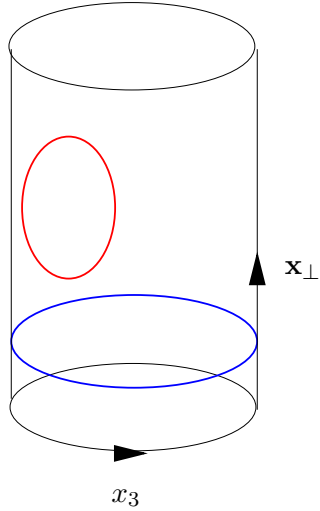, width=0.3\linewidth}\vspace{2cm}\vskip -2cm\hskip 9.4cm\hspace{-3cm} $x_3$\vskip -3.5cm  \hskip 8.8cm${\bf x}_{\perp}$\vskip 3cm
\caption{ Polyakov loop (along the compact $x_3$ direction) and Wilson loop (on the surface of the cylinder) in $S^1\otimes\mathbb{R}^3$ }
\label{cyl2}
\end{figure}   
which shows  two topologically distinct loops. The loop on the surface of the cylinder can be shrunk to a point, while the loop winding around the cylinder cannot.\\
\subsection{Residual Gauge Symmetries in QED}
I start with a brief discussion of electrodynamics with the gauge fields coupled to a charged scalar field as described by the Higgs model Lagrangian (\ref{himo}) (cf.\ \cite{LeNT940,LeNT941}). Due to the homotopic equivalence (\ref{srho}), we can proceed as in our discussion of 1+1 dimensional electrodynamics and classify gauge transformations according to their winding number and separate the gauge transformations into small and large ones with representative gauge functions given by Eq.\ (\ref{gfn}) (with $x$ replaced by $x_3$). If we strictly follow the Faddeev-Popov procedure, gauge fixing has to be carried out by allowing for both type of gauge transformations. Most of the gauge conditions employed do not lead to such a complete gauge fixing. Consider for instance within the canonical formalism with $A_0=0$ the Coulomb-gauge condition 
\begin{equation}
  \label{cbg}
 \mbox{div}{\bf A} =0,  
\end{equation}
and perform a large gauge transformation associated with the representative gauge function (\ref{gfn})
\begin{equation}
  \label{lgtqed}
 {\bf A}(x)\to {\bf A}(x)+{\bf e}_3 d_n\, \quad \phi(x)\to e^{ie x_3d_n}\phi(x)\, . 
\end{equation}
 The transformed gauge field \big(cf.\ Eq.\ (\ref{gt})\big) is shifted by a constant and therfore satisfies the Coulomb-gauge condition  as well.  
Thus, each gauge orbit ${\cal O}$  \big(cf.\ Eq.\ (\ref{GO})\big) is represented by infinitely many configurations each one representing a  suborbit $ {\cal O}_n$. The suborbits are connected to each other by large gauge transformations, while elements within a suborbit are connected by small gauge transformations.  The multiple representation of a gauge orbit  implies that the Hamiltonian  in Coulomb gauge contains a residual symmetry due to the presence of a residual redundancy. Indeed, the Hamiltonian in Coulomb gauge containing only the transverse gauge fields ${\bf A}_{tr}$ and their conjugate momenta  ${\bf E}_{tr}$ 
\big(cf.\ Eq.\ (\ref{hd})\big)
\begin{equation}
  \label{hdcg}  
 {\cal H} =   \frac{1}{2} (\mbox{\boldmath$E$}_{tr}^{2}+\mbox{\boldmath$B$}^{2})+ \pi^{*}\pi +(\mbox{\boldmath$D$}_{tr} \phi) ^{*}(\mbox{\boldmath$D$}_{tr} \phi )  + V(\phi)\, ,\quad H=\int d^3 x {\cal H}({\bf x})\, \end{equation}  
is easily seen to be invariant under the  discrete shifts of the gauge fields joined by discrete rotations of the Higgs field 
\begin{equation}
  \label{crs}
  [H,e^{iD_3 d_n}] = 0 .
\end{equation}
These transformations  are generated by the 3-component of Maxwell's displacement vector
$${\bf D}= \int d^3 x (\,{\bf E}+{\bf x}\,j^{\,0}\,)\, ,$$
with the discrete set of parameters $d_n$ given in Eq.\ (\ref{gfn}). At this point, the analysis of the system via symmetry properties is more or less standard and one can characterize the  different phases of the abelian Higgs model by  their realization of the displacement symmetry. It turns out that the presence of the residual gauge symmetry is necessary to account for the different phases. It thus appears that complete gauge fixing involving also large gauge transformations is not a physically viable option.\\  Like in the symmetry breakdown occurring in the non-abelian Higgs model, in this procedure of incomplete gauge fixing, the $U(1)$ gauge symmetry has not completely disappeared but the  isotropy group $H_{lgt}$ \big(Eq.\ (\ref{isot})\big) of the large gauge transformations (\ref{lgtqed}) generated by  $D_3$ remains. Denoting with $G_1$ the (simply connected) group of gauge transformations in (the covering space) $\mathbb{R}^1$ we deduce from Eq.\ (\ref{th0}) the topological relation
\begin{equation}
  \label{qedstr}
 \pi_1(G_1/H_{lgt})\sim \mathbb{Z}\, ,  
\end{equation}
which expresses the topological stability of the large gauge transformations. Equation (\ref{qedstr}) does not translate directly into a topological stability of gauge and matter field configurations. An appropriate Higgs potential is necessary to force the scalar field to assume a non-vanishing value. In this case the topologically non-trivial configurations are strings of constant gauge fields winding around the cylinder with the winding number specifying both the winding of the phase of the matter field and  the strength of the gauge field. If, on the other hand,  $V(\varphi)$  has just one minimum at  $\varphi=0$ nothing prevents a continuous deformation of a configuration to $A=\varphi=0$.  In such a case, only quantum fluctuations could possibly induce stability.\\ Consequences of the symmetry can be investigated without  such additional assumptions. In the  Coulomb phase for instance with the Higgs potential  given by the mass term $V(\phi)=m^2\phi\phi^{\star}$, the periodic potential for the gauge field zero-mode  
\begin{equation}
  \label{zemo}
a_3^0 = \frac{1}{V}\int d^3 x A_3(x)  
\end{equation}
can be evaluated   \cite{LNOT00}
\begin{equation}
  \label{effp}
V_{\mathrm{eff}}(a_3^0) = -\frac{m^2}{\pi^2 L^2}\sum_{n=1}^{\infty}\frac{1}{n^2}\cos (neLa_3^0)K_2(nmL)\, .
\end{equation}
The effective potential accounts for the effect of the thermal fluctuations on the gauge field zero-mode. It vanishes at zero temperature ($L \rightarrow \infty$). The  periodicity of $V_{\mathrm{eff}}$  reflects the residual gauge symmetry. For small amplitude oscillations $eLa_3^0\ll 2\pi$, $V_{\mathrm{eff}}$ can be approximated by the quadratic term, which in the small extension or high temperature limit, $mL=m/T\ll 1$, defines the Debye screening mass \cite{KAPU89,LEBE96} 
\begin{equation}
  \label{deb}
  m_D^2 = \frac{1}{3} e^2T^2 \, .
\end{equation}
This result can be obtained by standard perturbation theory. We note that this perturbative evaluation of $V_{\mathrm{eff}}$ violates the periodicity, i.e.~it does not respect the residual gauge symmetry. 
\subsection{Center Symmetry in ${\bf SU(2)}$ Yang-Mills Theory}
\label{sec:cesy2}
To analyze topological and symmetry properties of gauge fixed $SU(2)$ Yang-Mills theory, we proceed as above, although abelian and non-abelian gauge theories differ in an essential property. Since $\pi_1\big(SU(2)\big)= 0$, gauge transformations defined on $S^1\otimes\mathbb{R}^3$ are topologically trivial. Nevertheless, non-trivial topological properties emerge in the course of an incomplete gauge fixing  enforced by the presence of a non-trivial center \big(Eq.\ (\ref{center})\big) of $SU(2)$. We will see later that this is actually the correct physical choice for accounting of both the confined and deconfined phases. Before implementing  a gauge condition, it is useful to decompose the gauge transformations  according to their periodicity properties. Although the gauge fields have been  required to be periodic, gauge transformations may not. Gauge transformations preserve periodicity of gauge fields  and of matter fields in the adjoint representation \big(cf.\ Eqs.\ (\ref{matr}) and (\ref{FGT})\big) if they are periodic up to an element of the center of the gauge group 
\begin{equation}
  \label{cgt}
 U\left(x_{\perp},L\right) = c_{U}\cdot U\left(x_{\perp},0\right)\, .
\end{equation}
If fields in the fundamental representation are present with their linear dependence on $U$  \big(Eq.\ (\ref{matr})\big), their boundary conditions require the  gauge transformations $U$ to be strictly periodic $ c_{U}=1$. In the absence of such fields,  gauge transformations can be classified according to the value of $c_{U}$ \big($\pm 1$ in {SU(2)}\big).  An important example of an SU(2) \big(cf.\ Eq.\ (\ref{lpsu2z2})\big)  gauge transformation  {$u_{-}$}}
with  {$c=-1$  is
\beq
\label{uz-1}
u_{-} = e^{i\pi \mbox{\boldmath$\hat{\scriptstyle\psi}
$}\mbox{\boldmath$\scriptstyle\tau $}x_{3}/L}=\cos \pi x_3/L + i \mbox{\boldmath$\hat{\psi}
$}\mbox{\boldmath$\tau $}\sin \pi x_3/L   .
\eeq
Here {$\mbox{\boldmath$\hat\psi$}(x_{\perp})$ is a  unit vector in color space. For  constant $\mbox{\boldmath$\hat\psi$}$, it is easy to verify that the  transformed gauge fields   
\beqs
  A_{\mu}^{\,[u_{-}]} = e^{i\pi \mbox{\boldmath$\hat{\scriptstyle\psi}
$}\mbox{\boldmath$\scriptstyle\tau $} x_{3}/L}A_{\mu} 
e^{-i\pi \mbox{\boldmath$\hat{\scriptstyle\psi}
$}\mbox{\boldmath$\scriptstyle\tau $} x_{3}/L}-\frac{\pi}{gL}\mbox{\boldmath$\hat{\psi}
$}\mbox{\boldmath$\tau $}\delta_{\mu 3}
\eeqs
 indeed remain periodic and continuous. Locally, $c_U=\pm 1$ gauge transformations $U$ cannot be distinguished. Global changes 
induced by gauge transformations like (\ref{uz-1}) are detected by loop variables winding around the compact $x_3$ direction. The Polyakov loop, 
\beq
\label{PL}
P(x_{\perp}) =   P \exp\Bigl\{ig \int_{0}^{L} dx_{3} \,   A_{3}(x) \Bigr\} \, ,
\eeq
is the simplest of such variables and of importance in finite temperature field theory. The coordinate $x_{\perp}$ denotes the position of the Polyakov loop in the space transverse to $x_3$.
Under gauge transformations \big(cf.\ Eqs.\ (\ref{PI}) and (\ref{gaom})\big)
\beqs P(x_{\perp}) \rightarrow
U\left(x_{\perp},L\right)P(x_{\perp})
 U^{\dagger}\left(x_{\perp},0\right)\ .
\eeqs 
With  $x=(x_{\perp},0)$ and $x=(x_{\perp},L)$ labeling  identical points, the Polyakov loop is seen to distinguish $c_U=\pm 1$ gauge transformations. In particular, we have 
\beqs
\mbox{tr}\{P(x_{\perp})\} \rightarrow
\mbox{tr}\{c_{U} P(x_{\perp})\}
\stackrel{\mathrm{SU(2)}}{=} \pm \mbox{tr}\{P(x_{\perp})\} .
\eeqs 
With this result, we now can transfer the classification of gauge transformations to a classification of gauge fields. 
In $SU(2)$, the gauge orbits  {$\cal O$} \big(cf.\ Eq.\ (\ref{GO})\big) are  decomposed  according to   {$c=\pm 1$} into suborbits   {$\cal O_{\pm}$ }. Thus these suborbits are characterized by the sign of the Polyakov loop at some fixed reference point  $x_{\perp}^{0}$
\beq
\label{SO}
 A(x)\, \in \, {\cal O}_{\pm} \quad ,\quad \mbox{if}\quad \pm
\mbox{tr}\{P (x_{\perp}^{0})\}\ge 0 .
 \eeq 
Strictly speaking, it is not the trace of the Polyakov loop rather only its modulus  {$|\mbox{tr}\{P(x_{\perp})\}|$} which is invariant under all  gauge transformations. Complete gauge fixing, i.e.~a representation of gauge orbits ${\cal O}$ by exactly one representative, is only possible if the gauge fixing transformations are not strictly periodic. In turn, if  gauge fixing is carried out  with strictly periodic gauge fixing transformations ({$ U, c_{U}=1$})
 the resulting ensemble of gauge fields  contains one representative $A^f_{\pm}$ for each of the suborbits (\ref{SO}). The label $f$ marks the dependence of the representative on the gauge condition \big(Eq.\ (\ref{GC})\big). The (large) $c_U=-1$ gauge transformation  mapping  the representatives of two gauge equivalent suborbits onto each other are called {\em center reflections}
 \begin{equation}
   \label{zr}
   Z:  A^f_{+} \leftrightarrow A^f_{-} . 
\end{equation}
Under center reflections 
\begin{equation}
  \label{ZPO}
   Z : \quad \mbox{tr}\, P(x_{\perp}) \rightarrow -\mbox{tr}\, P(x_{\perp}).
\end{equation}
 The center symmetry is a standard symmetry within the canonical formalism. Center reflections commute with the Hamiltonian
 \begin{equation}
   \label{HZ}
[H,Z] = 0 \,.
\end{equation}
Stationary states in SU$(2)$ Yang-Mills theory can therefore be classified according to their  Z-Parity
\begin{equation}
   \label{HZN}
H |n_{\pm}\rangle = E_{n_{\pm}} |n_{\pm}\rangle \,, \quad  Z |n_{\pm}\rangle =\pm  |n_{\pm}\rangle \, .
\end{equation}
 {The dynamics of the  { Polyakov loop} is intimately connected to confinement.  The Polyakov loop is associated with the  {free energy} of a single heavy charge. In electrodynamics, the coupling of a heavy pointlike charge to an electromagnetic field is given by}
 \begin{eqnarray*}
\delta{\cal L} =\int d^4 x j^{\mu}A_{\mu} = e\int d^4 x \delta({\bf x-y})\,A_{0}(x)= e\int_{0}^{L} dx_{0}A_{0}(x_{0},{\bf y})\,, \end{eqnarray*}
 {which, in the Euclidean and after interchange of coordinate labels $0$ and $3$, reduces to the logarithm of the Polyakov loop.
The property of the system to confine can be formulated as a  {symmetry property}. The expected infinite free energy of a static color charge results in a vanishing ground state expectation value of the Polyakov loop} 
\beq \label{PEX}\langle 0|\mbox{tr}\, P(x_{\perp})|0\rangle = 0 \eeq
in the confined phase. 
 This property is guaranteed if  the vacuum is center  symmetric.
  The    {interaction energy} $V(x_{\perp})$  of two static charges separated in a transverse direction is, up to an additive constant,  given by the Polyakov-loop correlator} 
\beq\label{PCO}\langle 0|\mbox{tr}P(x_{\perp})\mbox{tr}P(0)|0\rangle = e^{-LV(x_{\perp})}.
 \eeq 
Thus, vanishing of the Polyakov-loop expectation values in the center symmetric phase indicates an infinite free energy of static color charges, i.e.~confinement. For non-zero Polyakov-loop expectation values,  the  free energy of a static color charge is finite and the system is deconfined. A non-vanishing expectation value is possible only if the center symmetry is broken. Thus,  in the transition from the confined to the  plasma phase, the center symmetry, i.e.~a discrete part of the underlying gauge symmetry, must be spontaneously broken. As in the abelian case, a complete gauge fixing, i.e.~a definition of gauge orbits including large gauge transformations may not be desirable or even possible. It will prevent a characterization of different phases by their symmetry properties. \\
As in QED,    non-trivial residual gauge symmetry transformations do not necessarily give rise   to  topologically non-trivial gauge fields. For instance, the  pure gauge obtained from the non-trivial gauge transformation (\ref{uz-1}), with constant  $\mbox{\boldmath$\hat{\psi}$}$,  $A_{\mu}=-\frac{\pi}{gL}\mbox{\boldmath$\hat{\psi}
$}\mbox{\boldmath$\tau $}\delta_{\mu 3}  $  is  deformed  trivially, along a path of vanishing action, into $A_{\mu}=0$. In this deformation, the value of the Polyakov loop  (\ref{PL}) changes continuously from $-1$ to $1$. Thus a vacuum degeneracy  exists with the value of the Polyakov loop labeling the gauge fields of vanishing action. A mechanism, like the Higgs mechanism,  which gives rise to the topological stability of excitations built upon the degenerate classical vacuum has not been identified.   
\subsection{Center Vortices}
\label{sec:cevo}
 Here, we again view the (incomplete) gauge fixing process as a  symmetry breakdown which is induced by the elimination of redundant variables. If we require the center symmetry to be  present after gauge fixing,  the isotropy  group formed by the center reflections must survive the ``symmetry breakdown''. In this way, we  effectively change the gauge group 
\begin{equation}
  \label{suso}
  SU(2) \to SU(2)/\mathbb{Z}(2) .
\end{equation}
Since $\pi_1\big(SU(2)/\mathbb{Z}_2\big) = \mathbb{Z}_2$, as we have seen (Eq.\ (\ref{pi1suz})), this space of gauge transformations contains topologically stable defects, line singularities in $\mathbb{R}^3$ or singular sheets in $\mathbb{R}^4$. Associated with such a  singular gauge transformation $U_{\mathbb{Z}_2}(x)$ are pure gauges (with the singular line or sheet removed) 
$$ A^{\mu}_{\mathbb{Z}_2}(x) = \frac{1}{ig}\,U_{\mathbb{Z}_2}(x)\,\partial ^{\mu}\,U_{\mathbb{Z}_2}^{\dagger}(x). $$ 
The following gauge transformation written in cylindrical coordinates $\rho,\varphi,z,t$
$$U_{\mathbb{Z}_2}(\varphi) = \exp{i\, \frac{\varphi}{2}\,\tau^3}$$
exhibits the essential properties of singular gauge transformations, the {\em center vortices},  and their associated singular gauge fields . $U_{\mathbb{Z}_2}$ is singular on the sheet $\rho = 0$ ( for all $z,t$). It has the property
$$U_{\mathbb{Z}_2}(2\pi) = -U_{\mathbb{Z}_2}(0) ,$$
i.e.~the gauge transformation is continuous in $SU(2)/\mathbb{Z}_2$ but discontinuous as an element of $SU(2)$. The Wilson loop detects the defect. According to Eqs.\ (\ref{wlop}) and (\ref{wlpg}), the Wilson loop, for an arbitrary path ${\cal C}$ enclosing the vortex, is given by 
\begin{equation}
  \label{wlcv}
W_{{\cal C}, \,\mathbb{Z}_2} = \frac{1}{2}\,\mbox{tr}\, \big\{U_{\mathbb{Z}_2}(2\pi)\, U_{\mathbb{Z}_2}^{\dagger}(0)\big\}= -1\, .
\end{equation}
The  corresponding pure gauge field has only one non-vanishing space-time component 
\begin{equation}
  \label{z2vo}
A^{\varphi}_{\mathbb{Z}_2}(x) = - \frac{1}{2g\rho}\tau^3\, ,\end{equation}
which displays the singularity. For calculation of the field strength, we can, with only one color component non-vanishing, apply Stokes theorem. We   obtain for the flux through an area of arbitrary size $\Sigma$ located in the $x-y$ plane 
$$\int_{\Sigma}\, F_{12} \rho d\rho d\varphi = -\frac{\pi}{g}\tau^3\, ,$$
and conclude
$$ F_{12}=  -\frac{\pi}{g}\tau^3 \delta^{(2)}(x).$$
This  divergence in the  field strength makes these fields irrelevant in the summation over all configurations. However, minor changes, like replacing the $1/\rho$ in $ A^{\varphi}_{\mathbb{Z}_2}$ by a function interpolating between a constant at $\rho=0$ and $1/\rho$ at large $\rho$ eliminate this singularity. The modified gauge field is no longer a pure gauge. Furthermore, a divergence in the action from the infinite extension can be avoided by forming closed finite sheets. All these modifications can be carried out without destroying the  property (\ref{wlcv}) that the Wilson loop is $-1$ if enclosing the vortex. This crucial property together with the assumption of a random distribution of center vortices yields an area law for the Wilson loop. This can be seen (cf.\  \cite{RELQ99}) by considering  a large  area ${\cal A}$ in  a certain plane containing a  loop of much smaller area ${\cal A}_W$. Given a fixed  number $N$ of intersection points of vortices with ${\cal A}$, the number of intersection points with ${\cal A}_W$ will fluctuate and therefore the value $W$ of the Wilson loop. For a random distribution of intersection points, the probability to find $n$ intersection points in  ${\cal A}_W$ is given by
$$p_n = \binom{N}{n}\Big(\frac{{\cal A}_W}{{\cal A}}\Big)^n\Big(1-\frac{{\cal A}_W}{{\cal A}}\Big)^{N-n}\, .$$
Since, as we have seen, each intersection point contributes a factor $-1$, one obtains in the limit of infinite ${\cal A}$ with the density $\nu$ of intersection points, i.e.\ vortices per area kept fixed, 
$$\langle W\rangle = \sum_{n=1}^{N} (-1)^n  p_n \to \exp \big( -2\nu {\cal A}_W\big).$$
As exemplified by this simple model, center vortices, if  sufficiently abundant  and sufficiently disordered,  could be responsible for confinement (cf.\ \cite{GREE03}).\\
It should be noticed that, unlike the   gauge transformation $U_{\mathbb{Z}_2}$, the associated pure gauge $A_{\mathbb{Z}_2}^{\mu}$ is not topologically stable. It can be deformed into $A^{\mu}=0$ by a continuous change of its  strength. This deformation, changing the magnetic flux, is not a gauge transformation and therefore the stability of $U_{\mathbb{Z}_2}$ is compatible with the instability of $A_{\mathbb{Z}_2}$.
In comparison to nematic substances with their stable  $\mathbb{Z}_2$ defects (cf.\ Fig.\ \ref{nemat2}), the degrees of freedom of Yang-Mills theories are elements of the Lie algebra and not group-elements and it  is not unlikely that the stability of $\mathbb{Z}_2$ vortices pertains only to formulations of Yang-Mills theories like lattice gauge theories where the elementary degrees of freedom are group elements. \\ 
 It is instructive to compare this unstable defect in the gauge field  with a topologically stable vortex. In a simple generalization \cite{NIOL73} of the non-abelian Higgs model  (\ref{GEGL}) such vortices appear. One considers a system containing  two  instead of one Higgs field with self-interactions of the type (\ref{SEIN})
\begin{equation}
  \label{nah2}
  {\cal L}_{m}=\sum _{k=1,2}\Big\{ \frac{1}{2} D_{\mu} \phi_k D^{\mu} \phi_k - \frac{\lambda_k}{4}(\phi_k^2-a_k^2)^2\Big\} -V_{12}(\phi_1\phi_2)\,,\quad \lambda_k>0\, .
\end{equation}
By a choice of the interaction between the two scalar fields which favors the Higgs fields to be orthogonal to each other in color space, a complete   spontaneous symmetry breakdown  up to multiplication of the Higgs fields with  elements of the center of $SU(2)$ can be achieved. 
The static, cylindrically symmetric Ansatz for such a ``$\mathbb{Z}_2$-vortex'' solution \cite{DVSC86}
\begin{equation}
  \label{scya}
  \phi_1 = \frac{a_1}{2}\tau^3,\quad  \phi_2 = \frac{a_2}{2}f(\rho)\, \big(\cos \varphi\, \tau^1+ \sin \varphi\, \tau^2\big),\quad A^{\varphi}= -\frac{1}{2g} \alpha(\rho)\tau^3 
\end{equation}
leads with $V_{12}\propto (\phi_1\phi_2)^2 $ to a system of equations for the functions $f(\rho)$ and $\alpha(\rho)$  which is almost identical to the coupled system of equations (\ref{eqhi}) and(\ref{eqgf}) for the abelian vortex. As for the Nielsen-Olesen vortex or the 't Hooft-Polyakov monopole, the topological stability of this vortex is ultimately guaranteed by the non-vanishing values of the Higgs fields, enforced by the self-interactions and the asymptotic alignment of gauge and Higgs fields. This stability manifests itself in the quantization of the magnetic flux \big(cf.(\ref{mgch})\big)  
\begin{equation}
  \label{mfz2}
  m= \int_{S^2} {\bf B}  \cdot  d \mbox{\boldmath $\sigma$} =-\frac{2\pi}{g} .
\end{equation}
In this generalized Higgs model, fields can be classified according to their magnetic flux, which either vanishes as for the zero energy configurations or takes on the value (\ref{mfz2}). With this classification, one can associate a $\mathbb{Z}_2$ symmetry similar to the center symmetry with singular gauge transformations connecting the two classes. Unlike center reflections (\ref{uz-1}), singular gauge transformations change the value of the action. It has been argued \cite{THOO78} that, within the 2+1 dimensional Higgs model, this  ``topological symmetry'' is spontaneously broken with the vacuum developing a domain structure giving rise to confinement. Whether this happens is  a dynamical issue as complicated as  the formation of flux tubes in Type II superconductors  discussed on  p.\pageref{Vmm}.  This spontaneous symmetry breakdown  requires  the center vortices  to condense  as a result of an attractive vortex-vortex interaction which makes the  square of the vortex mass  zero or negative. Extensions of such a scenario to pure gauge theories in 3+1 dimensions have been suggested \cite{SAMU79,KOVN01}. 
\subsection{The Spectrum of  the SU(2) Yang-Mills Theory}
Based on the results of Section \ref{sec:cesy2}  concerning the  symmetry and topology of Yang-Mills theories at finite extension, I will deduce properties of the  spectrum of the SU(2) Yang-Mills theory in  the confined, center-symmetric  phase. 
  \begin{itemize}
\item In the center-symmetric phase, 
$$ Z |0\rangle =  |0\rangle \, , $$ 
the vacuum expectation value of the Polyakov loop vanishes (Eq.\ (\ref{PEX})).
\item The correlation function of Polyakov loops  yields the interaction energy
$V$ of static color charges (in the fundamental representation)
\begin{equation}
\exp{\left\{-LV\left(r \right)\right\}}= \langle 0|T \left[ \mbox{tr}\,P\left(
x^{E}_{\perp}\right) \mbox{tr}\,P\left(0\right)\right]| 0\rangle 
\ ,
\qquad  r^{2}= \left(x^{E}_{\perp}\right)^{2} .
\label{SP24}
\end{equation}
\item Due to the rotational invariance in Euclidean space,  $x^{E}_{\perp}$ can be chosen to point in the time direction. After insertion of  a complete set of
excited states
\begin{equation}
\exp{\left\{-L V\left(r \right)\right\}}= \sum _{n_-}\left|
\langle n_-| \mbox{tr}\,P\left(0\right)| 0\rangle  \right|^{2}e^{-E_{n_-}  r} \ .
\label{SP25}
\end{equation}
In the confined phase, the ground state does not contribute  \big(Eq.\ (\ref{PEX})\big).  Since $P\left(x^{E}_{\perp}\right)$ is odd under reflections only odd excited states, 
$$ Z |n_-\rangle = -|n_-\rangle\, , $$
contribute to the above sum. 
If the spectrum exhibits a gap, 
$$E_{n_{-}}\ge E_{1_{-}}>0,$$
the potential energy
$V$ increases linearly with $r$ for large separations,
\begin{equation}
V\left(r \right) \approx \frac{E_{1}}{L} r \quad  \mbox{for}\quad r
\rightarrow \infty \quad \mbox{and} \quad L > L_c \ .
\label{SP26}
\end{equation}
\item  The linear rise  with  the separation, $r$, of two static 
charges \big(cf.~Eq.~(\ref{PCO})\big) is a consequence of  covariance and the presence of a gap in the states excited by the Polyakov-loop operator. The slope of the confining potential is the  string tension  $\sigma$. 
 Thus, in  Yang-Mills theory at finite extension,  the phenomenon
of confinement is connected to the presence of a gap in the spectrum of $Z-$odd states, 
\begin{equation}
E_- \geq  \sigma L \, , 
\label{LGP}
\end{equation}
which increases linearly with the extension of the compact direction.  When  applied to the vacuum, the Polyakov-loop operator generates states  which contain a gauge string winding around the compact direction. The lower limit (\ref{LGP}) is nothing else than the minimal energy necessary to create such a  gauge string in the confining phase. Two such gauge strings, unlike one,  are not  protected topologically from decaying into the ground state or $Z=1$ excited states. We conclude that the states in the $Z=-1$ sector contain $\mathbb{Z}_2$- stringlike excitations with excitation energies given by $\sigma L$.  As we have seen, at the classical level, gauge fields with vanishing action exist which wind around the compact direction. Quantum mechanics lifts the vacuum degeneracy and assigns to the corresponding states the energy (\ref{LGP}).        
\item $Z-$even operators in general will have non-vanishing vacuum expectation values and such operators are expected to generate the hadronic states with the gap determined by the lowest glueball mass $E_+=m_{gb}$ for sufficiently large extension $ m_{gb} L\gg 1$ . 
\item {
SU(2) Yang-Mills theory contains two sectors of excitations which, in the confined phase, are not    connected  by any physical process.
\begin{itemize} 
\item { The hadronic sector,  the sector  of $Z-$even states with a mass gap} (obtained from lattice calculations)  $E_{+} = m_{gb} \approx 1.5 \,\mbox{GeV}$ 
\item { The gluonic\,  sector, the sector  of $Z-$odd states with mass gap} $E_{-} = \sigma L$. 
\end{itemize}
\item 
 When compressing the system, the gap in  the $Z=-1$   sector decreases to} about $ 650$ MeV  {at} $L_{c}\approx 0.75\mbox{fm},\, (T_{c}\approx 270$\, MeV). According to SU(3) lattice gauge calculations, when approaching the critical temperature $T_{c} \approx 220$ MeV,   the lowest glueball mass  decreases. The extent of this decrease is controversial. The value   $m_{gb} (T_{c}) = 770$ MeV has been determined in \cite{FIHK93,GGHK94} while  in a more recent calculation \cite{ISSM02}  the significantly higher value  of $1250$ MeV is obtained for the glueball mass at $T_c$. 
\item
In the deconfined or plasma  phase, the center symmetry is broken. The expectation value of the Polyakov loop does not vanish. Debye screening of the fundamental charges takes place and formation of flux tubes is suppressed. Although the deconfined phase has been subject of numerous numerical investigations, some conceptual issues remain to be clarified. In particular, the origin of the exceptional  realization of the center symmetry is not understood. Unlike symmetries of nearly all other systems in physics, the center symmetry is realized in the low temperature phase and broken in the high temperature phase. The confinement-deconfinement transition shares this exceptional behavior with  the ``inverse melting'' process which has been observed in a polymeric system \cite{RAHK99}  and in a vortex lattice in high-$T_c$ superconductors \cite{ZELD01}. In the vortex lattice, the (inverse) melting into a crystalline state happens as a consequence of the increase in free energy with increasing disorder  which, in turn, under special conditions, may favor formation of a vortex lattice.    
Since nature does not seem to offer a variety of possibilities  for inverse melting, one might guess that a similar mechanism is at work in the confinement-deconfinement transition. A solution of this type would be provided if the model of broken topological $Z_2$ symmetry discussed in  section \ref{sec:cevo} could be substantiated. In this model the confinement-deconfinement transition is driven by the dynamics of the ``disorder parameter'' \cite{THOO78} which exhibits the standard pattern of spontaneous symmetry breakdown.    \\
The  mechanism driving the confinement-deconfinement transition must also be responsible for the disparity in the energies involved. As we have seen, glueball masses are of the order of 1.5 GeV. On the other hand, the maximum in the spectrum of the black-body radiation  increases with temperature and reaches according to Planck's law at $T = 220$~ MeV a value of $620$ MeV. A priori one would not expect a dissociation of the glueballs at such low temperatures. According to the above results concerning the $Z=\pm 1$ sectors, the phase transition may be  initiated by the gain in entropy through  coupling of the two sectors which results in  a breakdown of the center symmetry. In this case the relevant energy scale is not the glueball mass but the mass gap of the $Z=-1$ states which, at the extension corresponding to $220$ MeV, coincides with the peak in the energy density of the blackbody-radiation. 
\end{itemize}
\newpage
\section{QCD in Axial Gauge}
In close analogy to the  discussion of the various field theoretical models which exhibit topologically non-trivial excitations, I have described so far  $SU(2)$ Yang-Mills theory from a rather general point of view. The combination of symmetry and topological considerations and the assumption of a confining phase has led to intriguing conclusions about the spectrum of this theory. To prepare for more detailed investigations, the process of  elimination of redundant variables has to be carried out. In order to make the residual gauge symmetry (the center symmetry) manifest,  the gauge condition has to be chosen appropriately. In most of the standard gauges, the center symmetry is hidden and will become apparent in the spectrum only after a complete solution. It is very unlikely that approximations will preserve the center symmetry as  we have noticed in the context  of the perturbative evaluation of the effective potential in QED \big(cf.\ Eqs.\ (\ref{effp}) and  (\ref{deb})\big). Here I will describe $SU(2)$ Yang-Mills theory in the framework of axial gauge, in which the center reflections can be explicitly constructed and approximation schemes can be developed which preserve the center symmetry.    
\subsection{Gauge Fixing}
We now carry out the elimination of redundant variables and attempt to eliminate the 3-component of the gauge field   {$A_{3}(x)$}. Formally this can be achieved by applying the 
   gauge transformation
 $$ \Omega(x) =   P \exp\,ig \int_{0}^{x_{3}} dz \,   A_{3}
    \left(x_{\perp},z\right).$$
It is straightforward to verify that the gauge transformed 3-component of the gauge field indeed vanishes \big(cf.\ Eq.\ (\ref{FGT})\big)
  \begin{eqnarray*}
A_{3} \left(x\right) &\rightarrow& \Omega\left(x\right) \left(A_{3}\left(x\right)+\frac{1}{ig} \partial _{3} \right) \Omega^{\dagger} \left(x\right)=0. 
\end{eqnarray*}
However, this gauge transformation to axial gauge is  not quite legitimate. The gauge transformation is not periodic 
 $$ \Omega(x_{\perp},x_{3}+L)\neq \Omega(x_{\perp},x_{3}).$$   
In general,  gauge fields then do not remain periodic either under transformation with $\Omega$. Furthermore, with $A_3$ also the gauge invariant trace of the Polyakov loop (\ref{PL}) is incorrectly  eliminated by  {$\Omega$}. These shortcomings  can be cured, i.e.~
 periodicity can be  preserved and the loop variables 
 $\mbox{tr} \, P(x_{\perp})$ can be  restored  with the following modified gauge transformation
 {\begin{eqnarray}
  \Omega_{ag}(x) &=& \Omega_{D}\left(x_{\perp}\right) \big[P^{\dagger}(x_{\perp})\big]^{x_{3}/L}\, \Omega(x)\, .
\label{GFT} 
\end{eqnarray}}
The gauge fixing to axial gauge thus proceeds in three steps
\begin{itemize}
\item Elimination of the 3-component of the gauge field $A_{3}(x)$
\item Restoration of the Polyakov loops  $P(x_{\perp})$
\item  Elimination of the gauge variant components  of the Polyakov loops $P(x_{\perp})$ by diagonalization 
 {\beq
\label{GDP}
\Omega_{D}\left(x_{\perp}\right) P(x_{\perp})\Omega_{D}^{\dagger}\left(x_{\perp}\right) = e^{igL a_{3}(x_{\perp})\,\tau^{3}/2} \, .
\eeq}
\end{itemize}} 
\vskip .1cm
{\em Generating Functional}
\vskip .1cm
With the above explicit construction of the appropriate gauge transformations, we have  established that the 3-component of the gauge field indeed can be eliminated in favor of a diagonal $x^{3}$-independent  field $ a_{3}(x_{\perp})$. In the language of the Faddeev-Popov procedure, the axial gauge condition \big(cf.\ Eq.\ (\ref{GC})\big) therefore reads 
\beq
\label{axgc}
f[A] = A_{3} - \Big(a_{3}+\frac{\pi}{gL} \Big)\frac{\tau^{3}}{2} .\eeq
 The field $ a_{3}(x_{\perp})$  is compact, 
\beqs
\label{comvar}
  a_{3} = a_{3}(x_{\perp}),\quad - \frac{\pi}{gL}\le a_{3}(x_{\perp})\le \frac{\pi}{gL} \, . 
\eeqs
It is interesting to compare QED and QCD in axial gauge in order to identify  already at this level properties which are related to the non-abelian character of QCD. In QED the same procedure can be carried out with omission of the third step. Once more, a lower dimensional field has to be kept for  periodicity and gauge invariance. However, in QED  the integer part of $a_{3}(x_{\perp})$  cannot be gauged away; as winding number of the mapping $S^1 \rightarrow S^1$   it is protected topologically. 
In QCD, the appearance of the compact variable is ultimately due to the elimination of the gauge field $A_3$, an element of the Lie algebra, in favor of  $P(x_{\perp})$, an element of the compact Lie group. \\
With the help of the auxiliary field $a_{3}(x_{\perp})$, the generating functional for QCD in axial gauge is written as 
\begin{eqnarray}
\label{AGGF}
Z \left[J\right] &=& \int d[a_{3}]   
 d\left[A\right]\Delta_{f}\left[A\right]
\delta\left\{ A_{3} - \Big(a_{3}+\frac{\pi}{gL}\Big)\frac{\tau^{3}}{2}\right\} \, e^{i S \left[A\right]+i \int d^{4}x J^{\mu}  A_{\mu}} .
\end{eqnarray}
This generating functional contains as dynamical variables  the fields  $a_{3}(x_{\perp}), A_{\perp}(x)$
with
$$ A_{\perp}(x)=\{A_0(x),A_1(x),A_2(x)\}.$$
It is one of the unique features of axial gauge QCD that the   Faddeev-Popov determinant \big(cf.\ Eqs.\ (\ref{FPD}) and \ref{M})\big)
$$ \Delta_{f}\left[A\right] = | \det D_{3} | $$
  can be evaluated in closed form
\beqs
\label{det}
  \frac{\det D_{3}}{(\det \partial_{3})^{3}} = \prod_{x_{\perp}}\frac{1}{L^{2}}  \cos^{2} gLa_{3}(x_{\perp})/2 \, ,  
\eeqs
and absorbed   into the measure   
\beqs
\label{gfax1}
 Z \left[J\right]= \int D[a_{3}]   
 d\left[A_{\perp}\right]
 e^{i S \left[A_{\perp},a_{3}-\frac{\pi}{gL} \right]+i \int d^{4}x JA}. 
\eeqs
The measure 
\begin{eqnarray}
 D\left[a_{3}\right] &=& \prod_{x_{\perp}} \cos^{2}\left( gL
a_{3}(x_{\perp})/2 \right)\, \Theta \left[a_{3}(x_{\perp})^2-(\pi/g L)^2 \right]
 da_{3}\left(x_{\perp}\right) 
\label{HAM}
\end{eqnarray}
is nothing else than the Haar measure of the gauge group. It reflects the presence of variables ($a_3$) which are built from elements of the Lie group and not of the Lie algebra. Because of the topological equivalence of $SU(2)$ and $S^3$ \big(cf.\ Eq.\ (\ref{su2s3})\big) the Haar measure  is the volume element of $S^3$ 
$$d\Omega_3 = \cos^{2}\theta_1 \cos\theta_2 \,d\theta^1 d\theta^2 d\varphi\, ,$$
with the polar angles defined in the interval $[-\pi/2,\pi/2]$.
In the diagonalization of the Polyakov loop (\ref{GDP}) gauge equivalent fields corresponding to different values of $\theta_2$ and $\varphi$ for fixed $\theta_1$ are eliminated as in the example discussed above \big( cf.\ Eq.\ (\ref{s2o2})\big).  
 The presence of the Haar measure has far reaching consequences. 
\vskip .2cm
{\em Center Reflections}\vskip .1cm 
Center reflections {$Z$} have been formally defined in (\ref{zr}). They  are  {residual gauge transformations} which change the  sign of the Polyakov loop (\ref{ZPO}). These residual gauge transformations are loops in $SU(2)/\mathbb{Z}_2$  \big(cf.\ Eq.\ (\ref{lpsu2z2})\big) and, in axial gauge, are given by 
\beqs Z =  i   e^{i\pi \scriptstyle\tau^{1}/2} e^{i\pi
\tau^{3} x^{3}/L} .
\eeqs
They 
transform the gauge fields, and leave the action invariant
\bea
\label{CR}
 Z: \quad a_{3} \rightarrow -a_{3} ,\quad A_{\mu}^{3} &\rightarrow&
-A_{\mu}^{3},\quad \Phi_{\mu} \rightarrow  \Phi_{\mu}^{\dagger},\quad 
S[A_{\perp}a_{3}]  \rightarrow S[A_{\perp}a_{3}]\, .
\eea
The off-diagonal gluon fields have been represented in a spherical basis by the antiperiodic fields 
\beq
\label{CGL}
\Phi_{\mu}(x)= \frac{1}{\sqrt{2}}[A_{\mu}^{1}(x)+i A_{\mu}^{2}(x)]e^{-i\pi x^{3}/L}\, .
\eeq
We emphasize that, according to the rules of finite temperature field theory, the  bosonic gauge fields $A_\mu^{a}(x)$ are periodic in the compact variable $x_3$. For convenience, we have introduced in the definition of $\Phi$ an $x_3$-dependent phase factor which makes these field antiperiodic. With this definition, the action of center reflections simplify,  $Z$ becomes a (abelian) charge
conjugation  with the charged fields $\Phi_{\mu}(x)$ and the ``photons'' described by the neutral fields  {$A_{\mu}^{3}(x),
a_{3}(x_{\perp}) $}. 
Under center reflections, the trace of the Polyakov loop changes sign, 
\bea
\label{TRP}
\frac{1}{2}\, \mbox{tr}\, P(x_{\perp}) &=& -\sin \frac{1}{2} gLa_{3}(x_{\perp})\, .
\eea
 Explicit representations of center reflections are not known in other gauges.
\subsection{Perturbation Theory in the Center-Symmetric Phase}
The center symmetry protects the  $Z-$odd states with their large excitation energies (\ref{LGP}) from mixing with the  $Z-$even ground or excited states. Any approximation  compatible with confinement has therefore to respect the center symmetry. I will describe some first attempts towards the development of a perturbative but center-symmetry preserving scheme. {In order to display the peculiarities of the dynamics of the Polyakov-loop variables $a_{3}(x_{\perp})$ we disregard in a first step  their couplings   to the charged gluons $\Phi_{\mu}$ \big( Eq.\ (\ref{CGL})\big). The system of decoupled Polyakov-loop variables} is described  by the 
 Hamiltonian
\beq
h=\int d^{2}
x_{\perp}\left[-\frac{1}{2L}\frac{\delta^{2}}{\delta a_3({\bf x}_{\bot})^{2}} +\frac{L}{
2}\left[\mbox{\boldmath$\nabla$}a_3({\bf x}_{\bot})\right]^2\right] 
\label{h3}
\eeq
 and by the boundary conditions at  {$a_{3}=\pm\frac{\pi}{gL}$ for the ``radial'' wave function 
\beq
\label{bca3}
\hat{\psi}[a_{3}]\big|_{\mbox{boundary}}= 0\,  . \eeq
This system has a simple mechanical analogy. The Hamiltonian describes   a 2 dimensional array of degrees of freedom interacting harmonically with their nearest neighbors (magnetic field energy of the Polyakov-loop variables).  If we disregard for a moment the boundary condition,  the elementary excitations are ``sound waves'' which run through the lattice. This is actually the model we would obtain in electrodynamics, with  the sound waves representing  the massless photons. Mechanically we can interpret the boundary condition as a result of an infinite square well in which each mechanical degree of freedom is trapped, as is illustrated in  Figure \ref{coup}. 
\begin{figure}
\hskip 3cm\epsfig{file=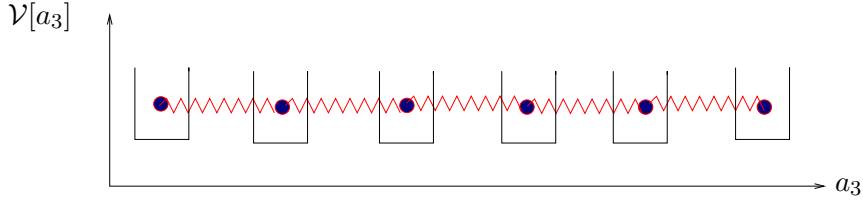, width=.6\linewidth}
$a_3$
\vskip -2.7cm \hskip1.7cm  ${\cal V}[a_3]$\vskip 2cm
\caption{ System of harmonically coupled  Polyakov-loop variables (\ref{h3}) trapped by the boundary condition (\ref{bca3}) in infinite square wells}
\label{coup}
\end{figure}
This infinite potential is of the same origin as the one introduced in Eq.\ (\ref{V}) to suppress contributions of fields beyond the Gribov horizon. Considered classically, waves with sufficiently small amplitude and thus with sufficiently small energy can propagate through the system without being affected by the presence of the walls of the potential. Quantum mechanically this may not be the case. Already the zero point oscillations may be changed substantially by the infinite square well. With discretized  space  (lattice spacing $\ell$)
and  rescaled dynamical variables 
\beqs
\label{resc}
 \tilde{a}_{3}(x_{\perp}) = gLa_{3}(x_{\perp})/2 \, ,
\eeqs
it is seen that for $\ell \ll L$ the electric field (kinetic) energy dominates. Dropping the nearest neighbor interaction, the  ground state wavefunctional is given by
\beqs
\hat{\Psi}_{0}\,[\,\tilde{a}_{3}] = \prod_{\vi} 
\left[ \left(
\frac{2}{\pi}\right)^{1/2}
 \cos \left[ \tilde{a}_{3}(\vi) \right] \right]   .
\label{35}
\eeqs
In the absence of the nearest neighbor interaction, the system does not support waves and the excitations remain localized. 
The states of  lowest excitation   energy are obtained by exciting a single 
degree of freedom at one site} $\tilde{{\bf x}}_{\perp}$ {into its first excited
state
\beqs
 \cos \left[ \tilde{a}_{3}(\tilde{\vi}) \right] \rightarrow \sin \left[2 \tilde{a}_{3}(\tilde{\vi}) \right] 
\eeqs
{with  excitation energy}
\beq
\Delta E = \frac{3}{8} \frac{g^{2} L}{\ell ^{2}} \, .
\label{ES}
\eeq
Thus, this perturbative calculation is in agreement with our general considerations and yields excitation energies rising with the extension $L$. From comparison with Eq.\ (\ref{LGP}), the string tension
$$\sigma =\frac{3}{8} \frac{g^{2} }{\ell ^{2}}$$
is obtained.
This value coincides with the  strong coupling limit of lattice gauge theory. However, unlike lattice gauge theory in the strong coupling limit, here no confinement-like behavior is obtained in QED. Only in QCD the Polyakov-loop variables $a_3$ are compact and thereby give rise to localized excitations rather than waves. It is important to realize that in this description of the Polyakov loops and their confinement-like properties we have left completely the familiar framework of classical fields with their well-understood topological properties. Classically the  fields $a_3=\mbox{const.}$ have zero energy. The quantum mechanical zero point motion raises this energy insignificantly in electrodynamics and dramatically for chromodynamics. The confinement-like properties are  purely quantum mechanical in origin. Within quantum mechanics, they are derived from the ``geometry'' (the Haar measure) of the kinetic energy of the momenta conjugate to the Polyakov loop variables, the chromo-electric fluxes around the compact direction.          

\vskip 0.1cm
{\em Perturbative coupling of gluonic variables}\vskip 0.1cm
In the next step,  one may include coupling of the Polyakov-loop variables to each other via the nearest neighbor interactions. As a result of this coupling, the spectrum contains bands of excited states centered around the excited states in absence of the magnetic coupling  \cite{LEMT95}. The width of these bands is suppressed by a factor $\ell^2/L^2$ as compared to the excitation energies (\ref{ES}) and can therefore be neglected in the continuum limit.  Significant changes occur by the coupling of the Polyakov-loop variables to the charged gluons $\Phi_{\mu}$.   
 { We continue to neglect the magnetic coupling    $ (\partial_{\mu}a_{3})^2$. The Polyakov-loop variables  {$a_{3}$}  appearing at most quadratically in the action can be integrated out  in this limit and the following effective action is obtained  
} 
\beq\label{EFA}
S_{\rm eff} \left[A_{\perp}\right]=
S\left[A_{\perp}\right]
+\frac{1}{2} M^{2}\sum_{a=1,2}
\int d^{4}x A^{a}_{\mu} \left(x\right)  A^{a,\mu} \left(x\right) \ .
\eeq
The antiperiodic  boundary conditions of the charged gluons, which have arisen in the change of field variables  \big(Eq.\ (\ref{CGL})\big)
reflect  the mean value of  {$A_{3}$} in the center-symmetric phase
$$ A_{3}=a_{3}+\frac{\pi}{gL},$$ 
while 
{ the geometrical ($g-$independent) mass}
\beq\label{GM}
M^{2}=\left(\frac{\pi^{2}}{3}-2\right) \frac{1}{L^{2}}
\eeq
{arises from their fluctuations.
Antiperiodic boundary
conditions  describe the appearance of  { Aharonov-Bohm fluxes} in the
elimination of the Polyakov-loop variables. 
 The original periodic charged gluon fields may be continued to be used if
the partial derivative {$\partial_{3}$} is replaced by the covariant one
$$
\partial_{3} \rightarrow \partial_{3}+\frac{i\pi}{2 L} [\tau^{3}\, , \quad . 
  \qquad 
$$ 
Such a change of boundary conditions is a phenomenon well known in quantum mechanics. It occurs for a point particle moving on a circle (with circumference $L$) in the presence of a  magnetic flux generated by a constant vector potential along the  compact direction. With the transformation of the wave function
$$\psi(x)\rightarrow e^{ieA x}\psi(x),$$
the covariant derivative
$$(\frac{d}{dx}-ieA)\psi(x) \rightarrow \frac{d}{dx}\psi(x)$$
becomes an ordinary derivative at the expense of a change in boundary conditions at $x=L$.   
 { Similarly, the charged massive gluons move  in a constant
color neutral gauge field  of  strength
 {$ \frac{\pi}{g L}$}  pointing in the spatial 3 direction. With   {$x_{3}$}  compact,  a color-magnetic flux is associated  with this gauge field,}}
\beq
\Phi_{\rm mag}= \frac{\pi}{g} \ ,
\label{ABF}
\eeq
  {corresponding to a  {magnetic field} of strength}
$$
B= \frac{1}{g L^{2}}\; .
\label{J13}
$$
\vskip .2cm
Also quark {boundary conditions are  changed under the influence of the}  {color-magnetic fluxes}
\beq\label{QBC}\psi\left(x\right) \rightarrow   \exp{\left[-i x_{3}\frac{\pi}{2L}
\tau^{3}\right]}\psi\left(x\right).\eeq
Depending on their color they acquire a phase of $\pm i$ when transported around the compact direction. Within the effective theory, the  Polyakov-loop correlator can be calculated perturbatively. As is indicated in the diagram of Fig. \ref{fd1}, Polyakov loops propagate only through their coupling to the charged gluons.\vskip .1cm
\begin{figure} 
\hspace{1cm}\epsfig{file=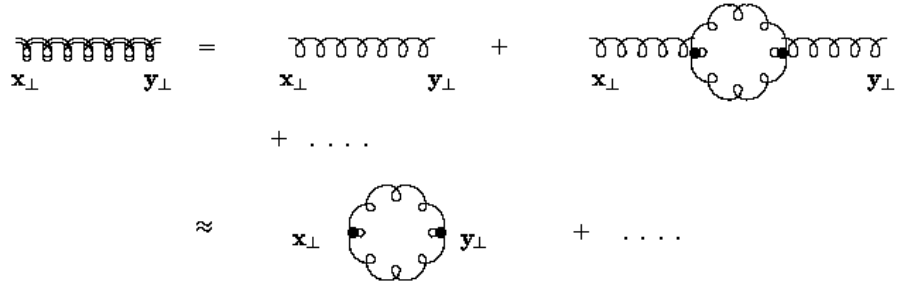, width=0.8\linewidth}
\caption{ One loop contribution from charged gluons  to the propagator of Polyakov loops (external lines)}
\label{fd1}
\end{figure}
Confinement-like properties are preserved   when coupling to the Polyakov loops to the charged gluons. The linear rise of the interaction energy of fundamental charges obtained in leading order persist. As a consequence of the coupling of the Polyakov loops to the charged gluons, the value of the string constant is now determined by the threshold for charged gluon pair production 
\beq
\label{SIGP}
 \sigma_{\rm pt} = \frac{2}{L^{2}}\sqrt{\frac{4\pi^{2}}{3}-2}\,\, ,  
\eeq
i.e.~the perturbative string tension vanishes in  limit $L \rightarrow \infty$. This deficiency results from the perturbative treatment of the charged gluons. A realistic string constant  will arise only if the threshold of a $Z-$odd pair of charged gluons increases linearly with the extension $L$ \big(Eq.\ (\ref{LGP})\big).\\ 
Within this approximation,  also the effect of dynamical quarks on the Polyakov-loop variables can be calculated by including quark loops besides the charged gluon loop in the calculation of the Polyakov-loop propagator (cf.\ Fig. \ref{fd1}). As a result of this coupling, the interaction energy of static charges ceases to rise linearly; it saturates for asymptotic distances at a value of 
\beqs
\label{CE6}
V(r) \approx 2m \ .
\eeqs
Thus, string breaking by dynamical quarks is obtained. This is a remarkable and rather unexpected result. Even though perturbation theory has been employed, the asymptotic value of the interaction energy is independent of the coupling constant $g$ in contradistinction to the $e^4$ dependence of the  Uehling potential in QED which accounts e.g. for the screening of the proton charge in the hydrogen atom by vacuum polarization \cite{PESC95}. Furthermore,  the quark loop contribution vanishes if  calculated with anti-periodic or periodic boundary conditions. A finite result only arises with the boundary conditions (\ref{QBC}) modified by the  Aharonov-Bohm fluxes. The $1/g$ dependence of the  strength of these fluxes \big(Eq.\ (\ref{ABF})\big) is  responsible for the coupling constant independence of the asymptotic value of $V(r)$. 
\subsection{Polyakov Loops in the Plasma Phase}
{If the center-symmetric phase would persist at high temperatures or small extensions, charged gluons with their increasing  geometrical mass (\ref{GM}) and the increasing strength of the interaction  (\ref{CGL})  with the Aharonov-Bohm fluxes, would decouple
$$ \Delta E \approx \frac{\pi}{L} \rightarrow \infty .$$ 
Only neutral gluon fields are periodic in the compact $x_3$ direction and therefore possess  zero modes. Thus at small extension or  high temperature $L \rightarrow0$, only neutral gluons would contribute to thermodynamic quantities. This is in conflict with results of lattice gauge calculations \cite{Reisz} and we therefore will assume that the  center symmetry is spontaneously  broken for  {$L\le L_c=1/T_{c}$}.   In the  high-temperature phase,  Aharonov-Bohm fluxes  must be screened and the geometrical mass must be reduced. Furthermore, with the string tension vanishing in the plasma phase, the effects of the Haar measure must be effectively suppressed and the Polyakov-loop variables may be treated as classical fields. On the basis of this assumption, I now describe the development of  a phenomenological treatment of the plasma phase \cite{EKLT98}. } For technical simplicity,  I will neglect the space time dependence of  {$a_{3}$} and describe the results for vanishing  geometrical mass  {$M$}. For the description of the high-temperature phase it is more appropriate to use the variables 
\beqs
  \chi= gLa_{3} +\pi 
\eeqs
with vanishing average Aharonov-Bohm flux. Charged gluons satisfy quasi-periodic boundary conditions
\begin{equation}
  \label{qpbc}
A_{\mu}^{1,2}(x_{\perp},x_{3}+L) =  e^{i\chi}A_{\mu}^{1,2}(x_{\perp},x_{3}).  
\end{equation}
Furthermore, we will calculate the thermodynamic properties by evaluation of the energy density in the Casimir effect \big(cf.\ Eqs.\ (\ref{SBC}) and (\ref{EP})\big). In the Casimir effect, the central quantity to be calculated is the ground state energy of gluons between plates on which the fields have to satisfy appropriate boundary conditions. In accordance with our choice of boundary conditions \big(Eq.\ (\ref{abc})\big), we assume the enclosing plates to extend in the $x_1$ and $x_2$ directions and to be separated in the $x_3$ direction. The essential observation for the following phenomenological description is the dependence of  the Casimir energy on the  boundary conditions and therefore on the presence of Aharonov-Bohm fluxes. 
The  Casimir energy of the charged gluons is obtained by summing, after regularization, the zero point energies  {  
\beq
\label{CE}
\varepsilon(L,\chi) = \frac{1}{2}\sum_{n=-\infty}^{\infty} \int \frac{d^2 k_{\perp}}{(2\pi)^{2}}\left[{\bf k}_{\perp}^{2} +  (2 \pi n+ \chi)^{2} \over {L^2} \right]^{1/2} = \;\frac{4\pi^2}{3L^4}B_{4}\Big(\frac{\chi}{2\pi}\Big)\eeq 
 }
with
$$B_{4}(x)= -\frac{1}{30}+ x^2(1-x)^2 .$$
Thermodynamic stability requires positive pressure at finite temperature and thus, according to Eq.\ (\ref{EP}), a negative value for the Casimir energy density. This requirement is satisfied if      
\beqs
\label{thst}
 \chi\leq 1.51 . 
\eeqs
For complete screening ({  $\chi = 0$ }) of  the Aharonov-Bohm fluxes, the expression for the pressure in  black-body radiation is obtained (the factor of two difference between Eqs.\ (\ref{SBC}) and (\ref{CE}) accounts for the two charged gluonic states). Unlike QED, QCD is not stable for vanishing Aharonov-Bohm fluxes. In QCD the perturbative ground state energy can be lowered by spontaneous formation of magnetic fields. Magnetic stability  can be reached if the strength of the Aharonov-Bohm fluxes does not decrease beyond a certain minimal value.  By calculating the Casimir effect in the presence of an external, homogeneous   {color-magnetic  field} 
\beqs
B^a_i =  \delta^{a3} \delta_{i3} B \, , 
\eeqs
this minimal value can be determined.
The energy of a single quantum state is given in terms of the oscillator quantum number $m$ for the Landau orbits, in terms of the momentum quantum number $n$ for the motion in the (compact) direction of the magnetic field, and by a magnetic moment contribution ($s=\pm1$)   
\begin{eqnarray*} 
E_{mns}  = \left[2gH(m+1/2) +  { { (2 \pi n + \chi)^2} \over {L^2} } 
+2sgH \right]^{1/2}
\ .
\nonumber \\
\label{regularenergy}
\end{eqnarray*} 
This expression shows that the destabilizing magnetic moment contribution $2sgH$ in the state with  
 $$ s=-1, m=0, n=0 $$
can be compensated by a non-vanishing   Aharonov-Bohm flux {$\chi$} of sufficient strength. For determination of the actual value of $\chi$, the sum over these energies has to be performed. After regularizing the expression, the Casimir energy density can be computed numerically.  The requirement of  {magnetic stability} yields a lower limit on  {$\chi$}. As Figure \ref{fi12} shows, 
\begin{figure}
\begin{center}
\begin{minipage}[b]{0.44\hsize}
\epsfig{file=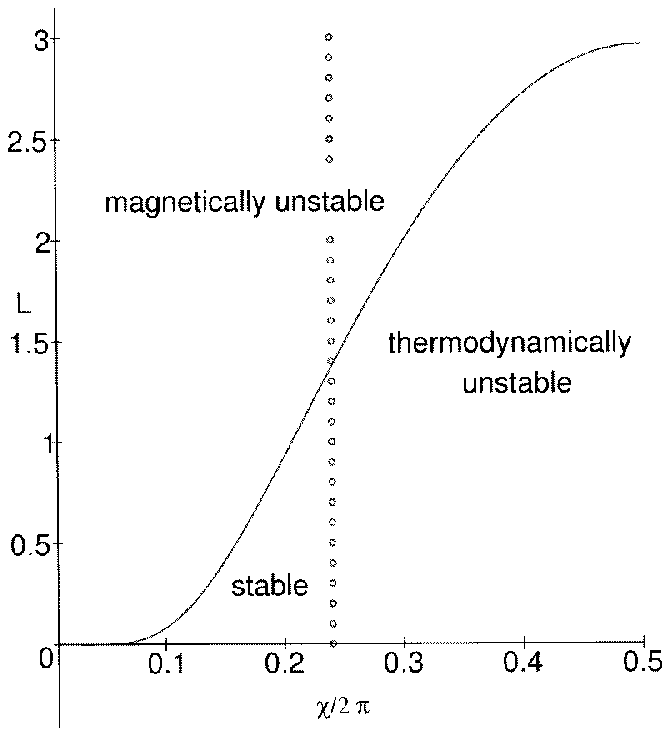,width=\hsize}
\end{minipage}
\begin{minipage}[b]{0.44\hsize}
\epsfig{file=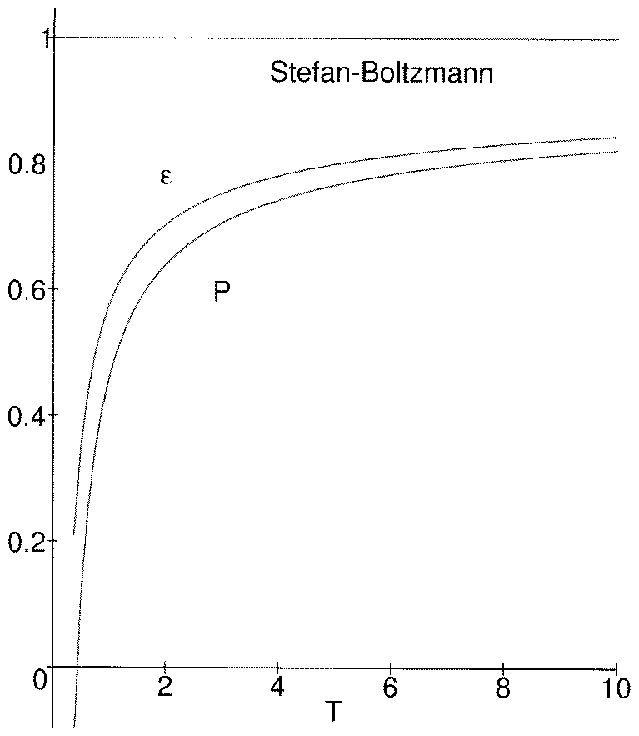, width=.93\hsize,angle=1}
\end{minipage}
\end{center}
\vspace*{0.1cm} \noindent
\caption{Left: Regions of stability and instability  in the $(L,\chi)$ plane. To the  right of the circles, thermodynamic instability; above the solid line, magnetic instability. Right: Energy density and pressure normalized to Stefan-Boltzmann  values vs.  temperature in units of  $\Lambda_{\rm MS}$}
\label{fi12}
\end{figure}
the Stefan-Boltzmann limit $\chi=0$ is  not compatible with magnetic stability  for any value of the  temperature. Identification of the Aharonov-Bohm flux with  the minimal
allowed values sets upper limits to energy density and pressure which are shown in Figure \ref{fi12}. 
 These results are reminiscent of lattice data  \cite{KALP00} in the slow 
 {logarithmic} approach of energy density and pressure 
 {
$$\chi(T)\ge \frac{11}{12}g^2(T),\; T \rightarrow \infty $$
to the Stefan-Boltzmann limit.

It appears that the finite value of the 
Aharonov-Bohm flux accounts for interactions present in the deconfined phase
fairly well; qualitative agreement with lattice calculations is also obtained for the  ``interaction measure'' {$\epsilon-3P$}. Furthermore, these limits on $\chi$ also yield a realistic  estimate for the 
 {change in energy density} {$-\Delta \epsilon$} across the phase transition. The phase  transition  is accompanied by  a change in strength of the 
Aharonov-Bohm flux from  the center symmetric value {$\pi$} to a value in the stability region. The lower bound   is determined by thermodynamic stability  
\beqs
 - \Delta \epsilon \geq \epsilon (L_{c},
  \chi=\pi)-\epsilon (L_{c}, \chi=1.51) = \frac{7\pi^{2}}{180}\frac{1}{L_{c}^{4}} .
\eeqs
For  establishing an upper bound,  the critical temperature must be specified. For {$T_{c} \approx 270$} MeV, 
\beqs
0.38\,\frac{1}{L_{c}^{4}}\le - \Delta \epsilon  \le 0.53\,\frac{1}{L_{c}^{4}}.
\eeqs  
These limits are compatible with the lattice result  {\cite{ENKR95}} 
\beqs
\Delta \epsilon=  -0.45\frac{1}{L_{c}^{4}} \ .
\eeqs
The picture of increasing  screening of the Aharonov-Bohm fluxes with increasing temperature seems to catch the essential physics of the  thermodynamic quantities. It is remarkable that the requirement of magnetic stability, which prohibits complete screening, seems to determine the temperature dependence of the Aharonov-Bohm fluxes and thereby to simulate the non-perturbative dynamics  in a semiquantitative way.   
 \subsection{Monopoles}
The discussion of the dynamics of the Polyakov loops has demonstrated that significant changes occur if compact variables are present.  The results discussed strongly suggest that confinement arises naturally in a setting where the dynamics is dominated by such compact variables.
 The Polyakov-loop variables $a_3(x_{\perp})$ constitute  only a small set of degrees of freedom in gauge theories. In axial gauge, the remaining degrees of freedom $A_{\perp}(x)$ are standard fields which, with interactions neglected, describe freely propagating particles. As a consequence, the coupling of the compact variables to the other degrees of freedom  almost destroys the confinement present in the system of uncoupled Polyakov-loop variables. This can be prevented to happen only if mechanisms are operative by which  all the gluon fields acquire infrared properties similar to those of the Polyakov-loop variables. In the  axial gauge representation it is tempting to connect such mechanisms to the presence of monopoles whose existence is intimately linked to the compactness of the Polyakov-loop variables. In analogy to the abelian Higgs model, condensation of magnetic monopoles  could be be a  first  and crucial element of a mechanism for confinement. It would correspond to the formation of the charged Higgs condensate $|\phi| = a$ (\ref{mani}) enforced by  the Higgs self-interaction (\ref{vp}). Furthermore, this magnetically charged medium should display excitations which behave as chromo-electric vortices. Concentration of the electric field lines to these vortices finally would give rise to a linear increase in the interaction energy of two chromo-electric charges with their separation as in Eq.\ (\ref{Vmm}). These phenomena actually happen in the Seiberg-Witten theory \cite{SEWI94}. The Seiberg-Witten theory is a supersymmetric generalization of the non-abelian Higgs model. Besides gauge and Higgs fields it contains fermions in the adjoint representation. It exhibits vacuum degeneracy  enlarged by supersymmetry and contains topologically non-trivial excitations, both monopoles and instantons. The monopoles can  become massless and when partially breaking the supersymmetry, condensation of monopoles occurs that induces confinement of the gauge degrees of freedom. \vskip .1cm  
In this  section I will sketch the emergence of monopoles in axial gauge and discuss some elements of their dynamics. 
 {Singular field arise in the last step of the gauge fixing procedure (\ref{GFT}), where  the variables characterizing the orientation of the Polyakov loops in color space are eliminated as redundant variables by diagonalization of the Polyakov loops. The diagonalization of group elements is achieved by the unitary matrix 
   \begin{eqnarray*}
\Omega_{D} =e^{i\mbox{\boldmath$\omega$}\mbox{\boldmath$\tau$}} = \cos \omega +i\tau \hat{\omega}\sin \omega\, ,
 \end{eqnarray*}
with $\omega(x_{\perp})$ depending on the Polyakov loop $P(x_{\perp})$ to be diagonalized. This diagonalization is ill defined   if
 \begin{equation}
  \label{mop1}
P(x_{\perp})=\pm 1 \, ,  
\end{equation}
 i.e.~if the Polyakov loop is an  element of the center of the group \big(cf.\ Eq.\ (\ref{center})\big). Diagonalization of an element in the neighborhood of the center of the group is akin to the definition of the azimuthal angle on the sphere close to the north or south pole.
 With $\Omega_{D}$ ill defined, the transformed fields 
\beas 
  A_{\mu}^{\prime} \left(x\right) &=& \Omega_{\olabbr{D}}\left(x_{\perp}\right) 
  \left[A_{\mu} \left(x\right)+
  \frac{1}{ig}\partial _{\mu}\right]\Omega_{\olabbr{D}}^{\dagger}\left(x_{\perp}\right) 
\label{am5a} 
\eeas
develop singularities. The most singular piece arises from the inhomogeneous term in the gauge transformation
\beqs 
  s_{\mu}\left(x_{\perp}\right) = \Omega_{\olabbr{D}}\left(x_{\perp}\right) 
  \frac{1}{ig}\partial _{\mu}\Omega_{\olabbr{D}}^{\dagger}\left(x_{\perp}\right). 
\label{am5b} 
\eeqs
For a given $a_{3}(x_{\perp})$  {  with orientation described by polar }$\theta(x_{\perp})$ { and azimuthal angles }$\varphi(x_{\perp})$  {  in color space, the matrix diagonalizing} $a_{3}(x_{\perp})$   { can be represented as}
\beqs 
  \Omega_{\olabbr{D}} =
  \left (\begin {array}{cc}
      e^{i\varphi } \cos(\theta/2) &
      \sin(\theta/2) \\\noalign{\medskip}
      - \sin(\theta /2) &
      e^{-i\varphi } \cos(\theta /2)
    \end {array} \right) 
\eeqs
and therefore the nature of the singularities can be investigated in detail.  
The condition for the Polyakov loop to be in the center of the group, i.e.\ at a definite}  { point on}\,$ S^{3}$\,  \big(Eq.\ (\ref{mop1})\big), determines in general  uniquely the corresponding position  in $\mathbb{R}^{3}$ and therefore the singularities form  { world-lines} in} 4-dimensional space-time.
The singularities are ``monopoles'' with topologically quantized charges. { $\Omega_{\olabbr{D}}$} { is determined only up to a gauge transformation}
$$\Omega_{\olabbr{D}}(x_{\perp})\rightarrow e^{i\tau^{3}\psi(x_{\perp})}\Omega_{\olabbr{D}}(x_{\perp})$$
  and is therefore an element of $SU(2)/U(1)$. The  mapping of a sphere\,$S^2$\,    around the monopole in $x_{\perp}$   space to $SU(2)/U(1)$ is topologically non-trivial 
$\pi_{2}[SU(2)/U(1)]=\mathbb{Z}$ \big(Eq.\ (\ref{th2su})\big). This argument is familiar to us from the discussion of the 't Hooft-Polyakov monopole \big(cf.\ Eqs.~(\ref{mgch}) and (\ref{wnu2})\big). Also here we identify the winding number associated with this mapping as the magnetic charge of the monopole.
\newpage
{\em Properties of Singular Fields}
\begin{itemize}
\item  {Dirac monopoles, extended to include color, constitute the  simplest  examples of singular fields (Euclidean} $x_{\perp}={\bf x}$)
\bea
\label{dimo}
  {\bf A}
 & \sim &\frac{m}{2gr}\Bigg\{\frac{1+ \cos\theta}{ \sin\theta}\,
  \hat{\mbox{\boldmath$\varphi $}}\tau^{3} + [
  (\hat{\mbox{\boldmath$\varphi $}}+i\hat{\mbox{\boldmath$\theta$}})
  e^{- i \varphi}(\tau^{1}-i\tau^{2})+\mbox{h.c.}]\Bigg\}.
\label{am18}
\eea
In addition to the pole at $r=0$, the fields contain a string in 3-space (here chosen along $\theta =0$) and therefore a sheet-like singularity in 4-space which emanates from the monopole word-line.
\item {Monopoles are characterized by} two charges, the 
``north-south'' charge for the two center elements of SU(2) \big(Eq.~(\ref{mop1})\big), 
\begin{equation}
  \label{mop2}
 z=\pm1 \, ,
\end{equation}
and the quantized strength of the singularity 
\begin{equation}
  \label{mop3}
  m=\pm 1,\pm2, .... \; .
\end{equation}
\item The topological charge (\ref{TPC}) is determined by the two charges of the monopoles present in a given configuration \cite{QURS99,FOTW99,JALE98}
  \begin{equation}
    \label{toma}
\nu=\frac{1}{2} \sum_{i} m_{i} z_{i}.    
  \end{equation}
Thus, after elimination of the redundant variables, the topological charge resides exclusively in singular field configurations.  
\item The action of singular fields is in general finite and  can be arbitrarily small for $\nu = 0$. The singularities in the  abelian and non-abelian  contributions to the field strength cancel since by gauge transformations singularities in gauge covariant quantities cannot be generated. 
\end{itemize}
\subsection{Monopoles and Instantons}
By the gauge choice, i.e.~by the diagonalization of the Polyakov loop by $\Omega_D$ in (\ref{GFT}), monopoles appear; instantons, which in (singular) Lorentz gauge have a point singularity \big(Eq.\ (\ref{singins})\big) at the center of the instanton, must possess according to the relation (\ref{toma}) at least two monopoles with  associated strings \big(cf.\ (\ref{dimo})\big). Thus, in axial gauge, an instanton field becomes singular on world lines and world sheets. To illustrate  the connection between topological charges and monopole charges (\ref{toma}), we consider the singularity content of instantons in axial gauge \cite{LENT04} and  calculate the Polyakov loop of instantons.   To this end, the  generalization of the instantons (\ref{singins}) to finite temperature (or extension) is needed. The so-called ``calorons''   are known explicitly \cite{HASH78}}
\beq
\label{pm7}
A_{\mu} =
\frac{1}{g}\bar{\eta}_{\mu\nu}\nabla_{\nu}\ln\left\{1+\gamma\frac{(\sinh u)/u}{\cosh u-\cos v}\right\}
\eeq
 { where} 
$$  u= 2\pi |{\bf x}_{\perp}-{\bf x}_{\perp}^{0}|/L\, ,\quad 
 v= 2\pi x_{3}/L \, ,\quad \gamma= 2(\pi \rho/L)^{2}. 
$$
The topological charge and the action are independent of the extension,
 { $$\nu = 1 ,\quad S =\frac{8\pi^2}{g^2} . $$} 
 The Polyakov loops can be evaluated in closed form 
\beq
\label{pm9}
P({\bf x}) = \exp\left\{i \pi \frac{\left({\bf x}_{\perp}-{\bf x}_{\perp}^{0}\right){\bf \tau}}{|{\bf x}_{\perp}-{\bf x}_{\perp}^{0}|} \chi(u)\right\}\, ,
\eeq
 { with}  
$$
\label{pm10}
\chi(u) = 1-\frac{(1-\gamma/u^{2})\sinh u+\gamma/u \cosh(u)}{\sqrt{(\cosh u+\gamma/u \sinh u)^{2}-1}}\, .
$$

 \begin{figure}
\hspace{3.5cm}\epsfig{file=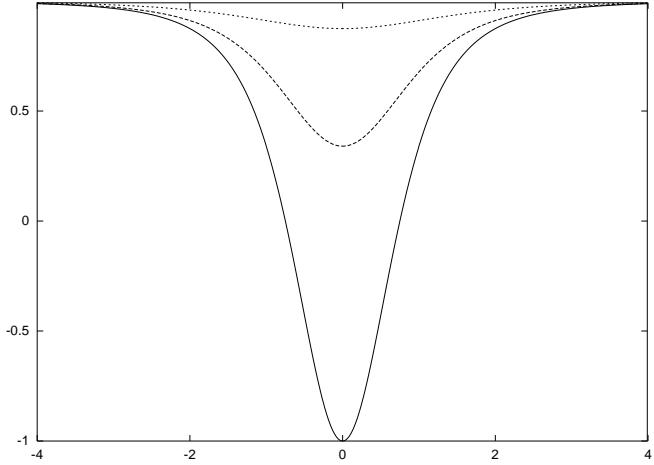, width=.4\linewidth,angle=-90}
\caption{ Polyakov loop (\ref{pm9}) of an instanton (\ref{pm7}) of ``size'' $\gamma=1$ as a function of time $t= 2\pi x_0/L$ for minimal distance to the center $ 2\pi x_1/L=0$ (solid line), $L= 1$ (dashed line), $L=2$ (dotted line),\; $x_2=0$ }
\label{polins}
\end{figure}
\vspace{0.cm}
\vskip.1cm 
As  Figure \ref{polins} illustrates, instantons contain a $z=-1$ monopole at the center and a $z=1$ monopole at infinity; these monopoles   carry the topological charge of the instanton. Furthermore, tunneling processes represented by  instantons connect field configurations of different winding number \big(cf.\ Eq.\ (\ref{WN})\big) but with the same value for the Polyakov loop. In the course of the tunneling, the Polyakov loop of the instanton may pass through or get close to the center element $z=-1$, it however always returns to its original value  $z=+1$. Thus, instanton ensembles in the dilute gas limit are not center symmetric and therefore cannot give rise to confinement. One cannot rule out that the $z=-1$ values of the Polyakov loop are  encountered more and more frequently with increasing instanton  density. In this way, a center-symmetric ensemble may finally be reached in the high-density limit. This however seems to require a fine tuning of instanton size and the average distance between instantons.   
\subsection{Elements of Monopole Dynamics}
\vskip .1cm
In axial gauge, instantons are composed of two monopoles. An instanton gas \big(Eq.\ (\ref{IE})\big) of finite density $ n_{\olabbr{I}}$  therefore contains field configurations with infinitely many monopoles.  The instanton  density in 4-space can be converted approximately  to a monopole density in 3-space  \cite{JALE98}  
$$  n_{\olabbr{M}} \sim \left(L  n_{\olabbr{I}}\rho\right)^{3/2}
  \quad , \quad \rho \ll L ,$$
\beqs
\label{pm12}
  n_{\olabbr{M}} \sim  L  n_{\olabbr{I}} \quad , \quad \rho \ge L \, . 
\eeqs
With increasing extension or equivalently decreasing temperature, the monopole density diverges for constant instanton density. Nevertheless,  the action density of an instanton gas remains finite. This is in accordance with our expectation that production of monopoles is not necessarily suppressed by large values of the action. Furthermore, a finite or possibly even divergent density of monopoles as in the case of the dilute instanton gas does not imply  confinement. \\ 
Beyond the generation of  monopoles via instantons, the system has the
additional option of producing one type ($z=+1$ or $z=-1$) of poles
and corresponding antipoles only. No topological charge is associated with such singular fields and their occurrence is not limited by the instanton bound \big(Eqs.\ (\ref{BB}) and (\ref{SD})\big) on the action as is the case for a pair of monopoles of  opposite $z$-charge.  Thus, entropy  favors the production of such configurations.
The entropy argument also applies in the plasma phase.  For purely kinematical reasons, a decrease in the monopole density must be expected as the above estimates within the instanton model show. This decrease is counteracted by the enhanced probability to produce monopoles when,  with decreasing $L$, the Polyakov loop approaches more and more the center of the group, as has been discussed above (cf.~left part of Fig.~\ref{fi12}). A finite  density of singular fields is likely to be present also in the deconfined phase. In order for this to be compatible with the partially perturbative nature of the plasma phase and with dimensional reduction to QCD$_{2+1}$,   poles and antipoles may have to  be 
strongly correlated with each other and to form effectively a gas of dipoles.\\
Since entropy favors proliferate production of monopoles and monopoles may be produced with only a small increase in the total action, the coupling of the monopoles to the  quantum fluctuations  must ultimately prevent unlimited increase in the number of monopoles. A systematic study of the relevant dynamics has not been carried out. Monopoles are not solutions to classical field equations. Therefore,  singular fields are mixed with quantum fluctuations even on the level of bilinear terms in the action. Nevertheless, two mechanisms can be identified which might limit the production of monopoles.
\begin{itemize} 
\item  The 4-gluon vertex couples pairs of monopoles to charged and neutral gluons  and can generate masses  for all the color components of the gauge fields. A simple estimate yields 
 \beqs
\label{pm5}
  \delta m^{2} =  -\frac{\pi}{ V}\sum_{{\scriptstyle i,j=1\atop\scriptstyle i<j}}^{N}
  m_{i}m_{j}|{\bf x}_{\perp i}-{\bf x}_{\perp j}| 
\eeqs
with  the monopole charges $m_i$ and  positions ${\bf x}_{\perp\,i}$. 
If operative also in the deconfined phase, this  mechanism would give rise to a  magnetic gluon mass. 
\item In general, fluctuations around singular fields generate an infinite action. Finite values of the action result only if the fluctuations\, $\delta \phi,\delta A^{3}$\,\,  satisfy the boundary conditions,    
$$
 \delta\phi(x)\, e^{2i\varphi(x_{\perp})} \quad \mbox{continuous along the strings}\, ,$$
$$ \delta\phi(x)\Bigg|_{\mbox{at pole}} = \delta A^{3}\Bigg|_{\mbox{at pole}} = 0\, .$$
For a finite monopole density, long wave-length fluctuations cannot simultaneously satisfy boundary conditions related to monopoles or strings which are close to each other. One therefore might suspect  quantum fluctuations with wavelengths 
\beqs
  \lambda \ge \lambda_{\mbox{max}} = n_{\olabbr{M}}^{-1/3} 
\eeqs
 {to be  suppressed.} 
\end{itemize}
We note that both mechanisms would also suppress the propagators of the quantum fluctuations in the infrared. Thereby, the decrease in the string constant by coupling Polyakov loops to charged gluons could be alleviated if not cured. 
\subsection{Monopoles in Diagonalization Gauges}
In  axial gauge, monopoles appear in the gauge fixing procedure \big(Eq.\ (\ref{GFT})\big) as defects in the diagonalization of the Polyakov loops. Although the choice was motivated by the distinguished role of the Polyakov-loop variables as order parameters, formally one may choose any quantity $\phi$ which, if local, transforms under gauge transformations $U$ as
$$\phi \to U\phi U^{\dagger}\, ,$$
where $\phi$ could be either an element of the algebra or of the  group. In analogy to (\ref{axgc}), the gauge condition can be written as 
\beq
\label{diga}
f[\phi] = \phi - \varphi\frac{\tau^{3}}{2}\, ,\eeq
with arbitrary $\varphi$ to be integrated in the generating functional. 
A simple illustrative example is \cite{THOO81}
\beq
\label{f12}
\phi=F_{12} .
\eeq
The analysis of the defects and  the resulting properties of the monopoles can be taken over with minor modifications from the procedure described above. Defects occur if
$$\phi=0$$
(or $\phi=\pm 1$ for group elements). The condition for the defect is gauge invariant. Generically, the three defect conditions determine for a given gauge field the  world-lines of the monopoles generated by the gauge condition (\ref{diga}). The quantization of the monopole charge is  once more derived from the topological identity (\ref{th2su}) which characterizes the mapping of a (small) sphere in the space transverse to the monopole world-line and enclosing the defect. The coset space, appears as above since the gauge condition leaves a $U(1)$ gauge symmetry related to the rotations around the direction of $\phi$ unspecified. With $\phi$ being an element of the Lie algebra, only one sort of monopoles appears. The characterization as $z=\pm 1$ monopoles requires $\phi$ to be an element of the group. As a consequence, the generalization of the connection between monopoles and topological charges is not straightforward. It has been established \cite{JAHN00} with the help of the Hopf-invariant \big(cf.\ Eq.\ (\ref{hoin})\big) and its generalization. \\
It will not have escaped the attention of the reader that the description of Yang-Mills theories in  diagonalization gauges is  almost in one to one correspondence to the description of the non-abelian Higgs model in the unitary gauge. In particular, the gauge condition (\ref{diga}) is essentially identical to the unitary gauge condition (\ref{ugc}). 
However, the physics content of these gauge choices is very different. The unitary gauge is appropriate if the Higgs potential forces the Higgs field to assume (classically) a value different from zero. In the classical limit, no monopoles related to the vanishing of the Higgs field appear in unitary gauge and one might expect that quantum fluctuations will not change this qualitatively. Associated with the  unspecified $U(1)$ are the photons in the Georgi-Glashow model.  In pure Yang-Mills theory, gauge conditions like (\ref{f12}) are totally inappropriate in the classical limit, where vanishing action produces defects filling the whole space. Therefore, in such gauges a physically meaningful condensate of magnetic monopoles signaling confinement  can arise only if quantum fluctuations change  the situation radically. Furthermore, the unspecified $U(1)$ does not indicate the presence of massless vector particles, it rather reflects an incomplete gauge fixing.   Other diagonalization gauges may be less singular in the classical limit, like the axial gauge. However,  independent  of the gauge choice, defects in the gauge condition have not been related   convincingly to physical properties of the system. They exist as as coordinate singularities and their physical significance  remains enigmatic. 
\section{Conclusions}
In these lecture notes I have described  the instanton, the 't Hooft-Polyakov monopole, and the Nielsen-Olesen vortex which are  the three paradigms of topological objects appearing in gauge theories. They differ from each other in the dimensionality of the core of these objects, i.e.~in the dimension of the submanifold of space-time on which gauge and/or matter fields  are singular. This dimension is determined by the topological properties of the spaces in which these fields take their values and dictates to a large extent the dynamical role these objects can play. 't Hooft-Polyakov monopoles are singular along a world-line and therefore describe particles. I have presented the strong theoretical evidence based on topological arguments that these particles  have been produced  most likely in phase transitions of the early universe. These relics of the big bang have not been and most likely cannot be observed. Their abundance has been diluted in the inflationary phase. Nielsen-Olesen vortices are singular on lines in space or equivalently on world-sheets in space-time. Under suitable conditions such objects occur in Type II superconductors. They give rise to various phases and a wealth of phenomena  in superconducting materials. Instantons become singular on a point in Euclidean 4-space and they therefore  represent tunneling processes. In comparison to  monopoles and vortices, the manifestation of these objects is only indirect. They cannot be observed but  are supposed to give rise to non-perturbative properties of the corresponding quantum mechanical ground state. \vskip .1cm
Despite their difference in dimensionality, these topological objects have many properties in common. They are all solutions of the non-linear field equations of gauge theories. They owe their existence and topological stability to vacuum degeneracy, i.e.~the presence of a continuous or discrete set of  distinct solutions with minimal energy. They can be classified according to a charge, which  is quantized as a consequence of the non-trivial topology. Their non-trivial  properties  leave a topological imprint on fermionic or bosonic degrees of freedom  when coupled to these objects. Among the topological excitations of a given type, a certain class is singled out by their  energy  determined by the quantized charge.
\vskip .1cm In these lecture notes I also have  described  efforts in the topological analysis of QCD. A complete picture about the role of topologically non-trivial field configurations has not yet emerged from such studies. 
With regard to the  breakdown of chiral symmetry, the formation of quark condensates and  other chiral properties, these efforts have met with success. The relation between the topological charge and fermionic properties appears to be at the origin of these phenomena. The instanton model incorporates  this connection explicitly by reducing the  quark and gluon degrees of freedom to instantons and quark zero modes generated by the topological charge of the instantons. However, a generally accepted topological explanation of confinement has not been achieved nor have  field configurations been identified   which are relevant for confinement. The negative outcome of such investigations may imply that, unlike mass generation by the Higgs mechanism, confinement does not have an explanation within the context of classical field theory. Such a conclusion  is supported  by the simple explanation of confinement in the strong coupling limit of lattice gauge theory. In this limit, confinement results  from the kinetic energy \cite{KOSU75} of the compact link variables. The potential energy generated by the magnetic field, which has been the crucial ingredient in the construction of the Nielsen-Olesen Vortex and the 't Hooft-Polyakov monopole, is negligible in this limit. It is no accident that,  as we have seen,  Polyakov-loop variables, which as group elements are compact, also exhibit confinement-like behavior.   \\ Apart from  instantons as the genuine topological objects,  Yang-Mills theories exhibit non-trivial topological properties related to the center of the gauge group. The center symmetry as a residual gauge symmetry offers the possibility to formulate confinement as a symmetry property and to characterize confined and deconfined phases. The role of the center vortices (gauge transformations which are singular on a two dimensional space-time sheet) remains to be clarified. The existence of  obstructions in imposing  gauge conditions is another non-trivial property of non-abelian gauge theories which might be related to confinement. I have described the appearance of monopoles as the results of such obstructions in so-called diagonalization or abelian gauges. These singular fields can be characterized by  topological methods and, on a formal level, are akin to the 't-Hooft Polyakov monopole. I have described the difficulties in developing a viable framework for formulating their dynamics which is supposed to yield confinement via a dual Meissner effect.
\section*{Acknowledgment}
I thank M. Thies, L.v. Smekal, and J. Pawlowski for discussions on the various subjects of these notes. I'm indebted to J. J\"ackel and F. Steffen for their meticulous reading of the manuscript and for their many valuable suggestions for improvement.   


\begin{thebibliography}{99}
\bibitem{GAUS833} C.~F. Gau\ss\ , Werke, Vol.~5, G\"ottingen, K\"onigliche Gesellschaft der Wissenschaften 1867, p. 605
\bibitem{DFN285} B.~A. Dubrovin, A.~T. Fomenko, and S.~P. Novikov, Modern Geometry, Part II. Springer Verlag 1985
\bibitem{FRAN97}T. Frankel, The Geometry of Physics, Cambridge University Press, 1997
\bibitem{TAIT98} P.G. Tait, Collected Scientific Papers, 2 Vols., Cambridge University Press, 1898/1900
\bibitem{MOFF69}H.~K. Moffat, The Degree of Knottedness of Tangled Vortex Lines, 
{\em J. Fluid Mech.} {\bf 35}, 117  (1969)
\bibitem{DIRA31}P.~A.~M. Dirac, Quantised Singularities in the Electromagnetic Field,
{\em Proc. Roy. Soc. A} {\bf 133}, 60  (1931) 
\bibitem{YAMI54} C.~N. Yang and R.~L. Mills, Conservation of Isotopic Spin and Isotopic Gauge Invariance 
Phys. Rev. {\bf 96}, 191 (1954) 
\bibitem{NIOL73} N.~K. Nielsen and P. Olesen, Vortex-Line Models for Dual Strings, 
{\em Nucl. Phys. B} {\bf 61}, 45 (1973)
\bibitem{DEGE66} P.~G. de Gennes, Superconductivity of Metals and Alloys, W.~A. Benjamin 1966 
\bibitem{TINK75} M. Tinkham, Introduction to Superconductivity, McGraw-Hill 1975
\bibitem{BFGL94} G. Blatter, M.~V. Feigel'man, V.~B. Geshkenbein, A.~I. Larkin, and V.~M. Minokur,
 {\em Rev.  Mod. Phys.} {\bf 66}, 1125  (1994)
\bibitem{NELS02} D. Nelson, Defects and Geometry in Condensed Matter Physics, Cambridge University Press, 2002
\bibitem{POFC95} C.~P. Poole, Jr., H.~A. Farach and R.~J. Creswick, Superconductivity, Academic Press, 1995
\bibitem{BOGO76} E.~B. Bogomol'nyi, The Stability of Classical Solutions,
{\em Sov. J. Nucl. Phys.} {\bf 24}, 449 (1976) 
\bibitem{JARO81} R. Jackiw and P. Rossi, Zero Modes of the Vortex-Fermion System,
 {\em Nucl. Phys.} B {\bf 252}, 343 (1991)
\bibitem{WEIN81} E. Weinberg, Index Calculations for the Fermion-Vortex System,
 Phys. Rev. D {\bf 24}, 2669 (1981)
\bibitem{NASE83} C. Nash and S. Sen, Topology and Geometry for Physicists, Academic Press 1983
\bibitem{NAKA90} M. Nakahara, Geometry, Topology and Physics, Adam Hilger 1990
\bibitem{MUNK2000} J.~R. Munkres, Topology, Prentice Hall 2000
\bibitem{JAHN00} O. Jahn, Instantons and Monopoles in General Abelian Gauges,
{\em J. Phys.} {\bf A33}, 2997 (2000) 
\bibitem{GAGR99} T.~W. Gamelin and R.~E. Greene, Introduction to Topology, Dover 1999
\bibitem{ARKH98} V.~I. Arnold, B.~A. Khesin, Topological Methods in Hydrodynamics, Springer 1998
\bibitem{THOU98} D.~J. Thouless, Topological Quantum Numbers in Nonrelativistic Physics, World Scientific  1998 
\bibitem{STEE51} N. Steenrod, The Topology of Fiber Bundels, Princeton University Press 1951
\bibitem{MORA92} G. Morandi, The Role of Topology in Classical and Quantum Physics, Springer 1992
\bibitem{MILL72}W. Miller, Jr., Symmetry Groups and Their Applications, Academic Press 1972
\bibitem{MERM79} N.~D. Mermin, The Topological Theory of Defects in Ordered Media,
{\em Rev. Mod. Phys}. {\bf  51}, 591  (1979)
\bibitem{BOEF01} V.~P. Mineev, Topological Objects in Nematic Liquid Crystals , Appendix A, 
in: V.~G. Boltyanskii and V.~A. Efremovich,Intuitive Combinatorial Topology, Springer 2001
\bibitem{CHAN92} S. Chandrarsekhar, Liquid Crystals, Cambridge University Press 1992
\bibitem{DGPR93} P.~G. de Gennes and J. Prost, The Physics of Liquid Crystals, Clarendon Press  1993
\bibitem{PSLW97} P. Poulin, H. Stark, T.~C. Lubensky and D.A. Weisz, Novel Colloidal Interactions in Anisotropic 
Fluids, Science {\bf 275} 1770 (1997)
\bibitem{GEGL72} H. Georgi and S. Glashow,  Unified Weak and Electromagnetic Interactions without Neutral Currents,
{\em Phys. Rev. Lett.} {\bf 28}, 1494 (1972)
\bibitem{WEYL} H. Weyl, Gruppentheorie und Quantenmechanik, Hirzel Verlag 1928.
\bibitem{JACK} R. Jackiw, Introduction to the Yang-Mills Quantum Theory,
 {\em Rev. Mod. Phys.} {\bf 52}, 661 (1980)
\bibitem{LeNT94} F. Lenz, H.~W.~L. Naus and M. Thies, QCD in the Axial Gauge Representation,
{\em Ann. Phys.} {\bf 233}, 317 (1994) 
\bibitem{LEWO01} F. Lenz and S. W\"orlen, Compact variables and Singular Fields in QCD,
in: at the frontier of Particle Physics, handbook of QCD  edited by M. Shifman, Vol. 2, p. 762, World Scientific 2001 
\bibitem{THOO74} G.'t Hooft, Magnetic Monopoles in Unified Gauge Models, {\em Nucl. Phys.} B {\bf 79}, 276 (1974)
\bibitem{POLY74}A.M. Polyakov, Particle Spectrum in Quantum Field Theory, {\em JETP Lett.} {\bf 20}, 194 (1974);  Isometric States in Quantum Fields, {\em JETP Lett.} {\bf 41}, 988 (1975)
\bibitem{LeNT940} F. Lenz, H.~W.~L. Naus, K. Ohta, and M. Thies, Quantum Mechanics of Gauge Fixing,
{\em Ann. Phys.} {\bf 233}, 17 (1994) 
\bibitem{VISH94} A. Vilenkin and E.~P.~S. Shellard, Cosmic Strings and Other Topological Defects, Cambridge University Press 1994
\bibitem{SGCS97} S.L.Sondhi, S.~M. Girvin, J.~P. Carini, and D. Shahar, Continuous Quantum Phase Transitions, {\em Rev.Mod.Phys.}{\bf  69}  315, (1997)
\bibitem{RAJA82} R. Rajaraman, Solitons and Instantons, North Holland 1982
\bibitem{JUZE75} B. Julia and A. Zee, Poles with Both Electric and Magnetic Charges in Nonabelian Gauge Theory,
  {\em Phys. Rev.} D {\bf 11}, 2227 (1975)  
\bibitem{TWGS76} E. Tomboulis and G. Woo, Soliton Quantization in Gauge Theories,
 {\em Nucl. Phys.} B {\bf 107}, 221 (1976);   
 J.~L. Gervais, B. Sakita and S. Wadia, The Surface Term in Gauge Theories, 
 {\em Phys. Lett}. B {\bf 63 B}, 55 (1999)  
\bibitem{CALL78} C. Callias,  Index Theorems on Open Spaces,
{\em Commun. Mat. Phys.} {\bf 62}, 213 (1978)
\bibitem{JARE76a} R. Jackiw and C. Rebbi, Solitons with Fermion Number $1/2$, 
{\em Phys. Rev.} D {\bf 13}, 3398 (1976)  
\bibitem{JARE76b} R. Jackiw and C. Rebbi,  Spin from Isospin in Gauge Theory,
 {\em Phys. Rev. Lett.}  {\bf 36}, 1116 (1976)
\bibitem{HATH76} P. Hasenfratz and G.~'t Hooft, Fermion-Boson Puzzle in a  Gauge Theory,
{\em Phys. Rev. Lett.}  {\bf 36}, 1119 (1976)  
\bibitem{KOTU90} E.~W. Kolb and M.~S. Turner, The Early Universe, Addison-Wesley 1990
\bibitem{PEAC99} J.~A. Peacock, Cosmological Physics, Cambridge University Press 1999
\bibitem{GRIB78} V.~N. Gribov,  Quantization of Non-Abelian Gauge Theories,
{\em Nucl. Phys.} B {\bf 139}, 1 (1978)
\bibitem{SING78} I.~M. Singer, Some Remarks on the Gribov Ambiguity,
{\em Comm. Math. Phys.} {\bf 60}, 7 (1978)  
\bibitem{WUYA75} T.~T. Wu and C.~N. Yang, Concept of Non-Integrable Phase Factors and Global Formulations of Gauge Fields, {\em Phys. Rev.} D {\bf 12}, 3845 (1975)   
\bibitem{BELA75} A.~A. Belavin, A.~M. Polyakov, A.~S. Schwartz and Yu.~S. Tyupkin, Pseudoparticle solutions of the Yang-Mills equations, {\em Phys. Lett}. B {\bf 59}, 85 (1975)
\bibitem{BJOR} J.~D. Bjorken, in: Lectures on Lepton Nucleon Scattering and Quantum Chromodynamics, W. Atwood {\em et al.}, Birkh\"{a}user 1982
\bibitem{JACK85} R. Jackiw, Topological Investigations  of Quantized Gauge theories, in: Current Algebra and Anomalies, edt. by S. Treiman et al., Princeton University Press, 1985
\bibitem{SCHW93} A.~S. Schwartz, Quantum Field Theory and Topology, Springer 1993
\bibitem{GRSW87} M.~B. Green, J.~H. Schwarz and E. Witten, Superstring Theory, Vol.~2, Cambridge University Press 1987
\bibitem{THOO76} G.~'t Hooft, Computation of the Quantum Effects Due to a Four Dimensional Quasiparticle,
 {\em Phys. Rev.} D {\bf 14}, 3432 (1976) 
\bibitem{ZINN02} J. Zinn-Justin, Chiral Anomalies and Topology, {\bf hep-th/0201220} 
\bibitem{ESPO98} G. Esposito, Dirac Operators and Spectral Geometry, Cambridge University Press 1998
\bibitem{SCSH96} T. Sch\"afer and E.~V. Shuryak, Instantons in QCD,
{\em Rev. Mod. Phys}. {\bf 70}, 323 (1998) 
\bibitem{CADG78} C.~G. Callan, R.~F. Dashen and D.~J. Gross,  Toward a Theory of Strong Interactions,
{\em Phys. Rev.} D {\bf 17}, 2717 (1978)  
\bibitem{DAFF76} V.~de Alfaro, S.~Fubini and G.~Furlan,
A New Classical Solution Of The Yang-Mills Field Equations,
{\em Phys. Lett.} B {\bf 65}, 163 (1976).
\bibitem{LENT04} F. Lenz, J.~W. Negele and M. Thies, Confinement from Merons, {\bf hep-th/0306105} to appear in {\em Phys. Rev. D}
\bibitem{MOTS92} H.~K. Moffat and A. Tsinober, Helicity in Laminar and Turbulent Flow,
{\em Ann. Rev. Fluid Mech.} {\bf 24}281,  (1992) 
\bibitem{DAVI01} P.~A. Davidson, An Introduction to Magnetohydrodynamics, Cambridge University Press, 2001 
\bibitem{WITT88} E. Witten, Some Geometrical Applications of Quantum Field Theory,  
in *Swansea 1988, Proceedings of the IX th International Congr. on Mathematical Physics, p.77
\bibitem{KAUF91} L.~H. Kauffman, Knots and Physics, World Scientific 1991
\bibitem{POLY88} A.~M. Polyakov, Fermi-Bose Transmutation Induced by Gauge Fields,
{\em Mod. Phys. Lett.} {\bf A 3}, 325 (1988)
\bibitem{SVET86} B. Svetitsky,  Symmetry Aspects of  Finite Temperature Confinement Transitions,
{\em Phys. Rep.} {\bf 132}, 1 (1986)
\bibitem{Toms} D.~J. Toms, Casimir Effect and Topological Mass,
{\em Phys. Rev. D} {\bf 21}, 928 (1980) 
\bibitem{LETH98} F. Lenz and M. Thies, Polyakov Loop Dynamics in the Center Symmetric Phase,
{\em Ann. Phys.} {\bf 268}, 308 (1998) 
\bibitem{KAPU89} J.~I. Kapusta, Finite-temperature field theory,
Cambridge University Press 1989
\bibitem{LeNT941} F. Lenz, H.~W.~L. Naus, K. Ohta, and M. Thies, Zero Modes and Displacement Symmetry in Electrodynamics,
{\em Ann. Phys.} {\bf 233}, 51 (1994) 
\bibitem{LNOT00} F. Lenz, J.~W. Negele, L. O'Raifeartaigh and M. Thies, Phases and Residual Gauge Symmetries of 
Higgs Models,  {\em Ann. Phys.} {\bf 285}, 25 (2000) 
\bibitem{LEBE96} M. Le Bellac, Thermal field theory, Cambridge
University Press 1996
\bibitem{RELQ99}H. Reinhardt, M. Engelhardt, K. Langfeld, M. Quandt, and A. Sch\"afke, Magnetic Monopoles,
 Center Vortices, Confinement and Topology of Gauge Fields,
{\bf hep-th/ 9911145} 
\bibitem{GREE03} J. Greensite, The Confinement Problem in Lattice Gauge Theory,
{\bf hep-lat/ 0301023}  
\bibitem{DVSC86} H.~J. de Vega and F.~A. Schaposnik, Electrically Charged Vortices in Non-Abelian Gauge Theories,
 {\em Phys. Rev. Lett.} {\bf 56},  2564 (1986) 
\bibitem{THOO78}G.~'t Hooft, On the Phase Transition Towards Permanent Quark Confinement,
{\em Nucl. Phys.} B {\bf 138}, 1 (1978) 
\bibitem{SAMU79} S. Samuel, Topological Symmetry Breakdown and Quark Confinement,
{\em Nucl. Phys.} B {\bf 154}, 62 (1979) 
\bibitem{KOVN01} A. Kovner, Confinement, $Z_N$ Symmetry and Low-Energy Effective Theory of Gluodynamicsagnetic,
 in: at the frontier of Particle Physics, handbook of QCD  edited by M.~ Shifman, Vol. 3, 
p. 1778,  World Scientific 2001
\bibitem{FIHK93} J. Fingberg, U. Heller, and F. Karsch,  Scaling and Asymptotic Scaling in the SU(2) Gauge Theory,
{\em Nucl. Phys.} B {\bf 392}, 493 (1993) 
\bibitem{GGHK94} B. Grossman, S. Gupta, U.~M. Heller, and F. Karsch, Glueball-Like Screening Masses in Pure SU(3) at Finite Temperatures,
{\em Nucl. Phys.} B {\bf 417}, 289 (1994)  
\bibitem{ISSM02} M. Ishii, H. Suganuma and H. Matsufuru, Scalar Glueball Mass Reduction at Finite Temperature in $SU(3)$ 
Anisotropic Lattice QCD, {\em Phys. Rev.} D {\bf 66}, 014507 (2002);
Glueball Properties  at Finite Temperature in $SU(3)$ Anisotropic Lattice QCD,
{\em Phys. Rev.} D {\bf 66}, 094506 (2002)  
\bibitem{RAHK99} S. Rastogi, G.~W. H\"ohne and A. Keller, 
Unusual Pressure-Induced Phase Behavior in Crystalline Poly(4-methylpenthene-1): Calorimetric and 
Spectroscopic Results and Further Implications,
{\em Macromolecules} {\bf 32} 8897 (1999) 
\bibitem{ZELD01} N. Avraham, B. Kayhkovich, Y. Myasoedov, M. Rappaport, H. Shtrikman, D.~E. Feldman, T. Tamegai, 
P.~H. Kes, Ming Li, M. Konczykowski, Kees van der Beek, and Eli Zeldov, 'Inverse' Melting of a Vortex Lattice,
{\em Nature} {\bf 411}, 451, (2001) 
\bibitem{LEMT95} F. Lenz, E.~J. Moniz and M. Thies, Signatures of Confinement in Axial Gauge QCD,
{\em Ann. Phys.} {\bf 242}, 429 (1995) 
\bibitem{PESC95} M.~E. Peskin and D.~V. Schroeder, An Introduction to Quantum Field Theory, Addison-Wesley Publishing Company, 1995
\bibitem{Reisz} T. Reisz, Realization of Dimensional Reduction at High Temperature,
{\em Z. Phys.} C {\bf 53}, 169 (1992) 
\bibitem{EKLT98} V.~L. Eletsky, A.~C. Kalloniatis, F. Lenz, and M. Thies, Magnetic and Thermodynamic Stability 
of $SU(2)$ Yang-Mills Theory, {\em Phys. Rev.} D {\bf 57}, 5010 (1998) 
\bibitem{KALP00}F. Karsch, E. Laermann, and A. Peikert, The Pressure  in 2, 2~+~1 and 3 Flavor QCD,
 {\em Phys. Lett}. B {\bf 478}, 447 (2000)  
\bibitem{ENKR95} J. Engels, F. Karsch and K. Redlich,  Scaling Properties of the Energy Density in $SU(2)$ Lattice Gauge Theory,
{\em Nucl. Phys}. B{\bf 435}, 295 (1995)
\bibitem{SEWI94} N. Seiberg, E. Witten,  Monopole Condensation, and Confinement in N~=~2 Supersymmetric QCD, 
 {\em Nucl. Phys.} B {\bf 426}, 19 (1994);  Monopoles, Duality and Chiral Symmetry Breaking 
 in N~=~2 supersymmetric QCD {\em Nucl. Phys.}B {\bf 431},  484 (1995) 
\bibitem{QURS99} M. Quandt, H. Reinhardt and A. Sch\"afke, Magnetic Monopoles and Topology of Yang-Mills Theory 
in Polyakov Gauge, {\em Phys. Lett}. B {\bf 446}, 290 (1999)  
\bibitem{FOTW99} C. Ford, T. Tok and A. Wipf, $SU(N)$ Gauge Theories in Polyakov Gauge on the Torus,
{\em Phys. Lett}. B {\bf 456}, 155 (1999) 
\bibitem{JALE98} O. Jahn and F. Lenz, Structure and Dynamics of Monopoles in Axial Gauge QCD,
{\em Phys. Rev}. D {\bf 58}, 85006 (1998)  
\bibitem{HASH78} B.~J. Harrington and H.~K. Shepard,  Periodic Euclidean Solutions and the Finite-Temperature Yang-Mills Gas,
  Phys. Rev. D {\bf 17}, 2122 (1978)
\bibitem{THOO81} G.~'t Hooft, Topology of the Gauge Condition and New Confinement Phases in Non-Abelian Gauge Theories,
{\em Nucl. Phys.} B {\bf 190}, 455 (1981) 
\bibitem{KOSU75} J. Kogut and L. Susskind, Hamiltonian Formulation of Wilson's Lattice Gauge Theories,
{\em Phys. Rev.} D {\bf 11}, 395 (1975) 
\end{thebibliography}
\end{document}